\documentclass[fleqn,usenatbib]{mnras}

\usepackage{newtxtext,newtxmath}

\usepackage[T1]{fontenc}

\let\VANthebibliography\thebibliography
\def\thebibliography{\DeclareRobustCommand{\VAN}[3]{##3}\VANthebibliography}

\usepackage{graphicx}	
\usepackage{amsmath}	
\usepackage{multicol}
\usepackage{subcaption}

\newcommand\HI{$\textrm{H}\scriptstyle\mathrm{I}$}

\newcommand{\ve}[1]{\mathbf{q1}}

\newcommand{\be}{\begin{equation}}      
\newcommand{\ee}{\end{equation}}      
      
\newcommand{\bef}{\begin{figure}}      
\newcommand{\eef}{\end{figure}}      
\newcommand{\bea}{\begin{eqnarray}}    
\newcommand{\eea}{\end{eqnarray}}

\newcommand{\av}[1]{\ensuremath{\left\langle q1 \right\rangle}}

\newcommand{\tve}[1]{\tilde{\boldsymbol{q1}}}

\def\bse{\begin{subequations}}
\def\ese{\end{subequations}}

\title[Mapping non-axisymmetric velocity fields of external galaxies]{Mapping non-axisymmetric velocity fields of external galaxies}

\author[F. Sylos Labini et al.]{
Francesco Sylos Labini,$^{1,2}$\thanks{E-mail: sylos@cref.it (FSL)}
Matteo Straccamore,$^{1,3}$
Giordano De Marzo$^{1,3}$
and S\'ebastien Comer\'on$^{4,5}$
\\
$^{1}$Centro Ricerche Enrico Fermi,  I-00184, Roma, Italia\\
$^{2}$INFN Unit\'a Roma 1, Dipartimento di Fisica, Universit\'a di  Roma Sapienza, I-00185 Roma, Italia\\
$^{3}$Dipartimento di Fisica, Sapienza, Universit\'a di  Roma,  I-00185, Roma, Italia\\
$^{4}$Departamento de Astrof\'isica, Universidad de La Laguna, E-38200, La Laguna, Tenerife, Spain\\
$^{5}$Instituto de Astrof\'isica de Canarias E-38205, La Laguna, Tenerife, Spain
}

\date{Accepted XXX. Received YYY; in original form ZZZ}

\pubyear{2023}

\begin{document}
\label{firstpage}
\pagerange{\pageref{firstpage}--\pageref{lastpage}}
\maketitle

\begin{abstract}
Disk galaxies are typically in a stable configuration where matter moves in almost closed circular orbits. However, non-circular motions caused by distortions, warps, lopsidedness, or satellite interactions are common and leave distinct signatures on galaxy velocity maps.   We develop an algorithm  that uses an ordinary least square method  for fitting a non-axisymmetric model to the observed two-dimensional  line-of-sight  velocity map of an external galaxy, which allows for anisotropic non-circular motions. The method approximates a galaxy as a flat disk, which is an appropriate assumption for spiral galaxies within the optical radius where warps are rare. In the outer parts of \HI{}      distributions, which may extend well into the warp region, we use this method in combination with a standard rotating tilted ring model to constrain the range of radii where the flat disk assumption   can be conservatively considered valid. Within this range, the transversal and radial  {  velocity} profiles,  averaged  in rings,  can be directly reconstructed from the velocity map.  
The novelty of the algorithm consists in using arc segments in addition to rings:  in this way spatial velocity anisotropies can be measured in both components, allowing for the reconstruction of 
 {  angularly resolved}  coarse-grained two-dimensional velocity maps.   We applied this algorithm  to 25 disk galaxies from the THINGS sample for which we can provide 2D maps of both velocity components. 
\end{abstract}


\section{Introduction}
\label{intro}

Galactic kinematics provide the observational foundation needed to study galaxy formation, evolution, and dynamics. For the Milky Way, the Gaia mission of the European Space Agency \citep{Gaia_2016} is now providing the most accurate information to date, consisting of very precise determinations of positions, proper motions, radial velocities, and distances for millions of stars, allowing for the reconstruction of their six-dimensional (6D) phase space distribution. {These data have recently revealed large-scale gradients in all velocity components and deviations from axi-symmetry (see, e.g., \cite{Antoja_etal_2018, Lopez-Corredoira_Sylos-Labini_2019, Antoja_etal_2021, Recio-Blanco_etal_2022, Katz_etal_2022, Wang_etal_2023}). 

For the case of external galaxies, the direct reconstruction of the {  full} 6D phase space distribution is not possible and the kinematic maps of the line of sight (LOS) velocity {  ($v_{los}$)} angular distribution must be interpreted under a set of hypotheses.} The simplest assumption is that the system is perfectly axisymmetric and emitters move on circular and stationary orbits around the galactic center. This situation is encoded in the   rotating disk model (RDM --- \cite{Warner_etal_1973}). The residuals between this model and observations retain information about radial and vertical motions and can be studied with appropriate techniques. However, the RDM does not describe the LOS velocity maps of the outskirts of many galaxies (see, e.g.,  discussions in \cite{Jorsater+vanMoorsel_1995,Zurita_etal_2004,Trachternach_etal_2008,Erroz-Ferrer_etal_2015}). Indeed, in the RDM case, the velocity map should be symmetrical with respect to the major axis of the galaxy's projected image. This symmetry is, however, often broken, and it is observed that the kinematic axis (i.e., the axis along which $v_{los}$ has the largest gradient) changes orientation with radius.
  
The rotating tilted-ring model {  (TRM)} is a more sophisticated method for analyzing the kinematics of galaxies {  \citep{Warner_etal_1973,rogstad_etal_1974}  that generalizes the RDM to more complex situations}. It allows for the orientation of the kinematic axis to change with radius, which corresponds to the fact that  the inclination angle $i$ and/or  the position angle  (P.A.) depend on radius. {  For this reason, from a technical point of view, the RDM can be interpreted merely as a  special case of the TRM where the inclination angle and the P.A. are kept constant. Specifically, the TRM} assumes that a galaxy can be approximated by a sequence of rings, where each ring is characterized by its own P.A., inclination angle, and rotational velocity.   The radial velocity is typically assumed to be zero in the analysis: however, as discussed below, this is not a necessary assumption {  and can be relaxed.}  Different software packages, such as {  {\tt rotcur}  \citep{Begeman_1987,Begeman_1989}   {\tt kinemetry}  \citep{Krajnovic_2006},  
{\tt Tirific} \citep{Jozsa_etal_2007,Kamphuis_etal_2015},
{\tt Barolo} \citep{DiTeodoro+Fraternali_2015},
{\tt Gbkfit} \citep{Bekiaris_etal_2016}, 
{\tt 2DBAT}  \citep{Oh_etal_2018},
{\tt MCMC} \citep{Oh_etal_2019},
allow one to fit the velocity field with a single velocity function in a set of annuli whose centers, position angles, and inclinations are allowed, if desired, to vary with radius. In its simplest inception the TRM has $N_{par} = 3N_r$ free parameters, where $N_r$ is the number of rings. The three free parameters, in each ring, are the inclination angle, the P.A. and the rotational velocity. In general, the number of free parameters can be larger because other quantities may be left free  as, {  for instance, by varying  coordinates for each ring, or smaller by fixing one of the three parameters to a single value for all rings. As we discuss below, it is also possible} in some cases to  set the radial velocity different from zero. 

If residuals between the observed and the modeled 
{  velocity fields are localized and small, they},  can  be interpreted as non-circular motions, and techniques can be used to extract this information from the data (see, e.g., \cite{Jorsater+vanMoorsel_1995,Zurita_etal_2004,Trachternach_etal_2008,Erroz-Ferrer_etal_2015}),  whereas if  {  residuals are global and large they}
are more likely errors of the model. }

{  One of the primary objectives in analyzing the velocity fields of external galaxies is to identify non-axisymmetric and radial motions that convey valuable information regarding spatial structures.}
 \cite{Barnes+Sellwood_2003} developed a technique based on the assumption that the plane of the disk is flat. They determined the rotation center, inclination, and P.A. by fitting a non-parametric circular flow pattern to the entire velocity map. However, this method averages over velocity distortions caused by local structures like spiral arms, and can give inaccurate mean orbital speeds if the galaxy has a warped geometry or if there are non-circular motions, such as a bar-like or oval distortion to the velocity field over a wide radial range. These problems occur when assuming a flat disk with an axisymmetric velocity field. {  For instance, the package {\tt rotcur} is ideal  to determine the mean orbital speed even when the plane of the disk may be warped if noncircular motions are small, but yields spurious variations of the parameters when the underlying flow contains non-axisymmetric distortions   (see discussion in, e.g., \cite{Spekkens+Sellwood_2007}).}
 
 Early attempts to   overcome these limitations and to reconstruct radial and non-axisymmetric motions in galaxies have relied on a combination of Fourier decomposition with the TRM by considering that non-axisymmetric distortions to the planar flow can always be described by a harmonic analysis \citep{Franx_etal_1994,Schoenmakers_etal_1997,Wong_etal_2004,Simon_etal_2005,Chemin_etal_2006,Gentile_etal_2007,Trachternach_etal_2008}.
 In particular, \cite{Schoenmakers_etal_1997} generalized the tool {\tt rotcur} to handle mildly non-circular flows and developed the tool {\tt reswri}.  This method, which is based on epicycle theory, is  valid for small departures from circular orbits and can only fit mildly elliptical streamlines. It  may, however, give misleading results if the observed non-circular motions are not small compared to the circular orbital speed. Additionally, the results may be affected by the implicit assumptions made about the galaxy's geometry. Indeed, the initial fit using the TRM may suppress real radial motions by interpreting them as spurious warps in the galactic geometry. Real radial motions in the TRM framework {  appear  as  a change of the P.A. (a variation of} the inclination angle plays a minor role as we discuss in what follows) {  so that warps can suppress the detection of real radial motions}. This implies that the method may not be able to accurately reconstruct radial motions that are instead interpreted as distortions of the galactic disk \citep{Schmidt_etal_2016}. Note that the intrinsic  degeneracy between the geometry of the disk and the presence of radial motions inevitably affect {  all methods}.  {  Radial flows have been found by using the TRM already two decades ago \citep{Fraternali_etal_2001}.   {  Recently \cite{DiTeodoro+Peek_2021} have studied a large sample of galaxies by  including radial velocities directly in the TRM fit and have shown that for circular orbits the degeneracy is not complete as the radial flows do not affect the observed velocities on the major axis}. However, it is not clear if this method is able to capture all of the radial velocities due to the degeneracy with warps.}

{   Following the methods developed by \cite{Barnes+Sellwood_2003}, which, as mentioned, assumes the galaxy to be a flat disk, \cite{Spekkens+Sellwood_2007} introduced a technique, {\tt VelFit}, for fitting a non-axisymmetric model to the velocity field of a galaxy, allowing for noncircular motions.  This was applied to NGC 2976, finding large radial velocities and suggesting the presence of a strong bar. Then, \cite{Sellwood+Sanchez_2010} applied {\tt VelFit} to five representative galaxies from the The \HI{} Nearby Galaxy Survey (THINGS) sample  \citep{Walter_etal_2008}  and found evidence of mild bars in NGC 2976 and NGC 7793 and a pronounced non-axisymmetric flow in the strongly barred galaxy NGC 2903. \cite{deNaray_etal_2012} further applied this technique to NGC 6503, concluding that it has regular gas kinematics that are well described by rotation only. The {\tt VelFit} method is encoded in the {\tt DiskFit} software \citep{Sellwood+Spekkens_2015}, which also estimates uncertainties in all fitted parameters.

More recently, \cite{Sellwood_etal_2021} developed a new tool, {\tt makemap}, which uses a modified bootstrap method to estimate the uncertainties in the velocity at each pixel (see also \cite{Oh_etal_2018,Oh_etal_2019}).}  This method was applied to a sub-sample of THINGS survey. The derived 2D velocity map, together with a map of the uncertainty in the estimated velocity at each pixel, is given as input to {\tt DiskFit} to obtain new rotation curves with better-determined uncertainties. These results typically do not show strong variations in the rotation amplitude and radial flows. However, this method was developed and applied specifically for the case of a flat disk with bar-like or oval distortions that may arise from either a non-axially symmetric halo or a bar in the luminous disk. The assumption of a flat disk is generally appropriate for spiral galaxy velocity fields measured within the optical radius, where warps are rare, and is therefore well-suited to interpreting kinematics derived from, for example, stellar spectroscopy. However, its application to the outer parts of \HI{} velocity fields, which typically extend well into the warp region, may be problematic.

In this respect, ideally, one would like to model galactic warps independently of observations of velocity fields.  However, this is not possible for galaxies with inclination angles significantly smaller than $90^\circ$. Thus,  the existence  of warps can only be proven independently of kinematic studies by observing edge-on galaxies \citep{Sancisi_1976,Reshetnikov+Combes_1998,Schwarzkopf+Dettmar_2001,Garcia-Ruiz_etal_2002,Sanchez-Saavedra_etal_2003} or the Milky Way \citep{Levine_etal_2006,Kalberla_etal_2007,Reyle_etal_2009}.   Edge-on observations are able to provide the shape and amplitude of warps {   if the warp is not purely along the line of sight; however it is very difficult to reconstruct the radial motion from these measurements. Indeed, radial flow signatures should be visible only along the minor axis in edge-on galaxies, but these are generally hard to be found}. Therefore, it is still a challenge to accurately reconstruct radial motions and control the effect of assumptions made on galactic geometry, particularly the shape and amplitude of warps. However, the characteristics of the observed warps can be taken into account to adjust the kinematic analysis accordingly. Warps are commonly found in the outer regions of spiral galaxies   and they generally start at $R_{25}$, the radius at which disk surface brightness in the blue band falls below 25 mag per square arcsec \citep{Briggs_1990}.  This situation implies that the assumption that the galaxy disk is flat within the optical radius is consistent with observations, while the outermost regions remain more difficult to analyze accurately precisely because of the degeneracy between the galaxy geometry and the velocity field characteristics. For example, \cite{Garcia-Ruiz_etal_2002}  {  found that 20 out of 26 galaxies in their sample were warped and that all galaxies with an extended \HI{} disk compared to their optical disk were warped. Thus warps are certainly present in the peripheries of disk galaxies: indeed,}  from a physical point of view, warps are more likely to be found in the outer regions of galaxies as these are less affected by the internal dynamics of the galaxy, such as the central bar or bulge, which have stronger gravitational pull; they are also more likely to be present in the outer regions of galaxies as they are more susceptible to the effects of external factors, such as interactions with other galaxies   (see, e.g., \cite{Reshetnikov+Combes_1998,Ann+Park_2006}). 
In addition, it is worth noticing that the warp angle, defined as the angle between the outermost detected point and the mean position of the plane of symmetry, typically does not exceed around  $ 10^\circ$, although there are some exceptions where values of around $20^\circ$  or larger have been found \citep{Garcia-Ruiz_etal_2002,Sanchez-Saavedra_etal_2003,Reshetnikov_etal_2016,Peters_etal_2017}.

In this paper we develop  {  an algorithm,  named {  the velocity ring model} (VRM), that uses an ordinary least square (OLS) method for fitting a non-axisymmetric model to the observed two-dimensional velocity field.} In its simplest formulation, {  it is } similar to that studied by \cite{Barnes+Sellwood_2003,Spekkens+Sellwood_2007,Sellwood_etal_2021} and it is based on the assumption  that the galactic disk is  flat {  and for this reason the ideal range of application is limited to within the optical disk where warps are rare}.   {  As for the RDM case we discussed above, the VRM can be interpreted merely as a  special case of the TRM where the inclination angle and the P.A. are kept constant. The novelty of the VRM  is that it  allows to characterize non-axisymmetric and   heterogeneous velocity fields,  in which both radial and transversal velocity  components may have more complex anisotropic patterns than  bar-like or oval distortions for which the method  introduced by \cite{Barnes+Sellwood_2003,Spekkens+Sellwood_2007,Sellwood_etal_2021}  was developed. In principle, the amplitude of the velocity gradients in both components can be arbitrary but must satisfy  the constraint  that the inner part of the disk is dominated by rotation.  }

{  In the peripheries of disk galaxies  where \HI{}    distributions  may show signs of warping, the VRM can be combined with the TRM to determine the range of radii where the warp is not present. However, it should be noted that this does not necessarily mean that a warp exists outside this range. As discussed below, this coupling of the two methods imposes a stringent requirement since a strong radial flow can be misinterpreted by the {  TRM as a warp and vice-versa}. 
In the outer regions of the galaxy where warps are more likely to occur, larger gradients in radial velocity may also be induced by tidal effects from nearby galaxies or other types of spatial structures. Due to the inherent degeneracy between non-circular motions and warps, it is impossible to determine the true origin of these variations, whether they are due to disk deformations, radial flows, or both. However, coupling the VRM with the TRM can help constrain the region where this degeneracy is present. }

As mentioned above, the assumption of a flat disk is generally reasonable inside the optical radius or as long as the variations of the orientation angles or gradients of radial flows are small enough. It is in this region where a refined version of the VRM, which we also develop in this work, can reconstruct an anisotropic velocity field with non-circular motions. In the VRM, a galactic disk is represented as a sequence of $N_r$ rings, each having the same inclination angle and P.A.. The velocity field in each ring is decomposed into a radial component $v_r$ and a transverse component $v_t$. The model's free parameters consist of $N_{par}=2N_r$ and the two global orientation angles. To increase the {  angular} resolution of the method in the analysis of a velocity field, we have developed a more sophisticated version called the VRM with arcs (VRMA). In this method, each ring of the VRM is partitioned into $N_a$ arcs, so the number of free parameters becomes $N_{par}=2N_rN_a$. This allows for the reconstruction of both velocity components in each arc, providing a 2D {  angularly resolved} coarse-grained representation of the galactic radial and transverse velocity fields. The kinematic maps obtained with the VRMA method can be compared with observations of structures such as bars and satellites to determine their kinematic influence.  {  We note that the use of arcs is not new: \cite{Kamphuis_etal_2013}  constructed detailed tilted ring models of several galaxies by  adding distortions resembling arcs or spiral arms and  in this way they have improved the TRM fits. However, differently from the VRMA case, the arcs used were ad-hoc for the particular galaxy studied.}

In this paper, the two methods VRM/VRMA are, together with the standard TRM, applied to the 2D velocity maps of 25 disk galaxies measured by THINGS which is a high-resolution sample for the study of galaxy kinematics.  Results are then compared  with those of previous studies as those by  \cite{deBlok_etal_2008,Schmidt_etal_2016,DiTeodoro+Peek_2021,Sellwood_etal_2021}. {  The goal of the comparison is to evaluate the accuracy of the method used in the current study in reconstructing the radial and transversal velocity fields of the galaxies. The analysis aims to understand the role that different assumptions used in the various methods may have played in the differences observed.}
 
The paper is organized as follow. In Sect.\ref{methods} we discuss the different methods to measure the velocity field of an external galaxy.  Then in Sect.\ref{results} we show the results for the case of the 25 THINGS  galaxy. Finally in Sect.\ref{conclusion} we draw our main conclusions.  {  In Appendix A we discuss some tests of the different methods with simple toy models and in  Appendix B we report the detailed figures for all galaxies but NGC 628 that is discussed in the main body of the paper. }


\section{Methods} 
\label{methods}

Observations of an external galaxy give  the component of the velocity  $v_{los}(r,\phi)$ of the galaxy along the {  LOS}. The angular coordinates $(r,\phi)$ refer to the position of a point in the projected image of the galaxy in the plane of the sky. The transversal and radial velocity components, $v_t$ and $v_r$, refer to the velocity of the galaxy in the plane of the galaxy, where $R$ and $\theta$ are the polar coordinates {  (assuming  the center of the galaxy to be origin of the coordinates)}. The transformation between the coordinates in the plane of the sky and those in the plane of the galaxy can be written as (see, e.g., \cite{Beckman_etal_2004})
\bea
\label{eq:transformation}
&&
\tan(\theta) =  \frac{\tan (\phi - \phi_0)}{\cos(i)} 
\\ \nonumber 
&&
R = r \frac{\cos (\phi - \phi_0)}{\cos(\theta)} \;, 
\eea
where $i$ is the inclination angle, $\phi_0$ is the P.A. (i.e., the angle of the galactic  major axis 
with  respect to the North galactic pole). 
With these definitions we can write  
\be
\label{eq:vlos} 
v_{los}  = \left[v_t  \cos(\theta) + v_r  \sin(\theta) \right] \sin(i)  \;, 
\ee
where we neglect motions perpendicular to the plane of the disk, i.e., vertical velocities,  and we have subtracted the systemic velocity (see discussion below). 
{   The TRM assumes that the galaxy disk can be approximated by a set of rings with inclination angle and P.A. that vary with radius, i.e., $i=i(R)$ and $\phi_0=\phi_0(R)$.}
In the case of the VRM, the $N_{r}$ rings have all the same inclination angle and P.A.: therefore, Eqs. \ref{eq:transformation}-\ref{eq:vlos} are the same equations as those used in the TRM, and the only difference is that the P.A. and inclination are not allowed to vary with radius. For this reason the VRM is just a specific subset of the TRM. In the VRM, the velocity field in each ring is decomposed into a radial  $v_r$ and a transverse component $v_t$. Thus, we have $v_t=v_t(R, \theta)$, $v_r=v_r(R, \theta)$, $i=\text{const.}$, and $\phi_0=\text{const.}$ In the case of the VRM,  the inclination angle is an external parameter and it is determined from the photometric map, while the P.A. is measured from the kinematic map. It should be stressed that while for low inclinations  it becomes progressively difficult to determine the global inclination angle from a photometric map, the inaccuracy in its measurement  corresponds to an offset in the {  amplitude of the velocities}.


\subsection{The rotating tilted ring model (TRM)} 
\label{TRM}

In the RDM, galaxies are considered to be axisymmetric systems in concentric circular rotation in a plane about a central axis. The velocity of this rotation, i.e., the rotation curve $v_c(R)$, varies with the radius from the galactic center, and is assumed to be determined by the radial distribution of mass within the galaxy. The natural template employed to fit an observed velocity field is a disk with circular velocities such as those measured along the kinematic axis of the projected image of a real galaxy. To find the best model, one minimizes the residuals, i.e., the difference between the real and model velocities, with respect to the free parameters of the model, which are the circular velocity of the disk divided into $N_r$ rings, plus the global inclination angle and the P.A. On the other hand, the TRM {  \citep{Warner_etal_1973,rogstad_etal_1974} } assumes that a galaxy can be described as a set of concentric rings of circular velocity $v_c(R)$, and of inclination angle $i$ and P.A. $\phi$. The three ring parameters $v_c, i$, and $\phi$ are solved through an iterative procedure. 

{  We have employed the code {\tt kinemetry}  \citep{Krajnovic_2006} that uses the TRM  to measure the rotation curve of a galaxy \footnote{  For the THINGS galaxies we were unable to find published rotation curves derived with the  {\tt kinemetry}  package, as such we have compared
  our determinations  with those by \cite{deBlok_etal_2008} who,   for a subsample of the THINGS galaxies, have measured the rotation curves  with {\tt rotcur}.}.   It uses a general least-squares fitting routine to fit the free parameters of the TRM. 
  The radii of the fit are chosen so that the first corresponds to 1 pixel, the second one to 2.1 pixels, and from then they are determined with the recursive expression $r_i = r_{i-1}+1.1\times (r_{i-1}-r_{i-2})$. The code gives as outputs the ratio between the apparent minor $b_p$ and major  $a_p$ semi-axis of a ring $q(R)=b_p/a_p= \cos(i(R))$, the P.A. of the ring $\phi(R)$, and the {  observed velocity $v_o(R)$ of the ring, which is related to the rotation velocity by }
\be
\label{vc} 
 v_c(R)=\frac{v_o(R)} {\sin(i(R))} \;. 
\ee
The error bars to $v_c$  that we report below are computed through the errors propagation from the errors $\delta q(R)$, $\delta \phi(R)$ and $\delta v_o(R)$  which are also given as outputs by the code {\tt kinemetry}. 
{    The error bars obtained from  {\tt kinemetry} are the formal fitting errors which often underestimate the true errors due to the way the velocity fields are constructed.}  Since different numbers of channels are masked, various parts of the velocity field contribute differently. This implies that the errors in a pixel/beam are not constant over a ring, making it challenging to propagate the proper error from the noise in the data-cube for a given ring.   
} 
For this reason, several authors report error bars on the circular speed that are based on the difference between the separately fitted circular speeds on the approaching and receding sides of the galaxy (see, e.g., \cite{deBlok_etal_2008}).

{  It is worth noting that previous studies of the THINGS galaxies have limited the analysis to galaxies with inclination angles greater than 40 degrees \citep{deBlok_etal_2008} or greater than 30 degrees \citep{DiTeodoro+Peek_2021}.   In the framework of the TRM, measuring the rotation curve of a nearly face-on galaxy is challenging because $v_c$ and the inclination angle {  become mathematically degenerate \citep{Begeman_1989}}: indeed, Eq. \ref{vc} implies that small changes in $i$ induce large variations in $v_c$ when $i$ is small.  Note that when analyzing galaxies with low inclination angles with {\tt kinemetry}, we did not constrain the inclination angle, and as a result, the degeneracy between $v_c$ and $i$ introduced large fluctuations.
The determination of the global inclination angle $i$ of the galaxy, that is used in the VRM analysis, is more reliable than the measurement of the inclination of a single ring: indeed, ellipse-fitting routines can provide reasonable constraints on the global inclination angle so that it is possible to determine $i$ independently of $v_c$ avoiding large variations in $v_c$ that might arise from fitting $i$ and $v_c$ simultaneously when $i$ is small. In this situation an error in the determination of $i$ corresponds to an offset in the  amplitude of the  {  velocity components} \citep{Begeman_1989}.  

As in what follows we will present the kinematic properties of galaxies with low inclination angles, it is worth recalling that for inclinations lower than $\sim 40^\circ$  the velocity dispersion of the gas can be an important contaminant: it can introduce uncertainty in the derived kinematics that can affect the measurements of the rotational velocity.      Indeed,  as rotation takes place in a 2D disk, the projected component of the rotation velocity decreases in amplitude with decreasing inclination (see Eq.\ref{eq:vlos}) whereas  the contribution due to random motions, which take place in 3D, remains the same by changing the inclination angle.  {    
However, it is important to take into account that random motions within the \HI{} distribution typically have magnitudes of approximately 10 km/s \citep{vanderKruit+Shostak_1984} and display isotropic behavior. These motions predominantly impact the determination of the line profile peak when the inclinations are low, specifically when $i < 10^\circ$ \citep{Begeman_1989} . Additionally, it is important to note that for inclinations below $40^\circ$, the $\chi^2$ plane for $i$ and $v_{rot}$ exhibits a shallow minimum, making it challenging to accurately determine the de-projected rotational velocity due to the degeneracy between $i$ and rotational velocities. As a result, there are significant uncertainties associated with determining the de-projected rotational velocity \citep{deBlok_etal_2008}.} We will discuss this issue in what follows.


\subsection{Estimating the radial velocity profile with the TRM} 
\label{TRM_VR} 

As discussed in Sect.\ref{intro}, different attempts of reconstructing radial and non-axisymmetric motions from the LOS velocity map of galaxies can be found in the literature. Most of these approaches are based on a Fourier decomposition aimed at taking into account the angular dependency of the velocity field. The idea is to rewrite the LOS velocity equation as a Fourier series \citep{Schoenmakers_etal_1997} as 
	 \begin{equation}
		v_{los}^{model}  (R, \theta)=\sin i(R)\sum_{k=1}^N\left[v_t^{(k)}\cos (k\theta)+v_r^{(k)}\sin (k\theta)\right] \;. 
		\label{eq:harmonic_dec}
	\end{equation}
 This allows for a decomposition of the velocity field into different angular modes, each one with a specific amplitude and phase. In Eq.\ref{eq:harmonic_dec} radial motions are taken into account by the terms in $\sin(k\theta)$, while non-axisymmetric motions are described by harmonics of order $k>1$. The first attempts made by  \cite{Schoenmakers_etal_1997} combined this Fourier decomposition with the TRM. In particular, they first fitted the velocity field by means of the TRM, which corresponds to only retaining the term $v_t^{(1)}$ in Eq.~\ref{eq:harmonic_dec}, and then they performed the Fourier decomposition of the residuals of such a fit. They found that higher-order Fourier terms correspond to non-axisymmetric spiral-like structures, while they obtained negligible radial velocities. Despite this method has been largely applied (see, e.g., \cite{Wong_etal_2004,Simon_etal_2005,Chemin_etal_2006,Gentile_etal_2007, Trachternach_etal_2008}), it presents several drawbacks since there is an intrinsic degeneracy between the geometry of the galaxy and the presence of radial motion.

Indeed, as pointed out by \cite{Schmidt_etal_2016}, since the initial fit is performed using the TRM, radial motions may be suppressed by spurious warps appearing in the galactic geometry due to their degeneracy. This method is not capable of reconstructing radial motions that are instead interpreted as distortions of the galactic disk. Moreover, residuals are analyzed by means of the epicycle theory \citep{Franx_etal_1994}, which is valid only for small deviations from perfect circular orbits. If conversely, one includes the radial velocity directly in the initial TRM fit without neglecting the $v_r^{(m)}$ terms, the situation changes and non-negligible radial flows of matter can be obtained. Analogous conclusions are drawn in \cite{DiTeodoro+Peek_2021}, where a 3D variation of the TRM with radial velocities is exploited. These results, differently from previous studies based on the analysis of the circular TRM residuals, thus suggest that radial motions could play a non-marginal role in the dynamics of galaxies. However, due to the degeneracy with warps, it is not clear if including them in the TRM allows to measure correctly the full value of the radial velocity component.

As mentioned in the Sect.\ref{intro},   
\cite{Spekkens+Sellwood_2007} developed a  model based on several assumptions: 
i) the non-circular motions in the flow stem from a bar-like or oval distortion to an axisymmetric potential; ii) only a bi-symmetric distortion to the potential is considered while higher harmonics are neglected; iii) the bar-like distortion drives non-circular motions about a fixed axis in the disk plane and iv) the disk is flat. In these approximations the model velocity has only the $m=2$ terms of  a Fourier series around a circle of
radius $r$ in the disk plane and it can thus analyze only a certain kind of asymmetric motion. 


\subsection{The velocity ring model (VRM)} 
\label{VRM}

Let us discuss how the determination of the VRM proceeds. The first step of the code that we have developed to implement the VRM\footnote{The code to compute the VRM is publicly available at the webpage: {\tt https://github.com/MatteoStraccamore/VRM$\_$VRMA}.} consists in selecting the points of the galaxy belonging to different rings (and sectors for the case of the VRMA --- see below). To this aim it is necessary to move from the plane of the sky to the plane of the galaxy by using Eq.\ref{eq:transformation}. Then the analysis  proceeds as follows.
{   
We first determine the global inclination angle from the optical or \HI{} maps, as reported in \cite{Walter_etal_2008}. This value generally coincides with that measured by the TRM in the inner galactic disks. We then take the values of the centers of the galaxies from the data in \cite{Walter_etal_2008,deBlok_etal_2008}. Using Eq. \ref{eq:transformation}, we transform the polar coordinates on the plane of the sky $(r,\phi)$ to the polar coordinates on the plane of the galaxy $(R, \theta)$. (In what follows, we use these coordinates to plot the de-projected maps to a face-on distribution). 
We subtract the value of the systemic velocity $v_{sys}$ {  (see below)}  from the data to obtain the LOS velocity $v_{los} = v_{obs} - v_{sys}$ in the rest frame of the galaxy. We select the inner disk of the system from the photometric \HI{} map emission and compare it with the optical disk. We rotate the map by an angle $\psi_0$ (reported in Tab. \ref{table_1}) such that the kinematic axis of the inner galactic disk aligns with the horizontal axis. The rotation by this angle is a convention to align all galaxies so that there is no offset in the angular variable when doing the fit. 
The angle $\psi_0$ is equivalent to the P.A. (modulus a rotation of $90^\circ$) if the kinematic axis does not change orientation along the galactic disk, 
otherwise $\psi_0$ corresponds to the P.A. of the inner disk.} (It is worth noticing that we have tested that changing the angle $\psi_0$ by $\pm 10^\circ$ results in the velocity field remaining within the error in the determination of $v_r(R)$ and $v_t(R)$ in a ring). Once this rotation has been performed, the angle $\phi_0$ in Eq. \ref{eq:transformation} can be set to zero. 
We then make the coordinate transformation in Eq. \ref{eq:transformation}.

{  At this point we can fit the VRM using the OLS method}.
{  First of all we remove  the systemic velocity. 
This is done by fitting the whole map using a single ring, thus exploiting the following expression 
\be
	v_{obs}(R, \theta) = [v_t \cos(\theta) + v_r \sin(\theta)] \sin(i) + v_{sys} \,,
\ee
where $ v_{sys}$ is the systemic velocity. 
In order to do so we exploit an OLS procedure in the variables $\cos(\theta)$ and $\sin(\theta)$. 
Once 
$v_{sys}$ has been determined we find $v_{los}(R, \theta)= v_{obs}(R, \theta) - v_{sys}$.  
The values we obtain for $v_{sys}$ are very close to those   
reported  in \cite{Walter_etal_2008}.
}

{  Let us now consider a ring with radius $R$ and thickness $\Delta R$ containing $n$ points of the map with angular coordinates $\theta_1 \ldots \theta_n$. We denote by $v_{los}^{exp}(R, \theta_1)\ldots v_{los}^{exp}(R, \theta_n)$ the LOS velocity at the $n$ points and by $\vec{v}_{los}^{exp}(R)$ the vector containing these $n$ experimental values}
\begin{eqnarray}
\label{vec_exp} 
\nonumber
\vec{v}_{los}^{exp}(R)  = &
 \left(
\begin{array}{c}
v_{los}^{exp}(R, \theta_1)\\
\vdots \\
v_{los}^{exp}(R, \theta_n) \\
\end{array}  
\right) \;.
\end{eqnarray}

{  
In the galactic plane, our model, given by Eq.~\ref{eq:vlos}, reads 
\[
	v_{los}(R, \theta_i) =\left[v_t(R) \cos(\theta_i) + v_r(R) \sin(\theta_i) \right] \sin(i) \
\]
and we want to determine the best parameters $v_t(R)$ and $v_r(R)$ which minimize the objective function $L(v_t(R), v_r(R))$ defined as 
\[
	L\left(v_t(R), v_r(R)\right)=\sum_{i=1}^n\left(v_{los}(R, \theta_i)-v_{los}^{exp}(R, \theta_i)\right).
\]
As our model is linear in the variables $\cos(\theta_i), \; \sin(\theta_i)$, we apply the OLS procedure to all the $N_{r}$ rings corresponding to the different values of $R$ thus obtaining the best parameter radial and rotational velocity of each of them. }
}


\subsection{The velocity ring model with arcs (VRMA)}

{  By relaxing the assumption of circular symmetry, it is possible to extend the VRM to consider non-axisymmetric motions induced by galactic structures such as the bar, spiral arms, and others.}
This is done by splitting each ring into $N_a$ arcs, each characterized by a different radial and transversal velocity. In this way we introduce also a dependency on the angular coordinate $\theta$, meaning that $v_t=v_t(R, \theta)$, $v_r=v_r(R, \theta)$, $i=\text{const}$ and $\phi_0=\text{const}$.  We call this approach the Velocity Ring Model with Arcs (VRMA). 

{  In the  case of the  VRMA algorithm, arcs are variable in number and  placed in a way that is a-priori uncorrelated with the shape of the galaxy thus allowing, in principle,  the identification of generic spatial velocity anisotropies. The number of free parameters is $N_{par}= 2\cdot N_r\cdot N_a$}. {  Note that by fitting the same galaxy with a variable number of arcs ($N_a=1, 2, 4, \ldots$), we can assess the extent of non-axisymmetric motions and determine the robustness of the results. Furthermore, specific convergence tests can be performed to establish the reliability of the reconstructed two-dimensional maps $v_t(R, \theta)$ and $v_r(R, \theta)$, as discussed below. }

{  In the very same way described for the VRM case, we can also fit the VRMA to the maps, since we simply have to apply the procedure to all the arcs we divide each ring into. Thus we divide each ring in $N_a$ arcs with equal angular width and we define for each of them the vector of empirical observations (see Eq.\ref{vec_exp}) where observations satisfy $\theta-\frac{2\pi}{2N_a}<\theta_i<\theta+\frac{2\pi}{2N_a}$ with $\theta$ being the angular position of the arc's center. We can then apply the same procedure described for the VRM to all the arcs inside each ring.
The VRM is thus a special case of the VRMA when one sets $N_a=1$. }
{  Note that  Eq.\ref{eq:vlos} implies that the transversal velocity is undetermined for $\theta=90^\circ$: however in the VRMA method we estimate the velocity components 
in arcs of finite angular extension and for this reasons there is no  degeneracy    for  any specific value of $\theta$. The reasoning applies 
to the case of the radial component for $\theta=0^\circ$. }

The profiles $v_t(R)$ and $v_r(R)$ correspond to the averages over rings. {  In the case of the analysis of the galaxies in our sample,} errors on the average velocity profiles are computed by making the rms value over  the $N_a$ arcs in each ring. To take into account the correlation introduced by the observational resolution  $\sigma$, the errors are given for an intermediate value of the number of arcs $N_{a} =4$ and 8, as for larger numbers of arcs the angular size of a cell can be comparable to  $\sigma$: in particular, errors may be underestimated in the inner disk as the angular size of a cell depends on its distance from the center and for small radii it can be comparable to the observational resolution. We have however checked that, for most of the cases,  the errors do not significantly depend on the number of arcs when this is increased to  16 and 32 {  (see discussion below).}


\subsection{Comparison of the different methods}
\label{overfitting} 

{   In order to compare results of the TRM and VRMA  we consider the histogram of their residuals $\rho$ that estimates the residuals probability density function (PDF),  where $\rho$ is defined as the difference between the observed velocity field and the reconstructed one. This comparison is meaningful if the number of  free parameters $N_{par} = 2 N_a^{VRMA} N_r^{VRMA}$ used in the VRMA is the same of that  used in a determination with the TRM, $N_{par} = 3 N_r^{TRM}$. We used $N_r^{TRM} = 100$ with {\tt kinemetry} that corresponds to  use $N_a^{VRMA}=3$ arcs and  $N_r^{VRMA}=50$ rings with the VRMA: below we use $N_a^{VRMA}=$2 and 4 to make the comparison as the actual number of free parameters may vary depending on the specific features of each galaxy. 

Fixing the number of free parameters to be the same and much smaller than the number of pixels for both methods allow a fair comparison of their performance. 
 We stress, however, that we have not used a rigorous method for such a {    comparison as our aim is to give a reasonable idea of their relative performances}.  

{  
However, the pixels in a radio map are not independent. In particular, the beam size of the THINGS data, typically 10", generally extends across several pixels (refer section 3.6 of \cite{Walter_etal_2008}). This correlation may be relevant to the overfitting problem, especially for the VRMA case where the angular size of cells is smaller than that of the rings in the TRM. Therefore, it is necessary to ensure that the angular size of the cells used by the VRMA algorithm is larger than  $\sigma$. In general, VRMA cells do not have the same angular size. In fact, it increases as a function of the distance from the center of the galaxy. Hence, we calculate the number of pixels per galaxy in the beam, denoted as $N_{pix}$, using the data from Tab.3 of \cite{Walter_etal_2008}. To avoid overfitting, we apply the following stringent criterion: if a cell has less than $2N_{pix}$ pixels (the factor of 2 arises because we are fitting 2 variables, thus requiring at least 2 resolution elements), it must be excluded from the fit. Generally, for real galaxies (as discussed below), we find that $N_{cell}^i>2N_{pix}$  for all cells, except for a small number of cells in (i) the very inner disk and (ii) the outermost regions of the galaxy, where the image boundaries are irregular. As a result, overfitting in the VRMA case should not affect the results for $N_a=2,4$ and $N_r=50$.}


\subsection{Convergence of the VRMA}
\label{convergence}

{  The VRMA divides the galaxy image into $N_{cells} = N_r N_a$ cells, and both velocity components $v^i_r$ and $v^i_t$ are estimated in each cell. However, the reconstruction is affected by noise that inevitably affect the measurement, so that each cell is characterized by the values of $v^i_r$ and $v^i_t$ with their fluctuations $\Delta v^i_r$ and $\Delta v^i_t$. While the VRMA can estimate the values of the velocity components in each cell, it does not provide a simple way to determine the values of their uncertainties. For this reason, to control effect of changing the resolution of the coarse-grained reconstructed map we have made tests with different number  of arcs, $N_a=1, 2, 4, 8, 16, 32$ and same number of rings $N_r=50$. These tests are intended precisely to control that the velocity field in a given angular region reconstructed with a certain number of arcs $N_a$ remains compatible with that reconstructed with a lower value of $N_a$.  When $N_a$ is  small enough the size of the cells in which the image is divided is larger than the observational resolution so that the spurious correlations $\sigma$ between pixels introduced by observational resolution do not affect the signal in different cells. 

Operationally, to control that the VRMA correctly reconstructs the velocity field on angular scales larger than  $\sigma$ we have adopted the following strategy. We start the analysis  by fixing the number of rings (we have typically chosen $N_r=50$) and with only a single arc, i.e., $N_a=1$;  then we double $N_a$ to 2, 4, 8, 16, 32, and we verify that in a given angular region that is covered by $N_c \ge 1$ cells with resolution $N_a$, the VRMA finds the same values of $v_r$ and $v_t$ independently of $N_a$. Only in this case do we conclude that the 2D maps of the transversal and radial velocity show  convergency with $N_a$ to the same spatial distribution of anisotropies. 

Quantitatively, we measure several moments of the velocity components and study their convergence by varying $N_a$ at fixed $N_r$.}
The first moment is the  dipole defined as  
\bea
\label{dipole} 
&&
D_t^i(R; N_a)=  \left(  \frac{v_t^i(R,\theta) -  v_t(R) } {  v_t(R) }   \right)
\\ \nonumber 
&&
D_r^i(R; N_a)=  \left(  \frac{v_r^i(R,\theta) -  v_r(R) }{  v_t(R) }      \right)
\eea
 {   where $v_t(R)$ and $v_r(R)$ are the profiles averaged over a ring and $v_t^i(R,\theta), v_r^i(R,\theta)$  for $i=1,..,i_{max}$ is the transversal/radial  velocity in the angular sector with 
\be
\frac{2 \pi (i-1)}{i_{max} }  \le \theta \le \frac{2 \pi i}{i_{max} } \;,
\ee
where the galaxy is angularly divided into two subregions, i.e. $i_{max}=2$. Note that  the radial velocity component is normalized to the average (in rings) transversal  velocity as the average (in rings) radial   velocity can be zero as we are interested  in the relative deviations from a perfectly rotating configuration. Similarly  to Eq.\ref{dipole}  we can define the quadrupole  $Q_t^i(R; N_a), Q_r^i(R; N_a)$  where $i_{max}=4$, and   the octopole $O_T^i(R; N_a),  O_r^i(R; N_a)$ where $i_{max}=8$.} 

Analyzing the same subregion by varying $N_a$ allows us to verify whether the 2D map converges, meaning that the signal-to-noise ratio in the whole subregion is larger than unity for the largest number of $N_a$ adopted (we use $N_a=32$ for the largest value of the number of arcs). For instance, the 2D transversal velocity map for $N_a=32$ converges if $O_T^i(R; 32) \approx O_T^i(R; 16) \approx O_T^i(R; 8)$ in the 8 regions $i=1,...,8$, where the symbol $\approx$ corresponds to a quantitative criterion such as the difference being {  approximately} smaller than 10\%. Note that an angular resolution $N_a>8$ is required to study the octopole moments convergence. Similarly, we have $N_a>4$ for the quadrupole moments and $N_a>2$ for the dipole.

{  It is important to emphasize that, as demonstrated by controlled numerical tests  discussed in what follows, the convergence of the octopole moments is a necessary condition but not a sufficient one to conclude that results reliable. The convergence of the octopole moments is necessary to test whether the velocity field reconstructed with the VRMA is trustworthy, in the sense that the noise that inevitably affects any measurement does not exceed the signal. However, the effects introduced by a geometric deformation   such as a galactic warp  are entirely different, as they result in a real (i.e., not due to noise) signal: in this situation the octopole moment converges anyway as long as the noise is smaller than the geometric-induced signal. In other words, if there is no convergence, then the noise in the measurement is certainly larger than the signal.  However, if there is convergence, then the effect of a geometric deformation can possibly  be present. 
}

\subsection{Controlled numerical tests} 
\label{tests}

It is necessary to conduct a quantitative analysis of the VRM/VRMA's ability to accurately recover the properties of a given system. To accomplish this, we have generated several toy models of galaxies with varying complexity in their velocity field.  {  These toy models can be seen as simple mathematical exercises to test the reconstruction methods and they are not intended to emulate the complexity real observations. 
{  The steps involved in creating the toy models are as follows. Firstly, we generate a disk of radius $R$ and finite thickness $h=R/10$ by randomly distributing $N_p$ particles within its volume, where $N_p=5 \times 10^5$. Then, we assign the three cylindrical components of velocity ($v_t, v_r, v_z$) to each particle, where $v_t, v_r$ are given by an analytical model and  $v_z$ is set to zero for simplicity (unless a random velocity dispersion is added to the velocity field). Next, we project this particle distribution onto the plane of the sky as seen by a randomly chosen observer (placed at infinity) defined by their inclination and P.A. This projection transforms the  coordinates in the plane of the galaxy ($R,\theta$)  into those  in the plane of the sky  ($r, \phi$). We also compute the line-of-sight  velocity field  $v_{los}$. Once the toy galaxy model has been generated following these steps, it is described by the positions ($r, \phi$) and LOS velocities ($v_{los}$) of the $N_p$ particles. This allows us to analyze the model using different methods such as TMR and VRMA, similar to a real case.}

As a first test, let us consider three different and simple cases of a flat disk. In the first case, $v_t=100$ km/s and $v_r=0$, in the second case, $v_t=0$ km/s and $v_r=100$, and in the third case, $v_t=v_r=100$ km/s. We reconstruct the velocity field with the VRM (i.e., by using the reconstructed values of $v_t$ and $v_r$ in Eq.\ref{eq:vlos}) and compare it with the input model. From Figure \ref{fig0} we can conclude that the method is able to fairly reconstruct the input properties of the velocity field for all angles $\theta$ in the plane of the galaxy.
\begin{figure}
\includegraphics[width=8cm,angle=0]{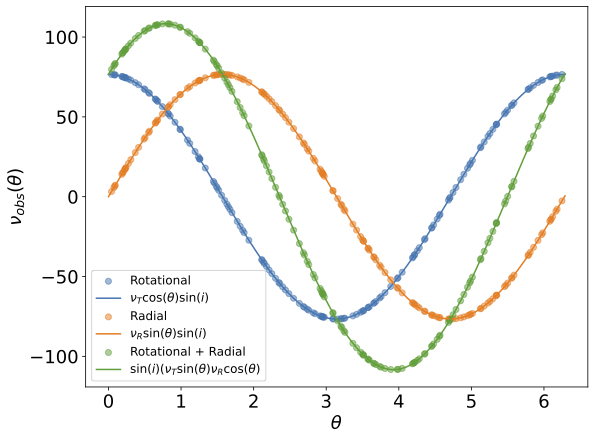}
\caption{ 
Behavior of the VRM reconstructed $v_{los}(\theta)$ (from Eq.\ref{eq:vlos}) in function of the angle $\theta$ 
for three different input velocity fields (also shown): purely rotational ($v_t=100$ km/s), purely radial ($v_r=100$ km/s) and a sum of rotational and radial with the same amplitude
($v_t=v_r=100$ km/s). }
\label{fig0} 
\end{figure}

 As a second example, we generated a flat disk with a radial dependent velocity field: in particular, $v_t(R)$ varies as for a simple exponential disk and $v_r(R)$ linearly increases with radius\footnote{Note that we have tested the methods with many different functional behaviors of $v_t(R)$  and $v_r(R)$. The example presented in Fig.\ref{class1} does not aim to encode any property of real galaxies, it just has very general variations of both velocity components allowing to illustrate the performances of the method.} . Results are reported in Fig.\ref{class1} for the case of an inclination angle of $50^\circ$. It can be seen (see panel (a)) that as long as $v_r$ is small (i.e., for small radii) the observed kinematic axis is oriented parallel to the projected major axis, as expected for a  rotating disk. However, when the amplitude of $v_r$ increases (i.e., for large radii) the kinematic axis' orientation changes with radius. The VRM analysis is able to reconstruct the intrinsic properties of this model with high accuracy: residuals in $v_{los}$, $v_r$, and $v_t$ have a dispersion of $\sigma_\rho < 0.5$ km/s in all cases, which is of the order of $1\%$ of the signal. 
{    It is worth noticing that the combination of a declining $v_t(R)$ and a linearly rising $v_R(R)$ can result in an unfortunate interplay that creates the appearance of a position angle  warp in the velocity field. However, it should be emphasized that the radial velocity does not actually impact the velocities along the major axis.} 
\begin{figure}
\includegraphics[width=\linewidth,angle=0]{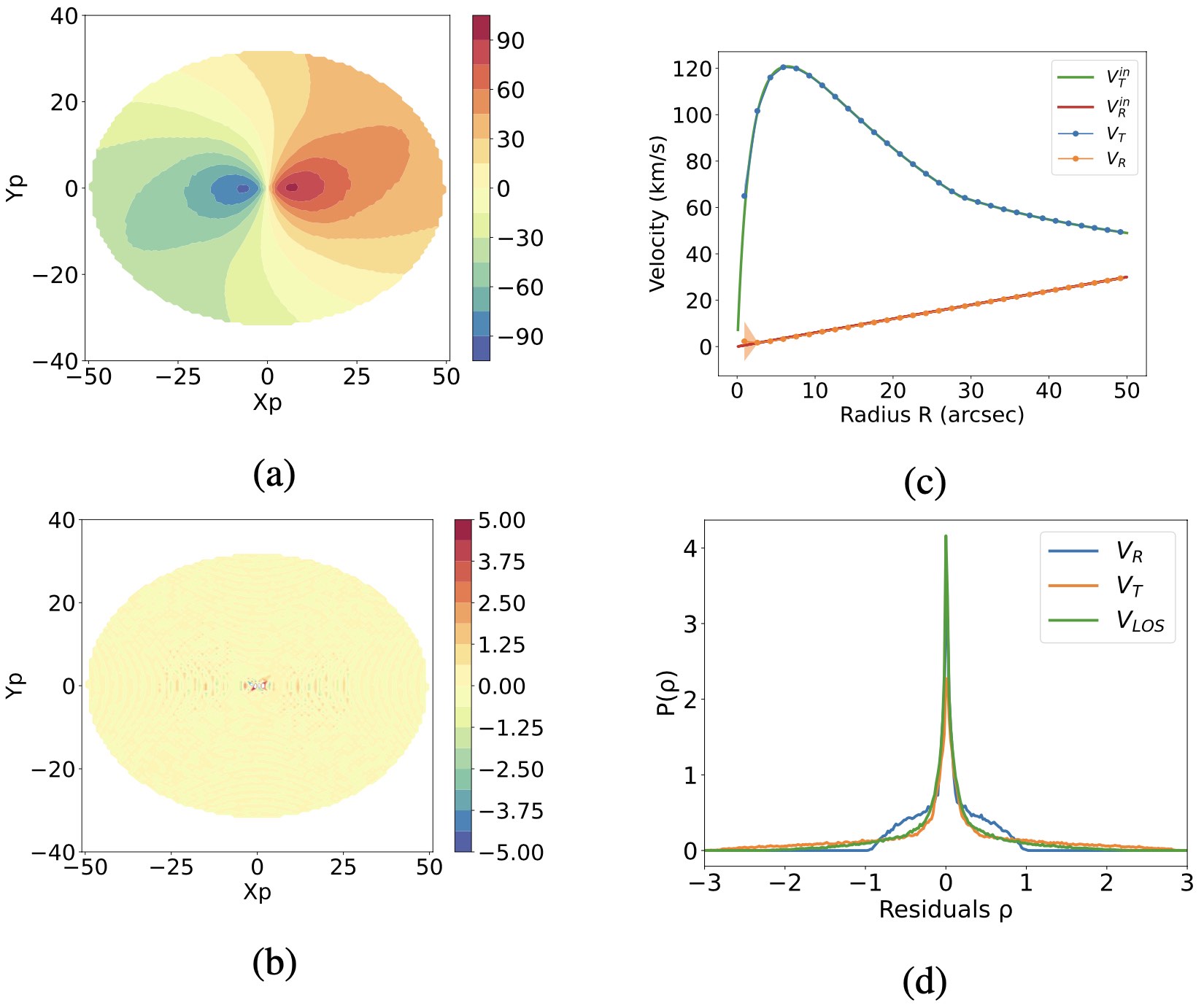}
\caption{  
A toy disk model in which  $v_t(R)$  varies as for a simple exponential disk  and $v_r(R)$ linearly grows with radius. The inclination angle is $i=50^\circ$. Panel (a): two-dimensional  line of sight velocity field, where $x_p$ and $y_p$ are the   angular coordinates in the plane of the sky (in which the disk has a  major semi-axis of 50 arcsec).   The colors correspond to the values of the line of sight velocity in km s$^{-1}$  (see vertical  bar). Panel (b)  two-dimensional residual field of the VRM:   colors correspond to the values of the line of the residual of the sight velocity in km s$^{-1}$  (see vertical  bar). Panel (c):  behaviors of the velocity profiles $v_r(R)$ and $v_t(R)$ as a function of the distance from the system center $R$ (in the plane of the galaxy) together with the input behaviors.  Panel (d):  histogram of the residuals $\rho$ (in km s$^{-1}$)  respectively in  $v_{los},\; v_r$ and $v_t$.
}
\label{class1} 
\end{figure}

{  Finally, to test the ability of the VRMA to recover the input properties of an anisotropic velocity field, we performed two different classes of tests. In the first class, we generated a toy model with an anisotropic velocity field (see the left panels of Figs. \ref{MW13a16}-\ref{MW13a12}) and reconstructed it using the VRMA with increasing resolution, i.e., $N_r=50$ and $N_a=8,16,32$. As shown by the behaviors of the octopole moments in Fig. \ref{MW13a12b}, there is good convergence of the reconstructed maps. However, we note that the variation in $v_t$ is rather significant in the $N_a=2$ and $N_a=4$ plots. This is due to a degeneracy between $v_t$ and $v_r$ when the arcs do not match the size of the fluctuations in $v_r$ . The situation can be clarified by using a larger number of arcs, i.e., by increasing the resolution of the maps. The analysis of this type of toy models motivated the introduction of the convergence test discussed in Sect. \ref{convergence}. To clarify the issue of resolution and to present convincing evidence that the analysis of real galaxies is not strongly affected by this type of degeneracy, we decided to include, for each galaxy, the coarse-grained velocity component maps with different resolutions (see discussion below). }

\begin{figure*}
\includegraphics[width=\linewidth,angle=0]{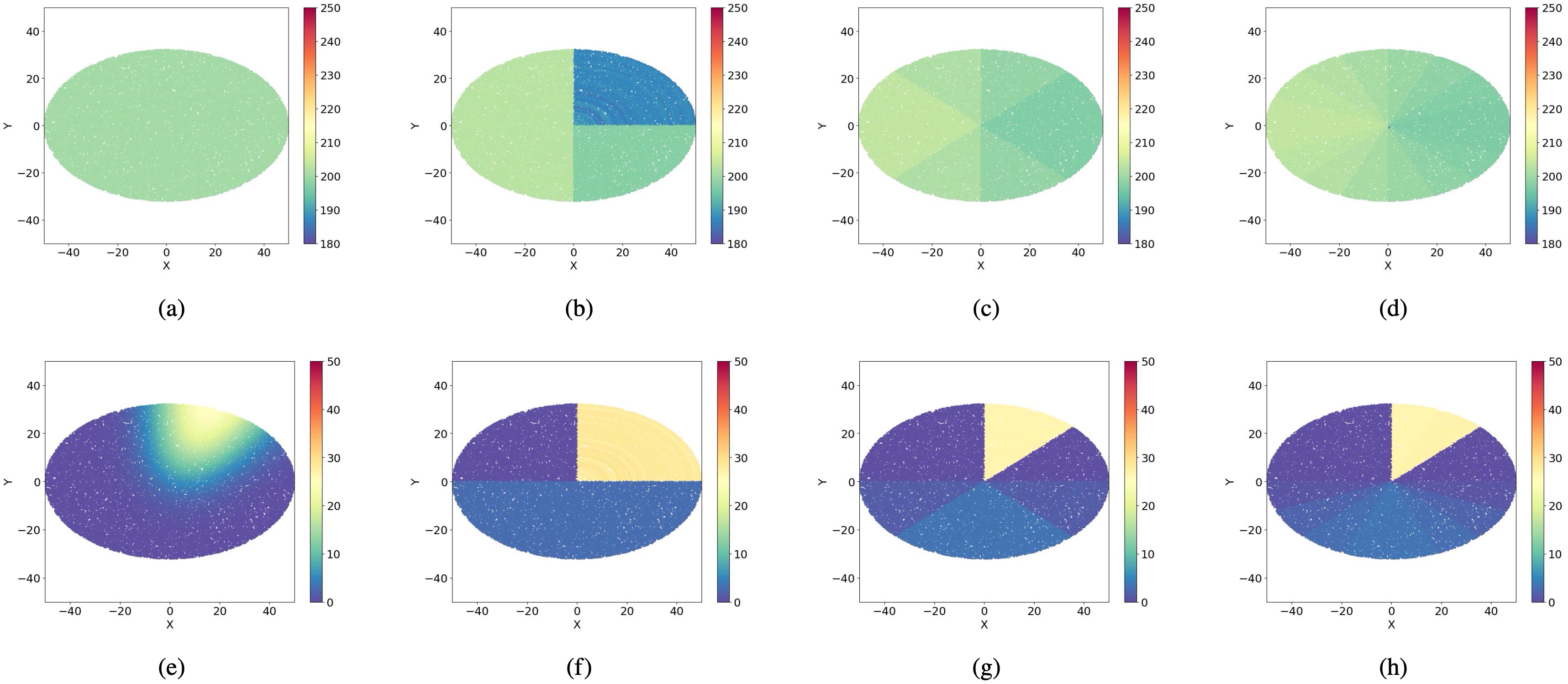}
\caption{  
In the top (bottom) panels  the transversal (radial) velocity map 
for the toy model discussed in the text (initial conditions in left panels) and the results of the VRMA with increasing resolution, i.e.
$N_r=50$ and $N_a=4,8,16$, are shown.  
}
\label{MW13a16} 
\end{figure*}
\begin{figure*}
\includegraphics[width=\linewidth,angle=0]{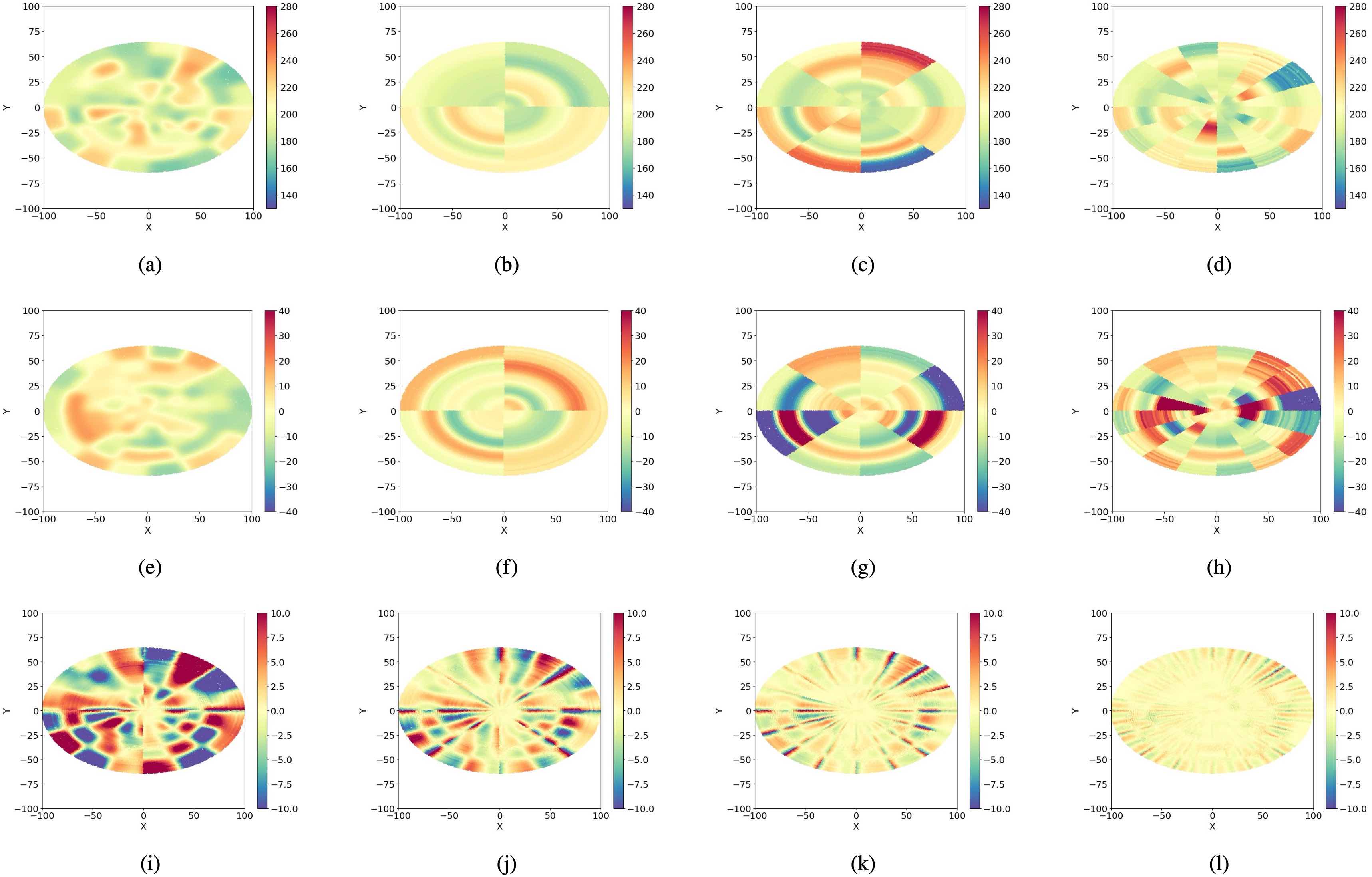}
\caption{
The first two rows as Fig.\ref{MW13a16}  but for a more complex velocity field. 
The last row shows the residuals for $N_a=4, 8, 16, 32$}
\label{MW13a12} 
\end{figure*}

\begin{figure}
\includegraphics[width=\linewidth,angle=0]{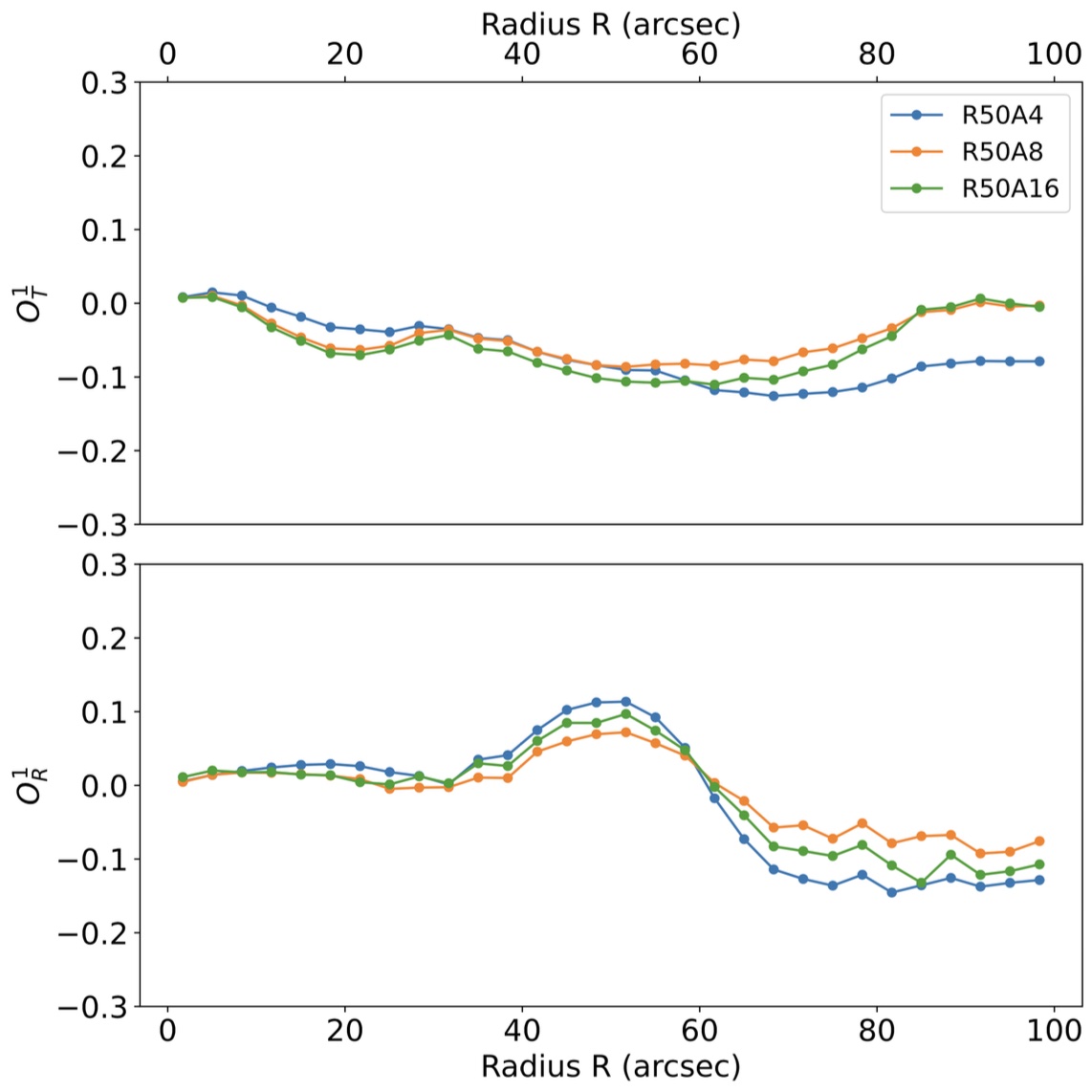}
\caption{Quadrupole moments (top panel transversal velocity, bottom panel radial velocity)
for the toy model shown in Fig.\ref{MW13a12}.}
\label{MW13a12b} 
\end{figure}
In the second class of tests we have considered the case in which one or both velocity components have a radial and a polar dependence. In particular, Figs.\ref{class2-4a}-\ref{class2-4b} presents the results of the following test: we have considered the velocity field of NGC 925 (see below for more details) reconstructed by means of the VRMA with 50 rings and 32 arcs as the input model and we have reconstructed it with the VRMA by using a lower resolution, i.e., 25 rings and 8 arcs. The agreement is very good, but for the very external rings.
\begin{figure}
\includegraphics[width=\linewidth,angle=0]{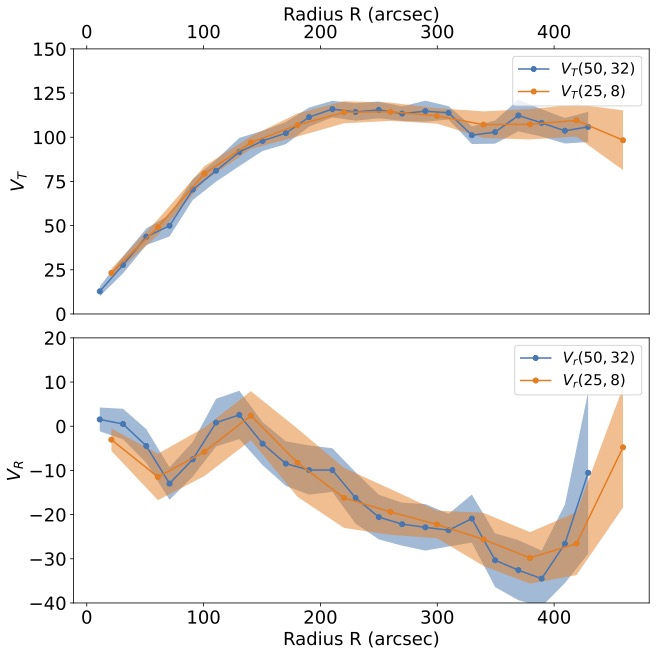}
\caption{  
In this test the input model was the  NGC 925 (see below for more details)  
VRMA velocity field with 50 rings and 32 arcs. We have reconstructed the velocity field with the VRMA using 25 rings and 8 arcs. 
The  figure shows the transversal and radial velocity profiles averaged in rings. 
}
\label{class2-4a} 
\end{figure}
\begin{figure}
\includegraphics[width=\linewidth,angle=0]{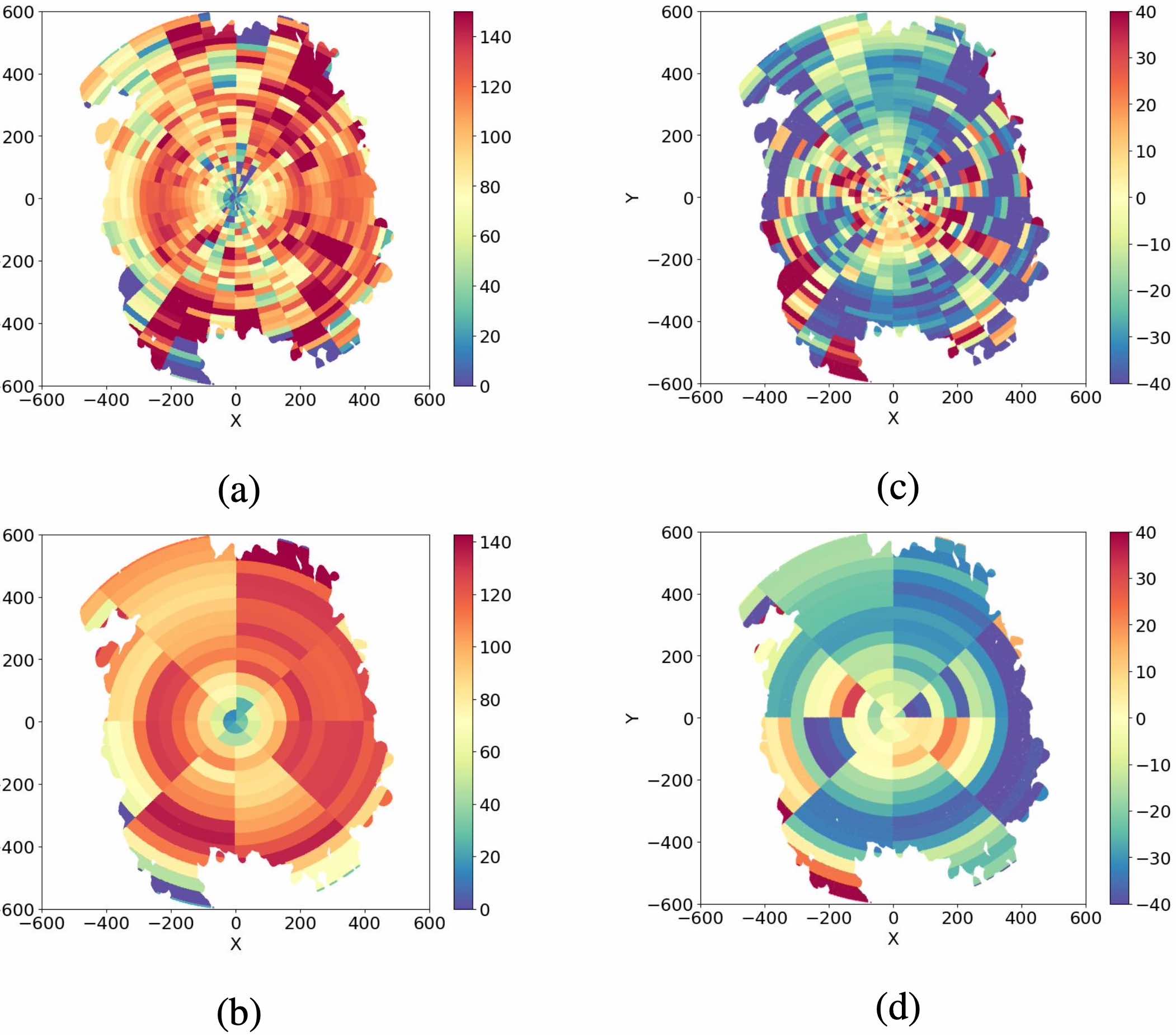}
\caption{  
Transversal and radial velocity 2D maps for the toy model of Fig.\ref{class2-4a}. 
The  four panels respectively show:  
(a) the input transversal velocity  map with 50 rings and 32 arcs;  
(b) the reconstructed transversal velocity  map with 25 rings and 8 arcs;
(a) the input radial  velocity  map;  
(b) the reconstructed radial velocity}
\label{class2-4b} 
\end{figure}

{  We refer the interested reader to  Appendix A where other tests with simple toy models are discussed.} 

\section{Results} 
\label{results}

\subsection{Sample selection}

The sample that we have used to study the kinematics of external galaxies was extracted from the THINGS survey, which is a high spectral and spatial resolution survey of \HI{}    emission of nearby galaxies obtained using the NRAO Very Large Array. The details of the survey are described in \cite{Walter_etal_2008}. Thanks to the high resolution of the \HI{}    images, this survey is a unique sample for the study of galaxy kinematics. The velocity resolution is 5.2 km/s or better, and the angular resolution is 6'' corresponding to a linear resolution of 58/435 pc for galaxies at a distance of 2/15 Mpc \citep{deBlok_etal_2008}. We present results for THINGS galaxies listed in Table \ref{table_1}, which are well-known and well-studied spiral galaxies. Note that we have not included NGC 3077, which is heavily interacting with M81, and NGC 4449, which is tidally disturbed and has a highly anisotropic velocity field, as well as IC 2574, which has substantial random motions. With respect to the analysis by \cite{deBlok_etal_2008}, we have also included NGC 628, NGC 3184, NGC 3351, NGC 3621, NGC 4214, NGC 5194, NGC 5236, and NGC 5457, which were excluded because of the small value of the inclination angle.

\begin{table}
\centering 
\begin{tabular}{|c|c|c|c|c|}
\hline 
Name                         &   $i $ ($^\circ$)   & $\psi_0$  ($^\circ$)   & $\delta v_{sys}$ (km s$^{-1}$) &  $R_{25}$ ('')\\
\hline 
NGC 628             &  25      & -110      & 0.7 & 295  \\
NGC 925             &  50      &  -20         & -4.5&319   \\
NGC 2366           &  65      & -130         & -4.6&130 \\
NGC 2403           &  60      & 145       &-1.4 &480 \\
NGC 2841           &  73      & 120          &-4.8 &210\\
NGC 2903           & 66       &  65          & 0  &360\\ 
NGC 2976           & 54       & -50            &  -4.8 &215 \\
NGC 3031           & 59       & -60           & 32.7 &656 \\
NGC 3184           &  29     & 90           & -0.9 &220 \\
NGC 3198           &  72     & 55         & -3.9 &215\\
NGC 3351           &  39     & 70       & -4.2 &215\\
NGC 3521  	    &  69      & -75      &  -4.0 &250\\
NGC 3621            &  62      &-75        & 7.4  &290\\
NGC 3627            &  61      & 90          & -10.9 &310\\
NGC 4214 	    &   38     & 190      &-1.14  &205  \\
NGC 4736             &   44    & -30         & 0.92  &230\\
NGC 4826             &   64    &  155        & 8.4 & 315 \\
NGC 5055	     &   51    &  170        &8.2     &360\\
NGC 5194             &   38    & 95          & 26.8 &230\\
NGC 5236             &   31    & 45           & 4.0   &380\\
NGC 5457             &   30    & -120      & 1.83 &720\\
NGC 6946            &  35      & 30         & 47.3  &340\\
NGC 7331            &  77      & 100       & 21.9  &275\\
NGC 7793 	    & 43       & -20         & 0.0   &315\\
DDO 154              &    70    &  45      & 0.6    &100    \\
\hline
\end{tabular}
\caption{  
Parameters of the galaxies in our sample: 
$i$ is the  inclination angle, 
$\psi_0$ is the orientation angle (see Sect.\ref{VRM} for details),
{    $\delta v_{sys}$  is the difference between our best fit  value of the systemic velocity and the estimation by Walter et al. (2008),
  and  $R_{25}$ is the optical radius (see text for details.)} 
  }
\label{table_1} 
\end{table}

\subsection{Results for individual galaxies}

{  In the following, we present our analysis of each galaxy in the sample. We provide detailed explanations of the calculations and the significance of the figures for the first galaxy, NGC 628. For the remaining galaxies, whose figures are reported in the Appendix B, we adopt the same approach and analysis. Note that in what follows, we will report the rotation curves we computed using the {\tt kinemetry} implementation of the TRM. In all cases, we used three free parameters per ring, i.e., inclination angle, position angle, and rotational velocity. The comparison of our TRM results for the inclination angle, position angle, and $v_c$ with those available in the literature, such as \cite{deBlok_etal_2008} and \cite{DiTeodoro+Peek_2021}, shows good agreement in general. To demonstrate this, for the subsample of THINGS galaxies analyzed by \cite{deBlok_etal_2008}, we also report their measurement of the rotation curve determined with {\tt rotcur}. We have smoothed the behaviors of the position angle and inclination angle, as large localized fluctuations, especially in the inner disks where measurements are more difficult, have little physical meaning in both models. Warps are typically detected in the outer regions of galaxies as they are more susceptible to the effects of external perturbations. Warps, if present, generally start beyond the optical radius $R_{25}$ \citep{Briggs_1990}.

For instance, the variation of the position angle of NGC 628 (see below) from the inner to the outer disk clearly supports the existence of a warp. This kind of deformation corresponds to smooth changes, moving from the inner to the outer disk, of the position angle and/or the inclination angle. For the case of coplanar disks, localized anisotropic variations of the radial velocity are better identified by means of the VRMA rather than indirectly by looking for changes in the orientation angles through the TRM. Finally, we show the PDF of residuals obtained with the TRM and VRMA with approximately the same number of parameters; in general, they are very similar.
}

 Non-axisymmetric  components are certainly present in the observed velocities of galaxies. The question is whether these are small-scale random motions or actual large-scale flows corresponding to large-scale variations of the circular velocity or continuous radial inflows. As shown by the tests with toy models presented in the previous section, the VRMA is able to reconstruct not only the amplitude of the transverse and radial velocity components but also their direction. However, in some cases, the reconstruction was not correct as $v_t$ and $v_r$ may be partially degenerate for small radii and angular width of the cells of the coarse-grained map. In this situation, a real variation of one component can give rise to a detected variation of the other one, which is an artifact of the reconstruction method. We have discussed that, because such degeneracies depend on the angular size of cells, the reconstruction can be considered accurate only if it converges by changing the resolution of the coarse-grained map. A quantitative analysis can be provided by looking at the moments {  introduced in Sect.\ref{convergence}.} In addition, one should keep in mind that despite the THINGS data being extremely well-resolved, the inner parts of the galaxies, i.e., at small radii, could potentially still be affected by beam smearing \citep{deBlok_etal_2008,Oh_etal_2008}. For these reasons, we have plotted, for each galaxy, the coarse-grained maps of the two velocity components with different resolutions (i.e., number of arcs with fixed number of rings): if, by varying the number of arcs, the reconstructed map converges, i.e., it shows the same signal in a given angular region, one can confidently conclude that the reconstruction has correctly measured the intrinsic properties of the velocity field.

As a final remark,  hereafter we limit our analysis to a brief description of the 2D velocity maps and their spatial anisotropies, pointing out  whether it is observed a correlation with spatial structures; however, we stress  a causal physical relationship remains highly questionable as long as there is no dynamical modeling. {  Such a task  goes beyond the scope of this work.}



\subsubsection{NGC 628}
\label{ngc628} 

NGC 628 is considered an archetypal example of a grand design spiral galaxy; it shows two spiral arms that are visible also in the \HI{}    intensity map. The TRM analysis finds that both the P.A. and the inclination angle are close to constant only for $R<300"$, a radius corresponding to the inner optical disk \citep{Schruba_etal_2011}. While the values of the P.A. of the inner disk from the distribution of \HI{}    determined by \cite{Kamphuis+Brigss_1992}, \cite{Daigle_etal_2006} and \cite{Aniyan_etal_2018} agree with each other and with its kinematical value, the inclination angle differs greatly between different authors. \cite{Kamphuis+Brigss_1992} reported a value of $i=6.5^\circ$ for the inner disk, similar to that of \cite{Aniyan_etal_2018} ($i=8.5^\circ$) and \cite{Walter_etal_2008} ($i=8^\circ$) whereas other authors report different values. In particular, \cite{Daigle_etal_2006} found $i=25^\circ$ (photometric) and $i=26.4^\circ$ (kinematical) while  \cite{deBlok_etal_2008}  found $i=15^\circ$ for the \HI{}    disk and $i=21^\circ$ for the optical disk. These differences {  arise from different estimation methods:  \cite{Kamphuis+Brigss_1992} assumed the value of $i$,  \cite{Daigle_etal_2006}  used  {\tt rotcur}  to find the kinematical  parameters (they found values in agreement with the photometric parameters  from the RC3 catalogue) and
\cite{Aniyan_etal_2018}   determined $i$ via kinematic fit to the \HI{}   data from the THINGS survey. {   Note that the amplitude of the transversal and radial velocity components depend on the value of the global inclination angle $i$.}}

{  The code {\tt kinemetry}  finds an inclination angle of $i=25^\circ$ for the inner disk (see panel (b) of Fig.\ref{NGC628}), a value that coincides with the photometric inclination angle reported in \cite{Daigle_etal_2006}. This is the value of the global inclination angle that we employed for the VRM analysis. } The difference in the amplitude of $v_c(R)$ between our TRM results and those of \cite{Aniyan_etal_2018} is due to the different values of the inclination angle used. A simple rescaling of the circular velocity by $\sin(8^\circ)/\sin(25^\circ)$ gives a similar rotation curve to that of \cite{Aniyan_etal_2018}, as the radial dependence of $v_c(R)$ is approximately the same.

The large change in the P.A. of about $90^\circ$ between 200'' and 400'' corresponds to a change in the orientation of the kinematic axis from the internal disk to the outermost regions.  Given that $R_{25} = 295''$ \citep{Schruba_etal_2011} this variation can be compatible with the presence of the warp in the outer region of the galaxy \citep{Briggs_1990}.   When limiting the analysis to $R<250''$ and considering that the inclination angle and P.A. are both constant, we find that the transversal velocity profile $v_t(R)$ obtained by means of the VRM is similar to the circular velocity $v_c(R)$ measured by the TRM. {  However, in the outer disk, there is a significant difference between $v_t(R)$ and $v_c(R)$ due to the change in both these angles.  Additionally, for $R<300"$, the radial velocity has a small amplitude, i.e. $v_r<10$ km s$^{-1}$, but it increases when the P.A. shows the large variation.}

The transversal ($O^i_T(R)$) and radial ($O^i_R(R)$) octopole moments (where $i=1, ..., 8$) show excellent convergence when determined with $N_r = 50$ and different numbers of arcs ($N_a = 8, 16,$ and $32$ --- see panel (c) of Fig.\ref{NGC628}) only in the range for $R < 300''$. {  At larger radii, the amplitude of fluctuations increases. We may thus conclude that, in the inner disk, the octopole kinematic maps of both $v_t(R)$ and $v_r(R)$ converge to well-defined anisotropy patterns corresponding to coherent and small amplitude streaming motions.  The transversal and radial velocity maps for $N_r = 50$ and $N_a = 1, 8, 16, 32$  are shown respectively in in the panels from (e) to (h)  and from (i) to (l)   of Fig.\ref{NGC628}. 

For  $R_c>300''$, we find that the inclination angle varies by approximately $30^\circ$ and the P.A. by approximately $70^\circ$. These variations may correspond to a very large warp in the disk's external regions, as assumed by the TRM, while the radial motions remain small.  However, it should be noted that  the $v_t$ and $v_r$ 2D reconstructed maps do not present the  symmetric features that are formed if a symmetric warp is present (see discussion in Sect.\ref{tests}).}

By assuming the disk remains relatively flat (i.e., with a variation of both angles smaller than $\approx 10^\circ$) the velocity field in the external regions may alternatively be interpreted as characterized by a large radial velocity component whose amplitude increases with radius and is characterized by large-scale gradients. Indeed, the transversal and radial octopole $O^i_T(R)$ and $O^i_R(R)$ do not converge well in the outermost regions for $N_r=50$ and $N_a=8, 16$, and 32: this can be evidence in favor of a rough velocity field. On the other hand, in the inner region $R<R_{25}\approx 300''$, they present low, i.e. $<10\%$, variations.

The inner disk fluctuations in the transversal and radial velocity fields do not show any systematic differences with respect to the direction of the kinematic axis (see panel (d) of Fig.\ref{NGC628}). This is consistent with the absence of structures, such as a bar, that break circular symmetry. Note that these fluctuations are computed by considering pixels with an angle in a range of $\theta \in [-20^\circ,20^\circ]$ parallel or perpendicular to the kinematic axis. In contrast, {  the  velocity fields in the outermost regions } show a clear breaking of circular symmetry, the origin of which is unclear, as discussed below.

{  
Panel (m) of Fig.\ref{NGC628} shows the number of pixel per cell $N_p^c$ in function of the radial distance and for two different resolutions, i.e. $N_a=4, 32$ and $N_r=50$.
The horizontal line corresponds to twice the number of pixels $N_{pix}$ in the beam area (see discussion in Sect.\ref{overfitting}). 
We note that both for $N_a=4$  and $N_a=32$  the number of pixel per cell is $N_p^c>2N_{pix}$ for all cells but those in the  innermost regions of the  disk and some sparse cells in the boundaries of the image. 
}
 The PDF of residuals (see panel (n) of Fig.\ref{NGC628}) shows that the VRMA 
case for $N_a=4$ and $N_r=50$ that has a number of free parameters  $N_{par} = 400$ is slighter more peaked than the residuals of {\tt kinemetry} that used $N_{par} = 300$ free parameters. Clearly by increasing $N_a$ we find that the residuals' dispersion further decreases: this is easily explained as due the increasing number of parameters used in the fit.

{  In conclusion, we detect no significant asymmetric motions in the inner part of the disk of NGC 628 whereas in the outer regions our analysis is unable to separate radial motions from a possible asymmetric warp. Indeed, the change of orientation of the kinematic axis going from the inner to the outer disk can be explained by a combination of a large warp and an increase in radial velocities.}

\begin{figure*}
\includegraphics[width=\linewidth,angle=0]{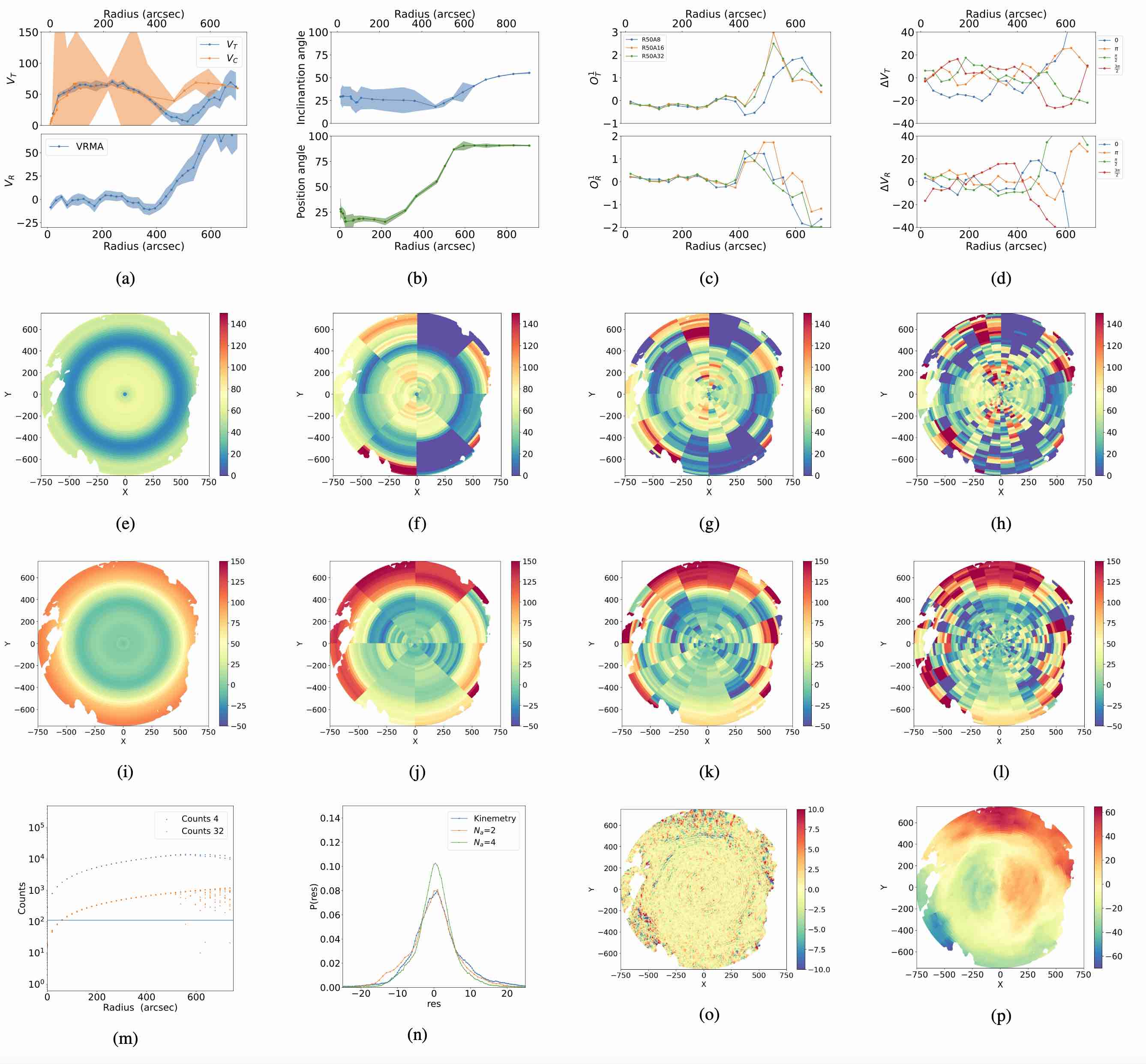}
\caption{NGC 628:
The orientation of the galaxy is such that the kinematic axis in the inner disk is oriented along the horizontal axis, with positive 
values of the ordinate corresponding to the receding velocity and negative values to approaching motions.  
 (a): upper panel transversal velocity profile $v_t(R)$ from the VRM and circular velocity from the TRM; 
 (a): bottom panel radial velocity profile $v_t(R)$ from the VRM; 
 (b): upper panel inclination angle  (error bars from the code {\tt kinemetry}; 
 (b): bottom panel position angle (error bars from the code {\tt kinemetry}; 
 (c): upper panel moment $O^1_T(R)$ for $N_r=50$ and $N_a=8,16,32$; 
 bottom panel moment $O^1_R(R)$ for $N_r=50$ and $N_a=8,16,32$;  
 (d): upper panels: transversal velocity difference along the kinematic axis;    bottom panels: radial velocity difference along the kinematic axis;   
 (e): transversal velocity map   for $N_r=50$ and $N_a=1$; 
 (f): transversal velocity map  for $N_r=50$ and $N_a=8$; 
 (g): transversal velocity map  for $N_r=50$ and $N_a=16$; 
 (h): transversal velocity map  for $N_r=50$ and $N_a=32$; 
 (i): radial velocity map   for $N_r=50$ and $N_a=1$; 
 (j): radial velocity map   for $N_r=50$ and $N_a=8$; 
 (k): radial velocity map  for $N_r=50$ and $N_a=16$; 
 (l): radial velocity map   for $N_r=50$ and $N_a=32$; 
{  (m): counts in cells for $N_r=50$ and $N_a=16, 32$;}
  (n): histogram of residuals with {\tt kinemetry} and with the VRMA with for $N_r=50$ and $N_a=2, 4$ ;
 (o):  residual velocity map    for $N_r=50$ and $N_a=16$. 
 (p):  LOS velocity map.
 All 2D maps are in the coordinates in the plane of the galaxy.  
 }
\label{NGC628} 
\end{figure*}

\section{Discussion and conclusion} 
\label{conclusion}

{  In this paper we have developed  an algorithm,  named  the velocity ring model (VRM) and its refinement  and the velocity ring model  with arcs  (VRMA), that uses an ordinary least square method for fitting a non-axisymmetric model to the observed two-dimensional velocity field.}
 It assumes that the galactic disk is flat and, {  for this reason, it}  is best applied within the optical disk where warps are rare.
 {  The novelty of the VRMA code  is that it allows the fitting of arcs instead of full (semi-) circles, which enables the reconstruction of coarse-grained two-dimensional  maps for both velocity components with an angular resolution that can be varied.}  
Conceptually, the VRMA can be considered a special case of the tilted ring model (TRM) \citep{Warner_etal_1973,rogstad_etal_1974}, with the inclination angle and P.A. held constant. It extends the method introduced by \cite{Barnes+Sellwood_2003} and developed by \cite{Spekkens+Sellwood_2007,Sellwood_etal_2021}, {  to characterize any spatially anisotropic and heterogeneous velocity field, even with large velocity gradients in 
both the transversal and radial velocity component. }

The assumption of disk flatness used in the VRMA algorithm is reasonable for studying velocity fields within the optical radius of a galaxy where warps are rare. However, it may not be suitable for studying the outer parts of galaxies where warps are more likely to occur. Warps are typically moderate geometric deformations that affect the distribution of neutral hydrogen in the peripheries of a galaxy. {  For this reason, the} VRMA algorithm can be combined with the TRM to constrain the flatness of a galaxy's disk in regions where warps may be present. However, it is not always straightforward to determine whether a galaxy is warped or characterized by significant radial flows. Each case must be analyzed carefully, taking into account the physical implications of both scenarios. 
{  For instance}, a warped geometry is more likely if the galaxy has a nearby perturber. The combined VRMA-TRM can help identify the region in which one of these effects is certainly present. Overall, while the VRMA algorithm is a useful tool for studying galactic kinematics,  it is important to consider the limitations of the assumption of disk flatness and to carefully analyze each case to determine the underlying physical processes.
By applying these methods to 25 galaxies from the THINGS sample  \citep{Walter_etal_2008} we have found that 

\begin{itemize}
\item Determinations of the transversal velocity component $v_t(R)$, averaged over a ring,  by the  VRMA are in good agreement with the circular velocity $v_c(R)$ obtained using the TRM.
\item  Within the optical disk, where the assumption of disk flatness is reliable, the average profiles of radial velocity $v_r(R)$, obtained by estimating them over a ring, are similar to those obtained by \cite{DiTeodoro+Peek_2021} who included the radial velocity directly in the initial TRM fit. Some small differences between the two methods may be attributed to their different assumptions regarding the galaxy's geometry.

\item Within this region, the VRMA method can be used to reconstruct 2D coarse-grained maps of both components $v_t(R,\theta)$ and $v_r(R,\theta)$ that may reveal  large-scale coherent flows in both components.  We have shown that these maps converge by varying their angular resolution. {  These maps reveal a wealth of information about galactic kinematics, including large-scale flows and continuous radial inflows. As a result, they provide crucial information for dynamical modeling purposes.}
\item The $v_t(R,\theta) $ and $v_r(R,\theta)$ 2D coarse-grained maps allow  to identify kinematic structures  corresponding to dynamical perturbations induced by bar-like or oval distortions, warps,  lopsidedness or  satellites.  Some evidences {  of apparent correlations between   kinematic features and real space structures are discussed in each individual case. However, only by developing a full dynamical model it may be possible to establish a real physical relation between kinematic and spatial structures. }
\item Some galaxies, most notably NGC 3031, NGC 5194, NGC 5236, but also, in a lesser way NGC 2366, show a 2D velocity maps in which there are clear features induced by the presence of  a satellite: in this case the relation between kinematic and spatial structures seem to be more straightforward. Whereas NGC 925, NGC 2903, NGC 3198, NGC 3351, NGC 3627, NGC 4214 show features that can be explained as due to a bar structure, but in these cases the connection needs to be studied in much more details. 
\item By analyzing toy models we have shown that $v_t(R,\theta) $ and $v_r(R,\theta)$ maps allow to identify axisymmetric warps if present as they correspond to symmetric and correlated variations of the transversal and radial velocity fields. However,  no axisymmetric warps have been found in the galaxies of our sample. On the other hand, non-axisymmetric warps are more difficult to be disentangled from a non-axisymmetric velocity field.
\item  {  Peripheries of the galaxies in our sample, beyond the optical radius, are characterized by different kinematic features}. For instance, NGC 628 shows a velocity field that can be interpreted as a large warp. On the other hand, NGC 925, NGC 2841, NGC 5055, and NGC 7793 display signatures that suggest the presence of a mild warp in their outer regions. However, an axi-symmetric warp is ruled out since it would induce an axi-symmetric gradient in the velocity field, which is not observed. Instead, these kinematic features may be entirely or partially attributed to radial flows. Overall, our results highlight the importance of accurately identifying and characterizing kinematic structures and perturbations in the outer regions of galaxies, which can provide valuable insights into the underlying dynamical processes that shape their evolution.

\end{itemize} 

In summary, the VRM/VRMA methods offer  reliable means of measuring the kinematics of galaxies providing coarse-grained {  and angularly resolved}  2D velocity maps. These methods facilitate the detection of various kinematic structures and perturbations.This approach has the potential to answer significant questions about galaxy formation and evolution, including the impact of bars, mergers, and other interactions on the observed kinematics of galaxies.

\section*{Acknowledgements}

We thank Roberto Capuzzo-Dolcetta, Michael Joyce,  Mart\'in L\'opez-Corredoira, Daniel Pfenniger and Hai-Feng Wang  
for useful comments and discussions.  
SC acknowledges funding from the State Research Agency (AEI-MCINN) of the Spanish Ministry of Science and Innovation under the grant "Thick discs, relics of the infancy of galaxies" with reference PID2020-113213GA-I00.
We warmly thank an anonymous referee for a number of very useful comments, suggestions and criticisms that have allowed us to improve the presentation of our results.

\section*{ Data availability} 
The data used in this paper are from the THINGS survey \citep{Walter_etal_2008}. 
All other data presented in this work concerning toy models has been generated by the authors and will be shared on reasonable request to the corresponding author.
The code to compute the VRM is publicly available at the webpage: {\tt https://github.com/MatteoStraccamore/VRM$\_$VRMA}.


\bibliographystyle{mnras}

\section*{Appendix A: Additional tests with toy models} 

{   In this Appendix we further study some simple toy models beyond those discussed in Sect.\ref{tests}. 
The first toy model  aims to test the effect of a velocity dispersion on the reconstruction method. It is indeed well known that beyond the transversal and radial motions 
there can be non-negligible vertical and horizontal velocity dispersions which tend to blur the image. The simplest way to take into account this dispersion is to add 
three-dimensional random motions to the toy disk models discussed previously, and varied the thickness of the disk from $h=R_{disc}/10$ to $h=R_{disc}/3$ so to have a more significant effect.  To each velocity component of the unperturbed toy model it is added a random value with  zero average extracted from a uniform distribution dispersion $A_N$}. We found that even for small inclination angles (i.e., $i = 10^\circ$) and large values of random motions (up to $\sim 30\%$ of the peak value of $v_t$, or $A_N=40$ km/s, which is larger than the maximum value of $v_r$), the VRM is able to correctly reconstruct the average behaviors of $v_t$ and $v_r$. There is only a small offset in the determination of $v_t$ at large radii (see Fig.\ref{class1b}). Furthermore, we find that, even for small values of the inclination angle (i.e., $i =10^\circ$) and for large values of random motions, i.e. up to $\sim 30\%$ of the peak value of $v_t$ (i.e., $A_N=40$ km s$^{-1}$ and thus larger than the maximum value of $v_r$), the VRM is able to correctly reconstruct the average behaviors of $v_t$ and $v_r$ (see Fig.\ref{class1b}): there is only a very small offset in the determination of $v_t$ at large radii.
\begin{figure}
\includegraphics[width=\linewidth,angle=0]{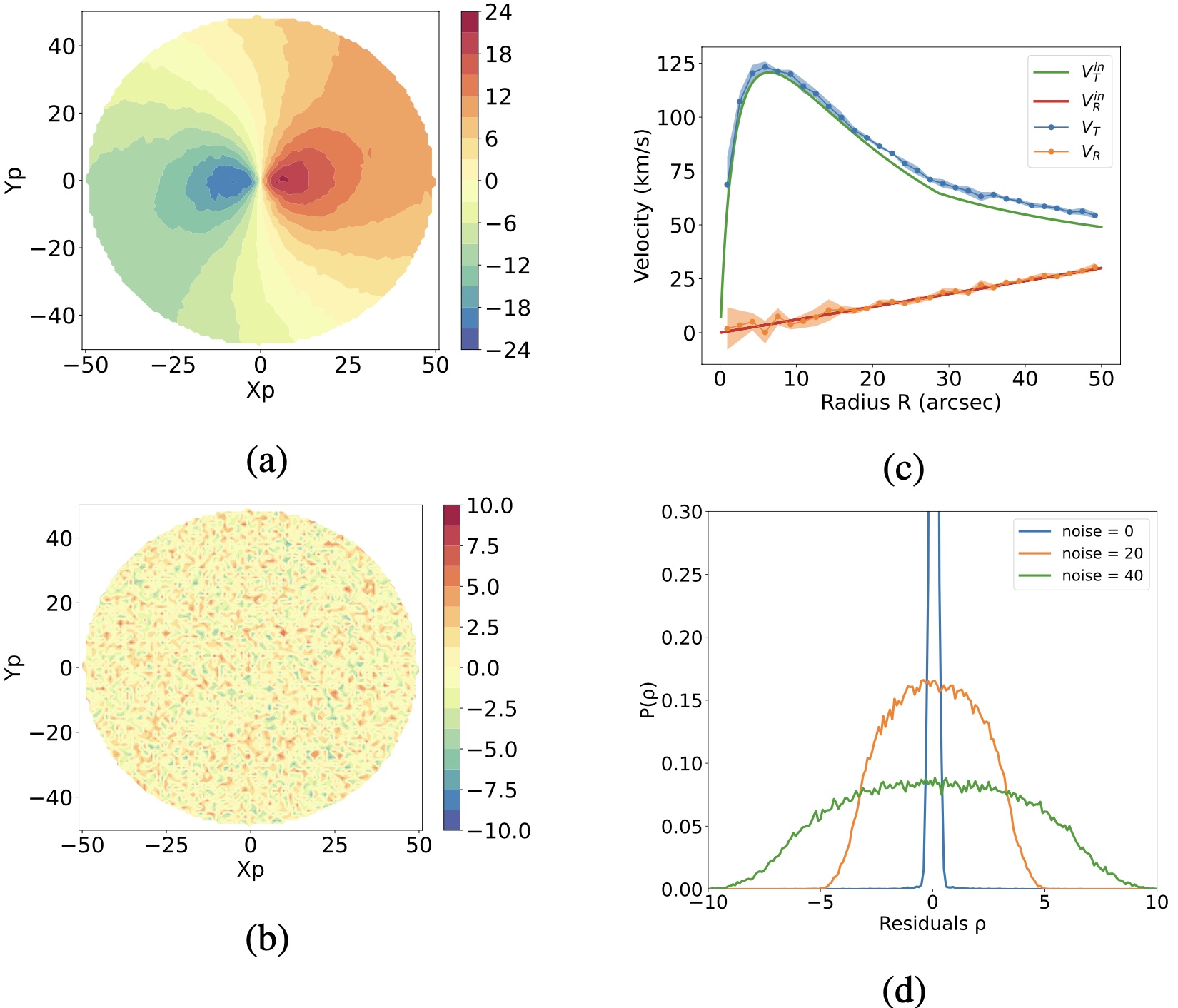}
\caption{As  Fig.\ref{class1}  but with the addition of a three dimensional random velocity. In this case the inclination angle is $i=10^\circ$ and the dispersion of the random motion is  $A_N=40$ km s$^{-1}$}
\label{class1b} 
\end{figure}

{  Let us consider the effect of an incorrect assumption on each of  the two methods, the VRMA and the TRM, with the help of simple toy models. 
The aim of this test, as others presented in this section, is to check the ability of the different methods to reconstruct the properties of a given distribution. As  discussed  above, the key role is played by the assumptions on which the methods are based on. In the first test, we apply the VRMA to the case where the disk is warped and the velocity field has only a rotational component. In the second test, we apply the TRM to the case where the disk is flat and the velocity field has both a non zero rotational and radial component. These tests can help us to better understand the limitations of the methods, especially in the peripheries of galaxies where warps, large radial flows, or a combination of both can be present. }

We report in Fig.\ref{class3dPA10} the results of the VRMA analysis of a toy galaxy with $v_r=0$, $v_t=200$ km/s and an axisymmetric curvature corresponding to a variation of the P.A. of $10^\circ$ passing from the inner to the outer regions of the disc. It can be seen that the warp manifests itself as a variation, in the outermost regions, of the radial velocity by approximately 30 km/s and the transverse velocity by approximately 20 km/s. 
\begin{figure*}
\includegraphics[width=\linewidth,angle=0]{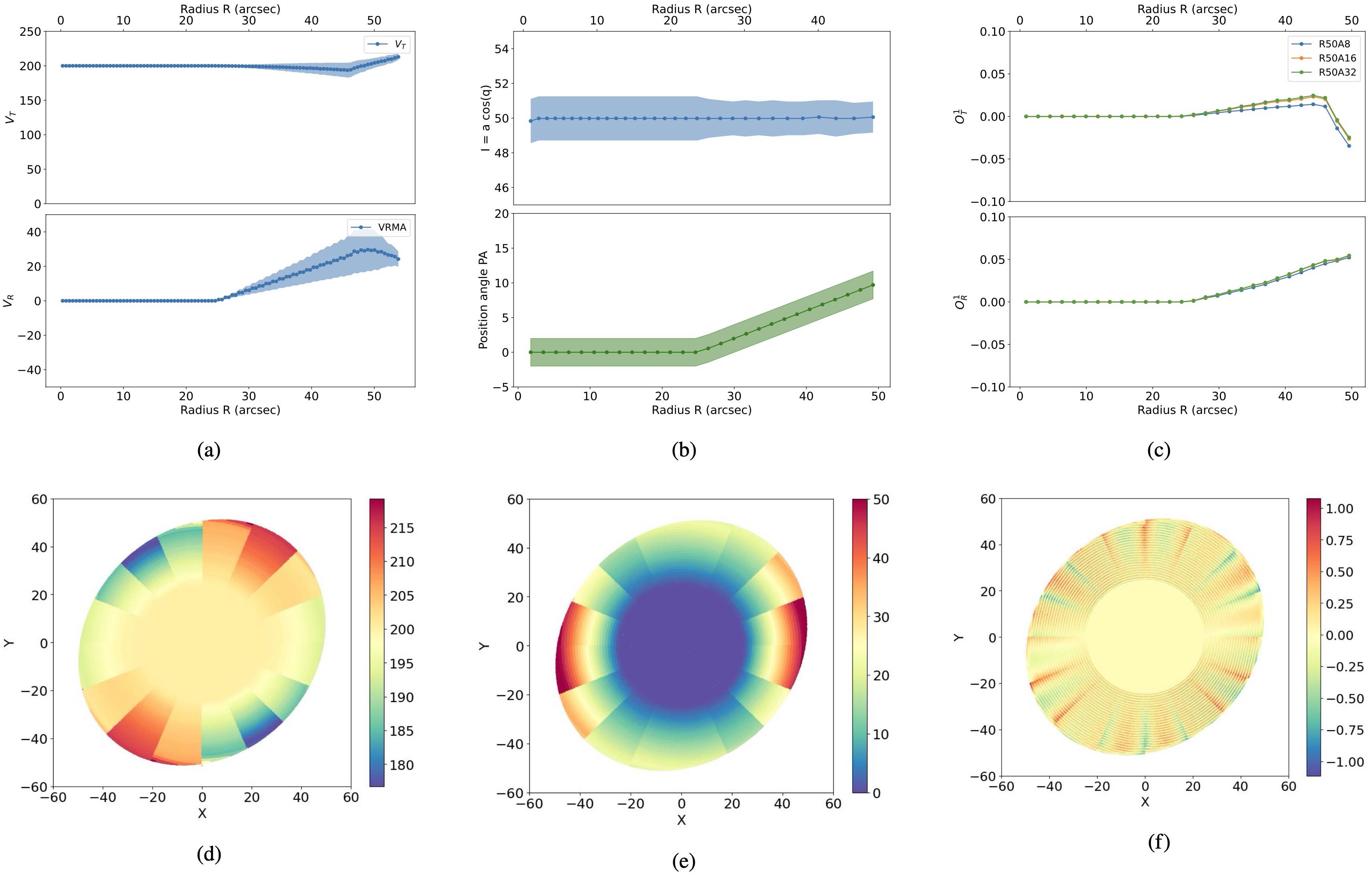}
\centering
\caption{ In this test we have analyzed with the VRMA a toy galaxy with $v_r=0$, $v_t=200$ km s$^{-1}$ and a warp  corresponding to a change of the P.A. of  $10^\circ$ going from the inner to the outer regions of the disk.   The six panels show respectively: 
(a) the transversal/circular and radial velocity profiles; 
(b) the inclination and position angle in function of the radius;  
(c) the octopole moments 
(d) the transversal velocity reconstructed map; 
(e) the radial velocity reconstructed map ; 
(f) the residual map.
}
\label{class3dPA10} 
\end{figure*}
\begin{figure*}
\includegraphics[width=\linewidth,angle=0]{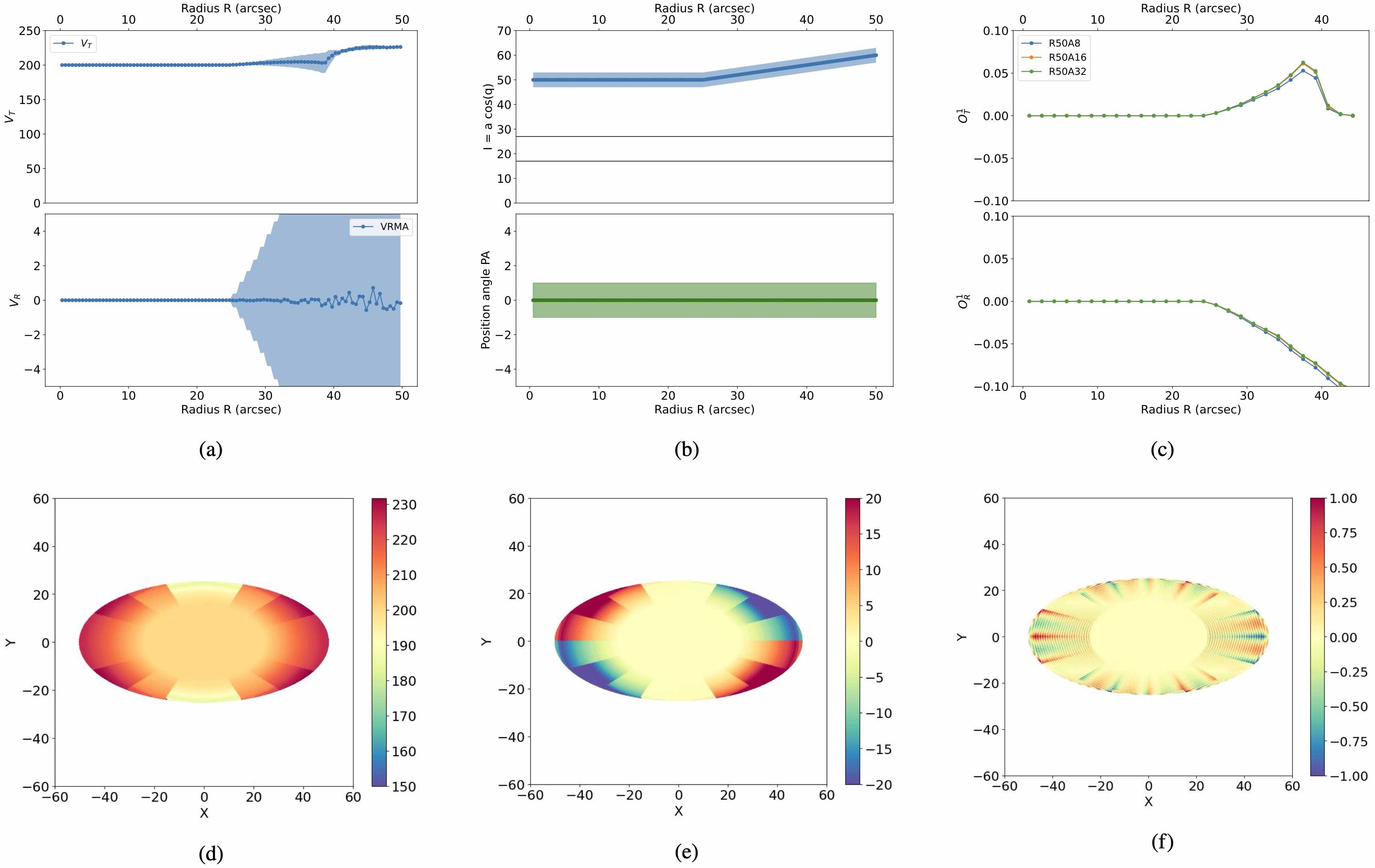}
\caption{ As Fig.\ref{class3dPA10} but in this case the warp corresponds to $\Delta i =10^\circ$.}
\label{di10} 
\end{figure*}
 When instead of varying the P.A. the inclination angle is changed going from the inner to the outer regions of the disk the radial velocity component remains unperturbed while the rotational velocity displays a radius-dependent behavior. The case where  $i$ increases by $\approx10^\circ$  is shown in Fig.\ref{di10}. The VRM, which assumes that the disk is coplanar, detects an azimuthally  averaged radial velocity that is close to zero, but with a significant dispersion. The transversal velocity displays a small anomalous growth when the inclination angle varies, which is {  the result of the fact that the VRMA keeps the inclination fixed to a single value.}  A clear feature is represented by the symmetric azimuthal variation of each of the two VRMA reconstructed components $v_t$ and $v_r$.

Finally, Fig.\ref{dPA10-di10} shows a case where the  motion is purely circular with $v_c=v_t=200$ km/s and the disk is warped in such a way that both the inclination angle and the P.A.  vary by $10^\circ$.   The VRM/VRMA  detects a change of the both velocity components of $\approx 30$ km/s.   
\begin{figure*}
\includegraphics[width=\linewidth,angle=0]{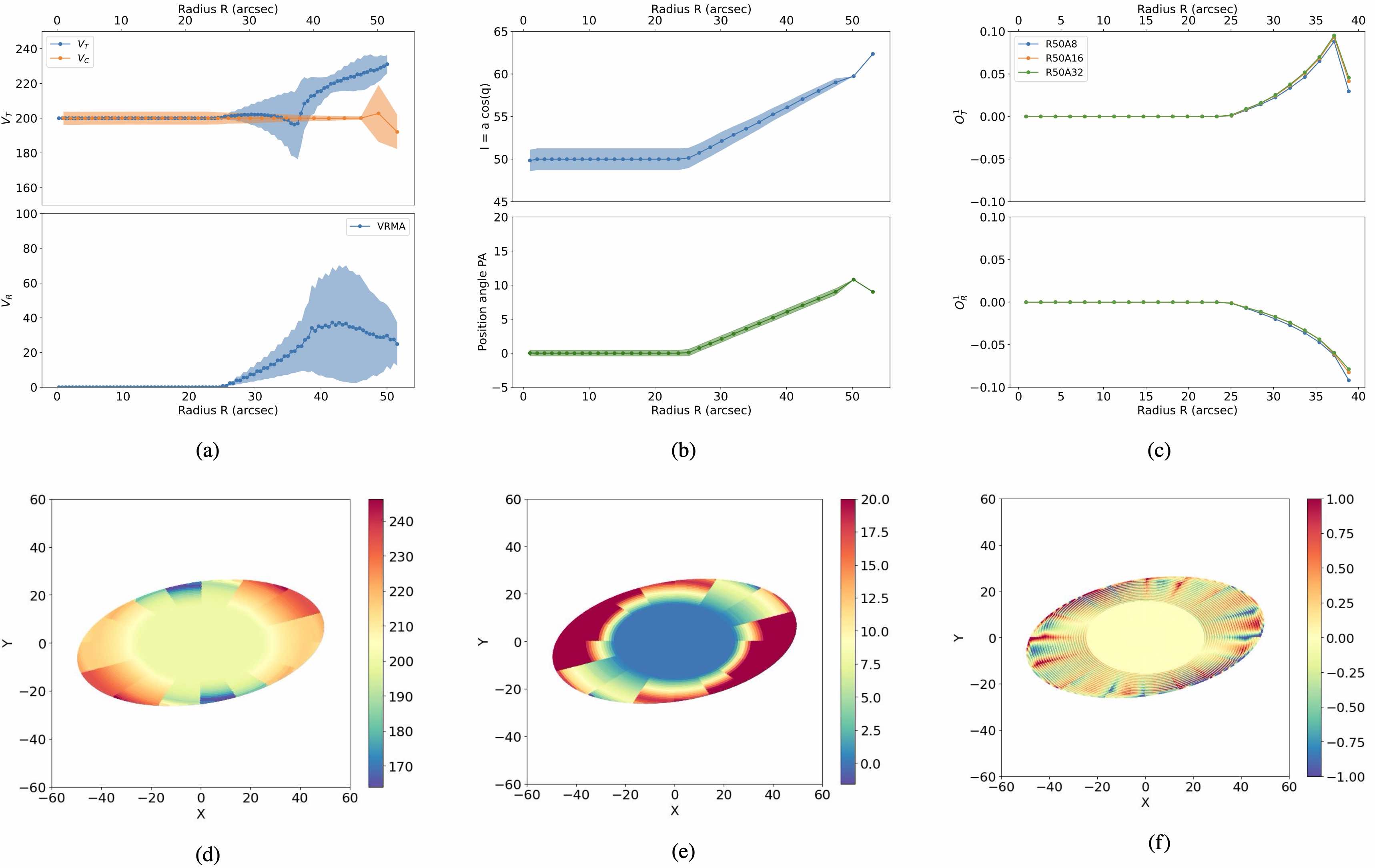}
\caption{ As Fig.\ref{class3dPA10} but in this case the warp corresponds to $\Delta i =10^\circ$ and $\Delta PA =10^\circ$.}
\label{dPA10-di10} 
\end{figure*}
It is worth stressing that in the examples in  Figs.\ref{class3dPA10}-\ref{dPA10-di10}, the  signal detected by the VRM/VRMA is spurious,{   i.e., it is induced by incorrect assumptions on the system geometry as  the disk is warped by construction whereas the VRM assumes it is flat. As discussed above,} this same problem affects all other methods (e.g., \cite{Barnes+Sellwood_2003,Spekkens+Sellwood_2007,Sellwood+Spekkens_2015,Sellwood_etal_2021}) which assume that the disk is flat when instead it  has a warped  geometry. 
{  In addition, note that the toy models in of Figs.\ref{class3dPA10}-\ref{dPA10-di10} correspond to symmetrically warped disks. This symmetry corresponds to the symmetric variation of the P.A. and of the inclination angle over the disk and induces azimuthally symmetric variations of both the velocity components $v_t$ and $v_r$ reconstructed by the VRMA.   However, if the warp is not symmetrical, the induced variations of the radial and transverse velocity components detected by VRMA will also not be symmetrical and thus in real cases it is more difficult to disentangle the degeneracy between a geometrical deformation and a radial flow. }

As we have seen from the above examples, the VRM/VRMA is able to accurately reconstruct the velocity components in a system that satisfies its assumptions, namely for a flat disk. However, if the system is not flat, the basic assumption of the VRM is violated, as seen in the example in Figs.\ref{class3dPA10}-\ref{dPA10-di10} where the warp is misinterpreted as a non-zero radial velocity. Similarly, the TRM, when applied to a toy model that is a flat disk with non-zero radial velocity
{  with the constrain that radial motions are not allowed to be varied from zero in the TMR fit, it clearly } detects zero radial motions and a warp that is, however, an artifact.  Fig.\ref{DiskRot+Rad_varying} shows an example where the transversal velocity dominates over the radial one at all radii but the  outermost ones.  We note that as the input radial velocity component increases, the circular velocity reconstructed by the TRM only slightly deviates from the input transversal velocity while the P.A. and inclination angles significantly vary from their input values. This highlights the limitations of the TRM in cases where the radial velocity component is non-negligible which are induced by the intrinsic degeneracy between radial flows and geometric deformations of the disk.
\begin{figure}
\includegraphics[width=\linewidth,angle=0]{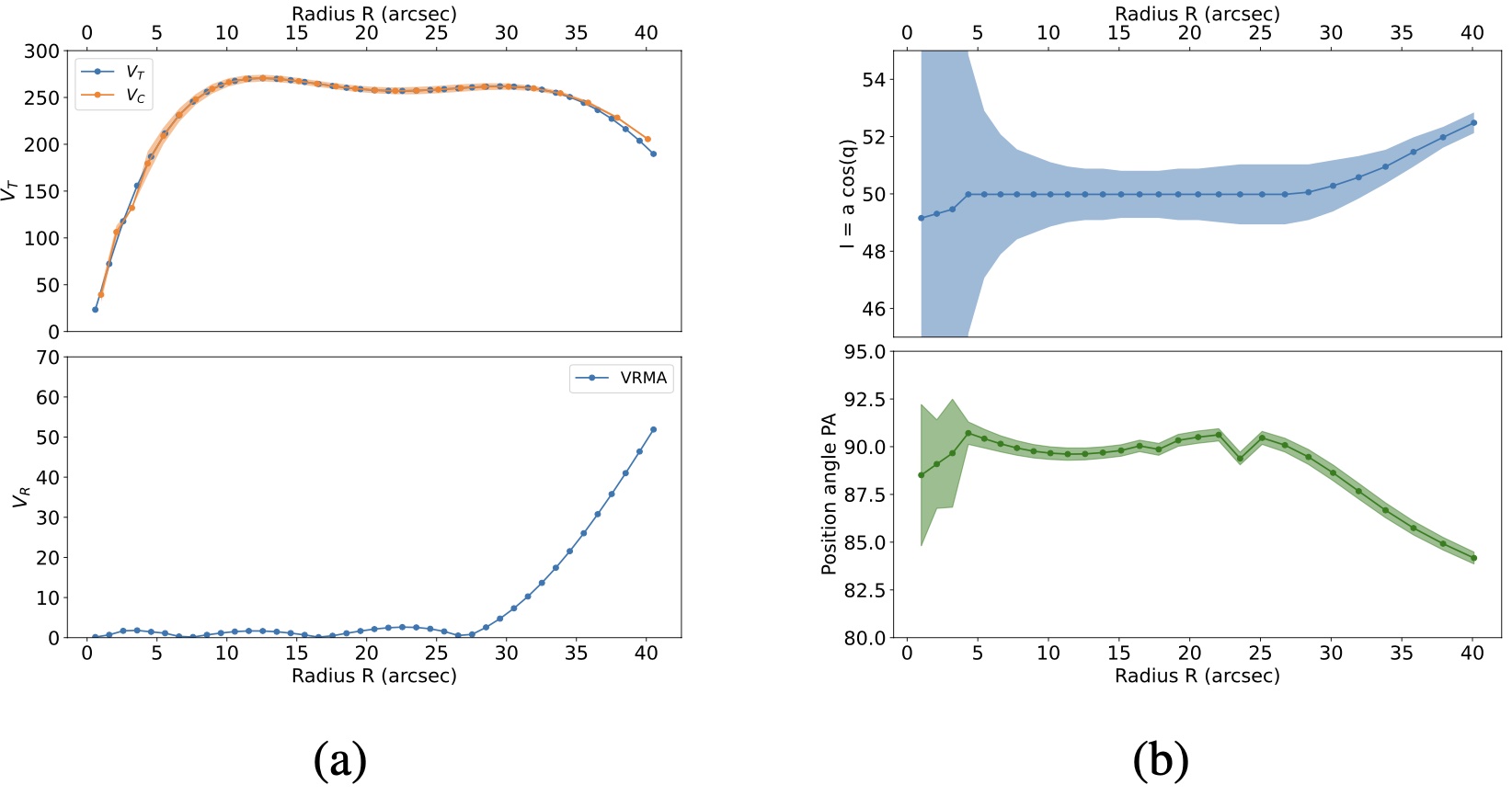}
\caption{A toy disk model in which $v_t (R)$ varies as for a simple exponential disk and $v_r(R)$ linearly grows with radius.
The upper left panel shows the transversal velocity component measured with the VRMA and the circular velocity with the TRM.
The bottom left shows the radial velocity component measured with the VRMA.
The upper right panel shows the inclination angle and the bottom right panel the P.A., both  measured with the TRM.} 
\label{DiskRot+Rad_varying} 
\end{figure}

\section*{Appendix B: Additional figures of the THINGS galaxies} 

{  In this appendix we report the analysis of the galaxies of the THINGS sample that we have considered except NGC 628 
that we have discussed in Sect.\ref{ngc628} }.


\subsubsection*{NGC 925}
 NGC 925 has a distinct bar structure as well as loosely wound spiral arms  {  (see panel (f) of Fig.\ref{NGC925})}. The bar is elongated along the kinematic axis. The TRM analysis shows that the P.A. remains relatively constant for radii smaller than 500", a radius larger than $R_{25}\approx 320"$, while the inclination angle varies by approximately 25 degrees between the inner and outer regions of the galaxy. For our analysis, we adopt an intermediate inclination angle of 50 degrees, which is similar to the values measured for the \HI{}    disk by \cite{deBlok_etal_2008} and the kinematical value by \cite{Daigle_etal_2006}. We find that $v_t(R) \approx v_c(R)$ across the whole disk: the reason for this is that the P.A. is close to constant while the change of the inclination angles  has a minor effect on the determination of $v_t(R)$.  The radial velocity has a small amplitude, $< 40$ km s$^{-1}$,  in the inner region, but reaches $\sim 50$ km s$^{-1}$ in the outermost disk where the inclination angle displays a change. {  These large values of $v_r$ in the peripheries of the disk can be real features of the velocity field but they can 
can also be artifacts due to the presence of a warp.  } { However, in this case the warp must be asymmetrical as no symmetric features in the 2D maps of $v_t$ and $v_r$ are visible (see Sect.\ref{tests}). The octopole moments converge well at all radii, despite the fact that they show relatively large changes in the velocity field of $40\%$ to $50\%$ of the average value in the outer regions. This convergence supports the conclusion that the anisotropies in the outermost regions are real and due to a rough velocity field. In contrast, fluctuations in the inner disk are small, indicating a quiet velocity field, with the exception of very small scales where the transversal velocity tends to zero.  Both the transversal two-dimensional velocity map (see panel (e) of Fig.\ref{NGC925}) and the radial velocity map (see panel (f) of Fig.\ref{NGC925}) show moderately coherent regions. In particular, both components are, in the inner disk, smaller than their average value over the ring, in the direction perpendicular to the kinematic axis. As such axis is aligned with the galaxy's bar, such anisotropies seem to have a simple physical interpretation in terms of the presence of a bar.%

In conclusion, the analysis of NGC 925 suggests that the inner regions of the galaxy conform well to a flat and differentially rotating disk, while the outer regions are characterized by a rough velocity field, possibly due to a combination of warp and radial velocities. {  Note that although spatial velocity anisotropies appear to be correlated with spatial structures, a causal relationship remains highly questionable as long as there is no dynamical modeling.} The results obtained using the VRM analysis are consistent with those obtained using other methods in the literature and support the conclusion that this galaxy is in an evolutionary stage, possibly as a result of accretion events or an out of equilibrium configuration.

\begin{figure*}
\includegraphics[width=\linewidth,angle=0]{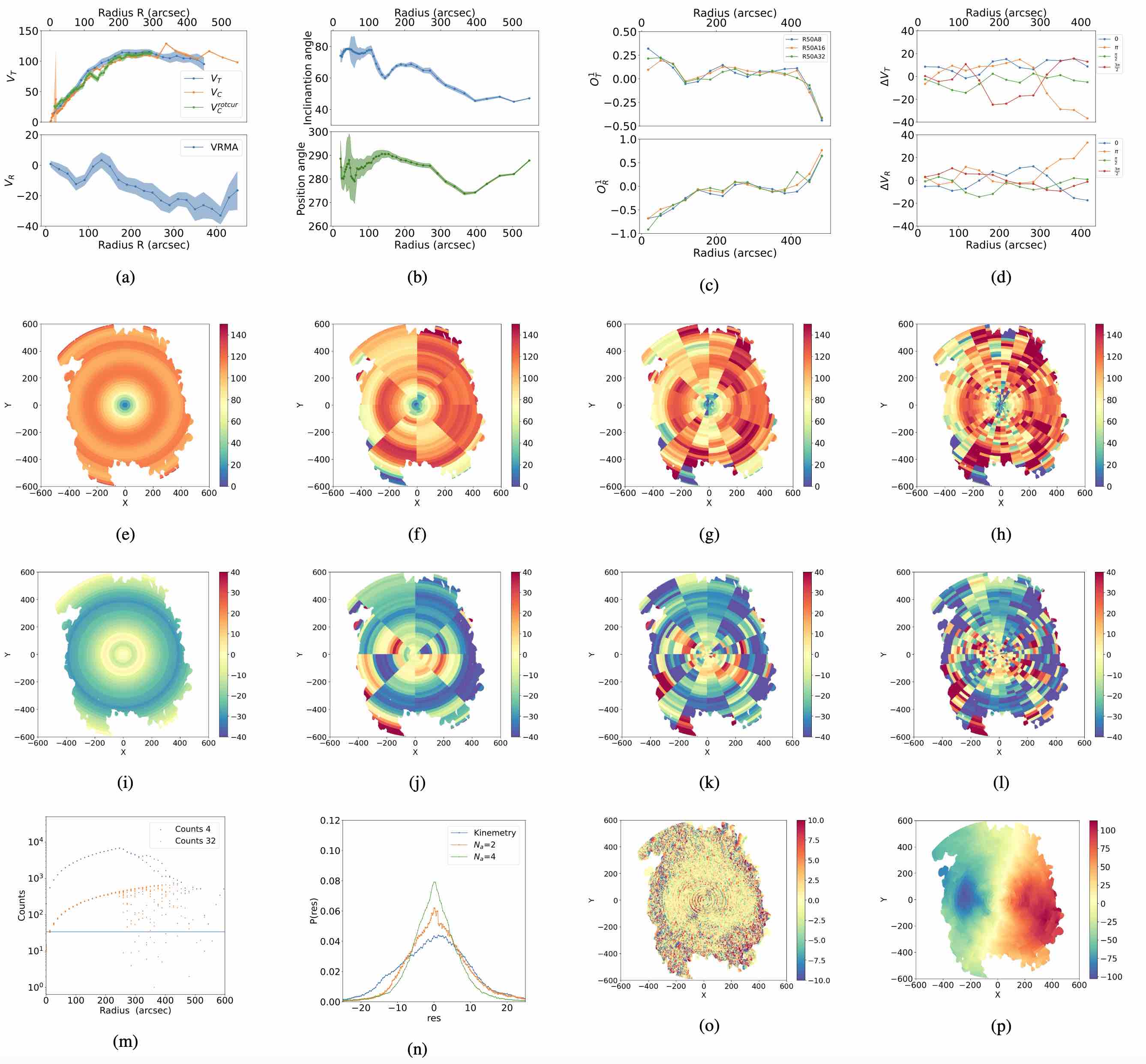}
\caption{Analysis of NGC 925: symbols and lines are as  Fig.\ref{NGC628}. }
\label{NGC925} 
\end{figure*}


\subsubsection*{NGC 2366}

{  NGC 2366 is a barred irregular dwarf galaxy that has a neighboring companion (NGC 2363). Results for the P.A. and inclination angle obtained by the TRM show that the P.A. does not vary by more than $10^\circ$ between the inner ($R_{25}\approx 130''$) and outer disk, while the inclination angle shows large fluctuations in the inner disk (as seen in Fig.\ref{NGC2366}).}
As the P.A displays a moderate variation, we find that $v_t(R) \approx v_c(R)$. It is known that NGC 2366 is characterized by the presence of noncircular motions \citep{Walter_etal_2008,deBlok_etal_2008,Trachternach_etal_2008,Oh_etal_2011}. We measure a radial velocity which is $<10$ km s$^{-1}$ in the inner disk; in the outermost regions of the galaxy, the amplitude becomes larger and it is negative.
The behaviors of $O^i_T(R)$ and $O^i_R(R)$ converge well in the inner disk, and show maximal variations of $50\%$ at very small radii (where the average transversal velocity is very small) and at larger ones. The 2D maps of $v_t(R)$ and $v_r(R)$ show moderate-amplitude, coherent anisotropy patterns. In the galaxy's periphery, there are some asymmetric velocity fluctuations that do not seem to be correlated with the kinematic axis. In particular, the radial velocity field displays a positive anisotropy in the direction of the small companion NGC 2363 (for $\theta \approx \pi/4$) as well as on the opposite direction. 

\begin{figure*}
\includegraphics[width=\linewidth,angle=0]{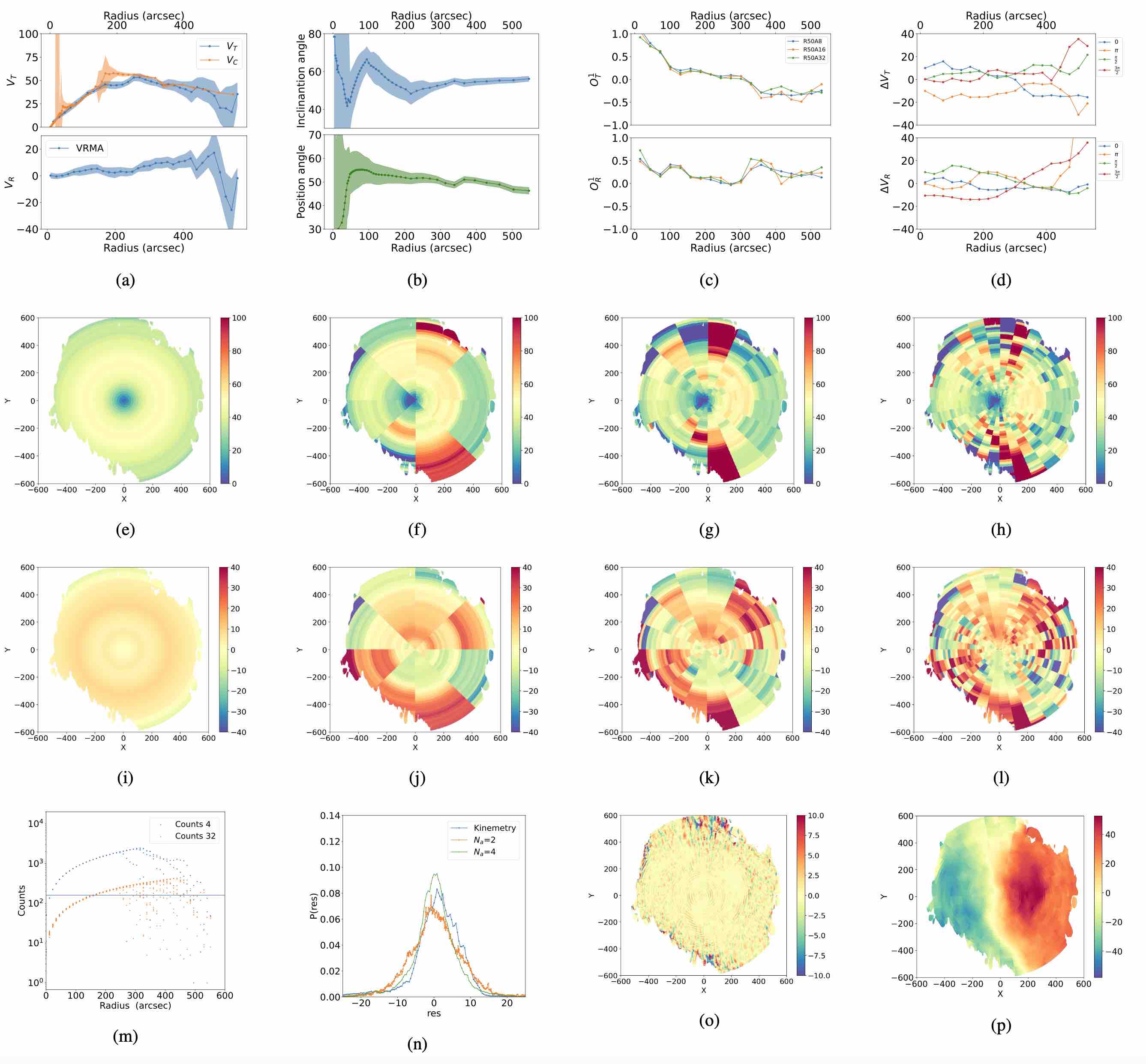}
\caption{Analysis of NGC 2366: symbols and lines are as  Fig.\ref{NGC628}.}
\label{NGC2366} 
\end{figure*}


\subsubsection*{NGC 2403}

{  NGC 2403 is a late-type Sc spiral galaxy whose rotation curve has been determined by many groups (see \cite{deBlok_etal_2008} and references therein). } The TRM finds that both the P.A. and the inclination angle have small variations with radius, i.e. $< 10^\circ$ both in the inner and outer disk as $R_{25}\approx 500''$ (see Fig.\ref{NGC2403} --- note that in the bottom part of panel (a) it is reported the behavior of $v_r(R)$ as measured by \cite{DiTeodoro+Peek_2021}:  this same profile will be reported below for all galaxies it is available.). Because both angles are nearly constant, we find $v_t(R) \approx v_c(R)$. Our determination of the radial velocity for NGC 2403, with an amplitude of $<10$ km s$^{-1}$, is in good agreement with previous studies, such as that of \cite{DiTeodoro+Peek_2021}. Any residual differences between our results and theirs can be attributed to the different assumptions and methods used to determine the radial velocity. The VRM method, which we use, assumes that the disk is coplanar, while the method used by \cite{DiTeodoro+Peek_2021} makes use of the TRM, which allows for variations in the P.A. and inclination angle with radius. This can lead to differences in the radial velocity measurements,  when the angles vary with radius  thus mixing true radial motions with  distortions of the galactic disk.  
{  In particular, \cite{DiTeodoro+Peek_2021} find a variation of about $10^\circ$ of the P.A. at small radii that can give rise to a radial velocity difference of $\sim 10$ km s$^{-1}$.}
The behaviors of $O^i_T(R)$ and $O^i_R(R)$ converge, showing maximal variations of $20\%$. The velocity field of this galaxy is very quiet, as also shown by the 2D maps of $v_t(R)$ and $v_r(R)$ which do not display large amplitude coherent fluctuations. The galaxy is dominated by regular rotation, a conclusion reached by \cite{Schoenmakers_etal_1997} who found that a Fourier harmonic analysis of the \HI{}    velocity field shows that non-circular motions are not important. However, the transversal velocity is systematically smaller by a small amplitude along the direction perpendicular to the kinematic axis than along parallel to it. While it is not clear whether this galaxy is barred or not \citep{Daigle_etal_2006}, these results point towards the possible breaking of circular symmetry that could be connected to a bar-like structure.

\begin{figure*}
\includegraphics[width=\linewidth,angle=0]{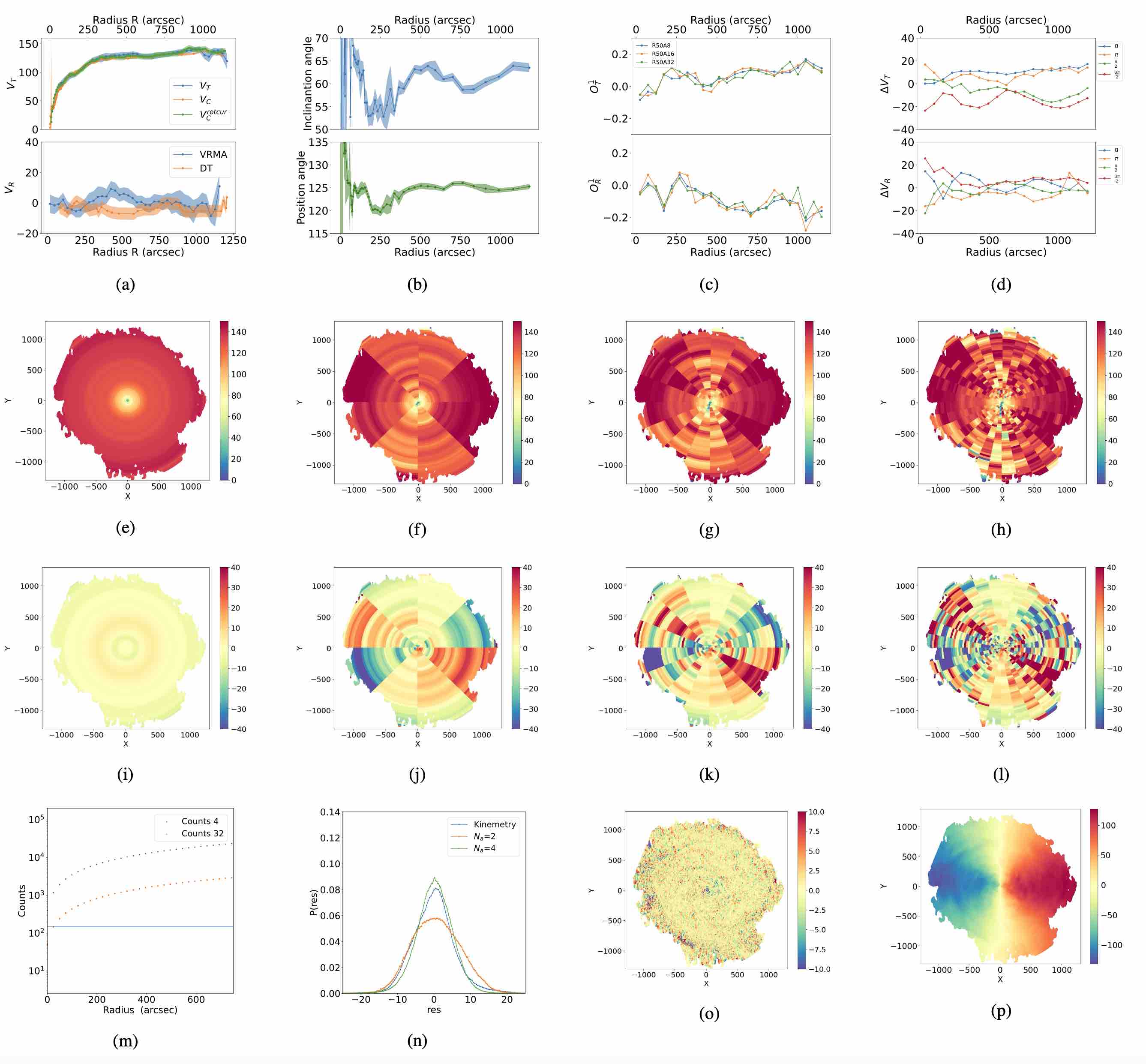}
\caption{Analysis of NGC 2403: symbols and lines are as  Fig.\ref{NGC628}.}
\label{NGC2403} 
\end{figure*}


\subsubsection*{NGC 2841}

{  NGC 2841 is an early-type (Sb) spiral. }The TRM shows that both the P.A. and inclination angle exhibit monotonic variations of about $ \approx15^\circ$ going from the inner to the outer regions of the galaxy, i.e. for $R>R_{25}\approx 300''$  (see Fig.\ref{NGC2841}).  The variation of the two orientation angles from the inner to the outer disk can be interpreted as a mild warp \citep{Sellwood_etal_2021} but of course, in view of the degeneracy between geometrical deformations and radial flows, also as true radial velocity field if a flat disk is assumed.  These trends are responsible for the small difference between $v_t(R)$ and $v_c(R)$ for $R>400''$, whereas they agree at small radii. The radial velocity measured by the VRMA, at small radii agrees with that of \cite{DiTeodoro+Peek_2021}, while at larger radii the two determinations differ by $\approx 10$ km s$^{-1}$ as an effect of the variation of the angles. The behaviors of $O^i_T(R)$ and $O^i_R(R)$ converge, showing some fluctuations at large radii, but with maximal variations being smaller than 20\%. This indicates that the velocity field is relatively quiet. In the inner disk, the transversal velocity along the direction perpendicular to the kinematic axis is larger than that in the direction parallel to it, while the radial velocity displays the opposite trend. %

\begin{figure*}
\includegraphics[width=\linewidth,angle=0]{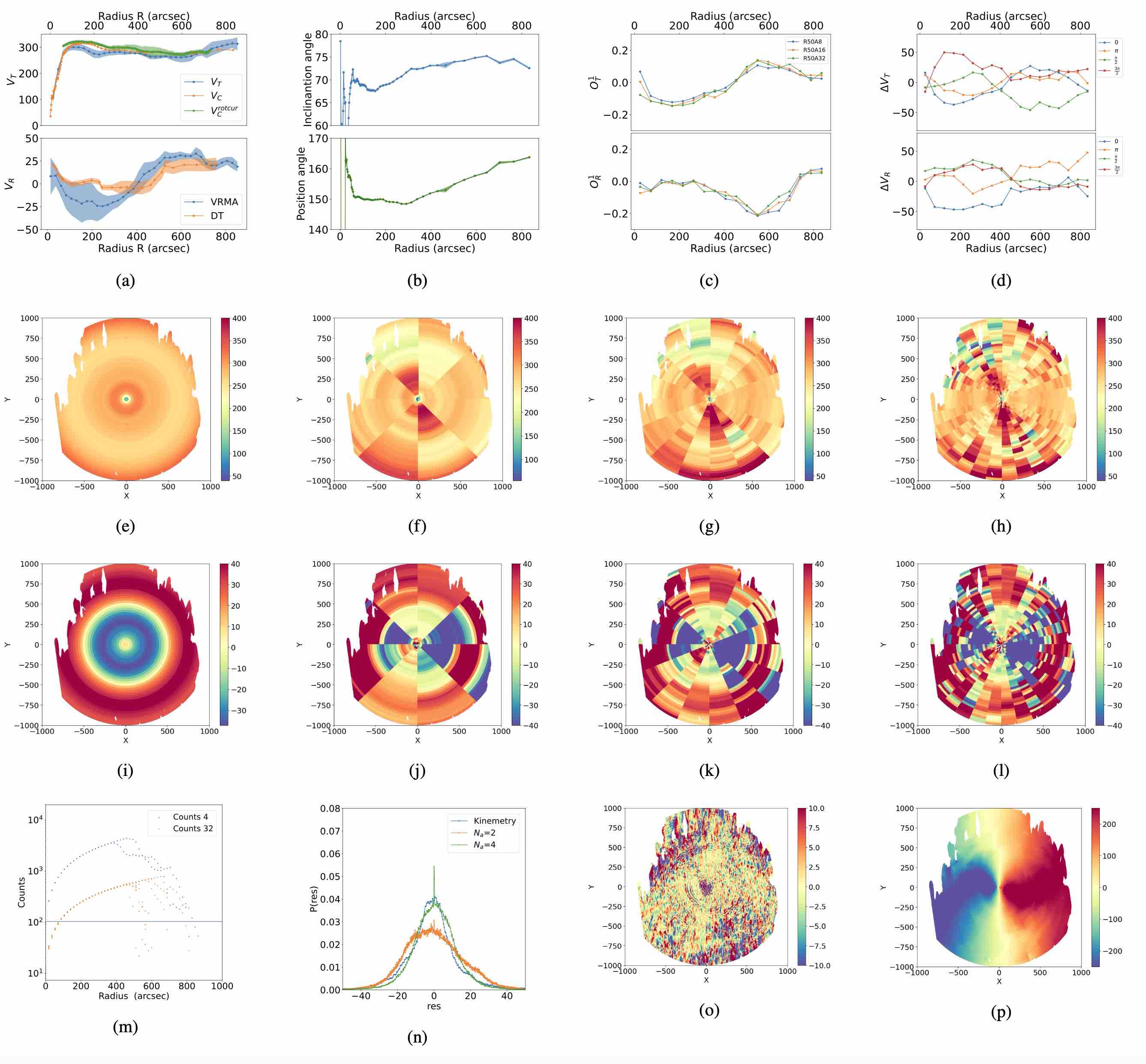}
\caption{Analysis of NGC 2841: symbols and lines are as  Fig.\ref{NGC628} }
\label{NGC2841} 
\end{figure*}


\subsubsection*{NGC 2903}

{  NGC 2903 is an isolated barred spiral galaxy.} The TRM detects an inclination angle with small variation in the inner disk, {  i.e., for $R<R_{25}\approx 360''$}, while in the outer regions of the galaxy, $i(R)$ varies by $< 10^\circ$, whereas the P.A. is almost constant across the whole disk (see Fig.\ref{NGC2903}). Given these small variations, we find that $v_t(R) \approx v_c(R)$. The radial velocity is $v_r < 30$ km s$^{-1}$ and agrees with the measurements by \cite{DiTeodoro+Peek_2021}. The behaviors of $O^i_T(R)$ and $O^i_R(R)$ converge and display variations that do not exceed $10\%$ of the average transversal velocity value. The 2D maps of $v_t(R)$ and $v_r(R)$ reveal a quiet velocity field, and the residual maps have fluctuations of a few km s$^{-1}$. Both the transversal and radial velocity along the direction perpendicular to the kinematic axis are larger than in the direction parallel to it. This asymmetry seems to be correlated with the direction of the bar that is oriented approximately orthogonally to the kinematic axis. {  It is worth stressing that also \cite{Sellwood+Sanchez_2010}  found evidences of a pronounced non-axisymmetric flow in this galaxy.} 
\begin{figure*}
\includegraphics[width=\linewidth,angle=0]{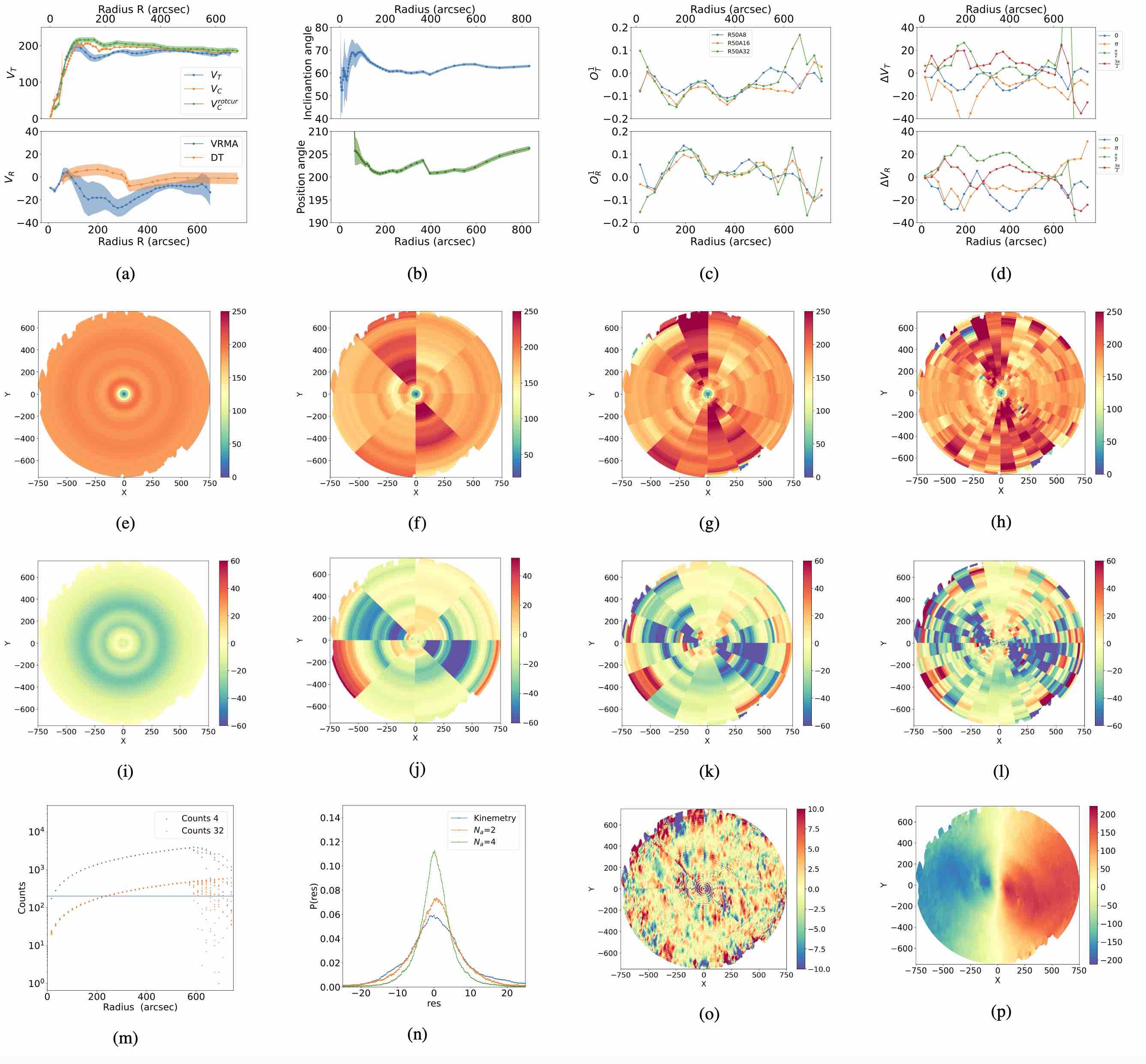}
\caption{Analysis of NGC 2903: symbols and lines are as  Fig.\ref{NGC628}.} 
\label{NGC2903} 
\end{figure*}


\subsubsection*{NGC 2976}

{  NGC 2976 is a peculiar dwarf galaxy}. The TRM determines the inclination angle with fluctuations of around $20^\circ$, while the P.A. changes by $20^\circ$ around $70''$ (see Fig.\ref{NGC2976}). These changes are small, and thus we find $v_t(R) \approx v_c(R)$ in the inner disk $R<300''$, which again shows that the VRM is robust with respect to the effect of small-amplitude changes of the angles as measured by the TRM. {  In particular, the change of the P.A. occurs in the very inner disk where the determination is more noisy.} The radial velocity in the inner disk, $R<220''$, is small, $v_r<10$ km s$^{-1}$, and approximately coincides with the measurement of \cite{Spekkens+Sellwood_2007}. The behaviors of $O^i_T(R)$ and $O^i_R(R)$ converge at all radii, showing small variations of $< 10\%$ in the inner disk, where the velocity field is quiet. However, in the outer regions, there are some large fluctuations of order one, where the velocity is more strongly anisotropic. Correspondingly, the 2D velocity maps present small amplitude anisotropies in the inner regions and  large amplitude ones outside the optical radius. There are no symmetric anisotropy patterns with respect to the kinematic axis in either the transverse or radial velocity. Note that this galaxy, as a member of the M81 Group, shows signs of strong tidal interactions in its \HI{}        distribution and this can explain the relatively large anisotropies in for $R>R_{25}$ which, however, involve a very small fraction of the galaxy's mass represented by a outermost parts of the \HI{}     mass.

\begin{figure*}
\includegraphics[width=\linewidth,angle=0]{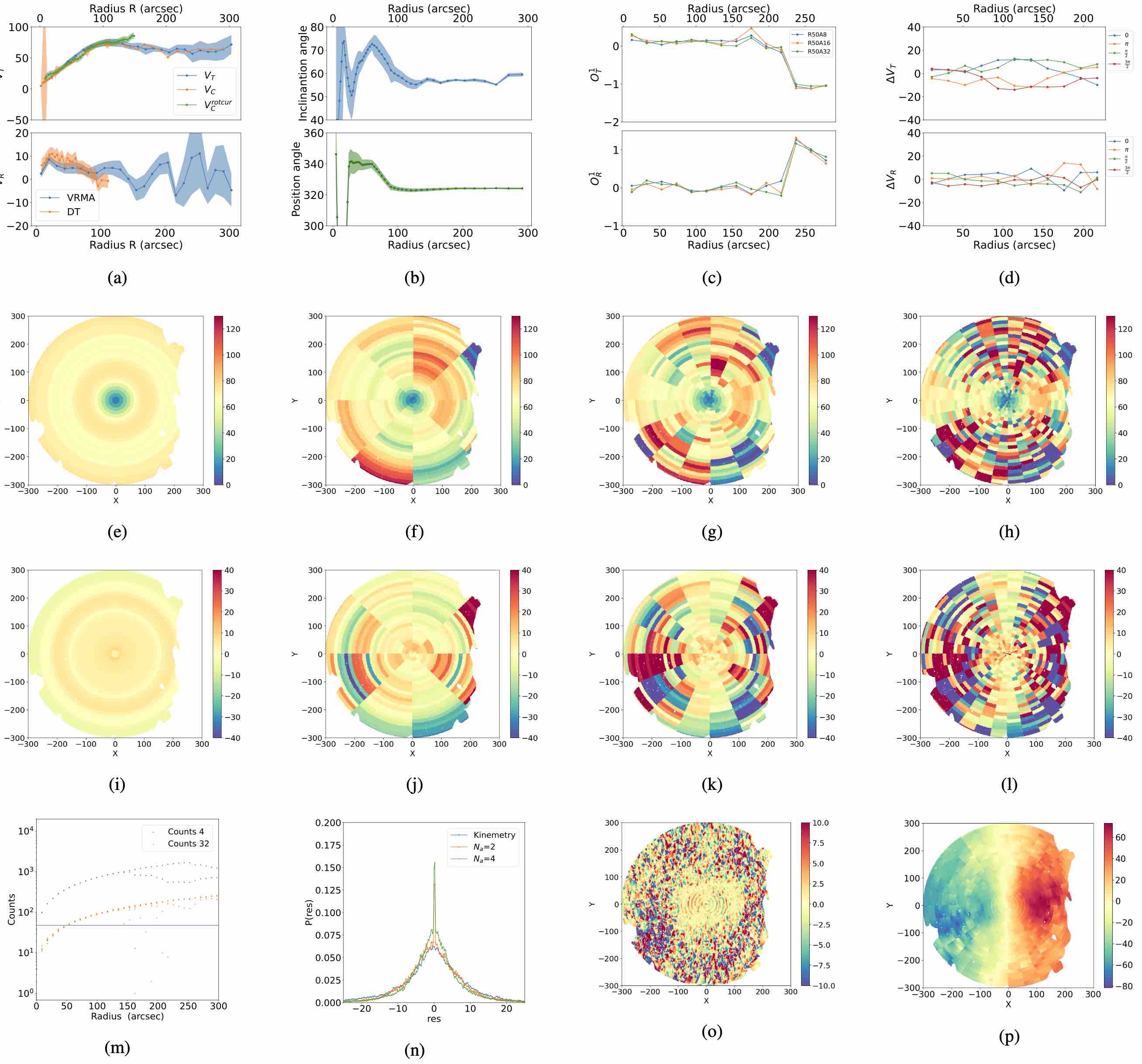}
\caption{Analysis of NGC 2976: symbols and lines are as  Fig.\ref{NGC628}.  }
\label{NGC2976} 
\end{figure*}


\subsubsection*{NGC 3031}

{   NGC 3031 is a grand-design spiral and together with NGC 3034 and NGC 3077 forms an interacting system. The global inclination angle of  $i=59^\circ$ \citep{deBlok_etal_2008} approximately corresponds to the inclination angle measured by the TRM in the middle region of the disk, i.e. $400''<R<1000''$ (where the optical radius is $R_{25} \approx 660''$). }Some fluctuations of the order of $20^\circ$ are present both at small radii where the determination of the angles is more problematic. Moreover, the TRM finds that the P.A. is $\approx 150^\circ$ with a small increasing trend in the outermost periphery of the galaxy (see Fig.\ref{NGC3031}). As these variations are moderate, we find, for $R<1500''$, that $v_t(R) \approx v_c(R)$, which again shows that the VRM is robust with respect to the possible effect of small-amplitude warps or the influence of relatively large radial velocities. Indeed, by using the VRM, we find that the radial velocity is small in the inner disk and becomes larger in the outermost regions. The determination of $v_r$ by means of the VRM differs from that of \cite{DiTeodoro+Peek_2021} for $R>500''$ (but agrees at smaller radii) because of the variation of the angles. The behaviors of $O^i_T(R)$ and $O^i_R(R)$ converge well for $R<1000''$, with small amplitude fluctuations ($<20\%$). However, at larger radii, the fluctuations in the reconstruction of the 2D properties affect the signal. The 2D maps reveal a quiet velocity field in the inner disk for $R<800''$, with small amplitude coherent motions associated with the prominent spiral arms. However, in the outer disk for $R>800''$, i.e. outside the optical radius, the \HI{}    gas is no longer in regular rotation around the galaxy and the motion of the gas is starting to become dominated by tidal interaction processes within the group. As a result, the velocity field is characterized by large-amplitude anisotropy patterns, with the most evident one in the direction (i.e., $\theta \approx 3/4 \pi$) of the close dwarf galaxy UGC 5336 \citep{Croxall_etal_2009}, that can be seen in both the transversal and radial velocity distribution.

\begin{figure*}
\includegraphics[width=\linewidth,angle=0]{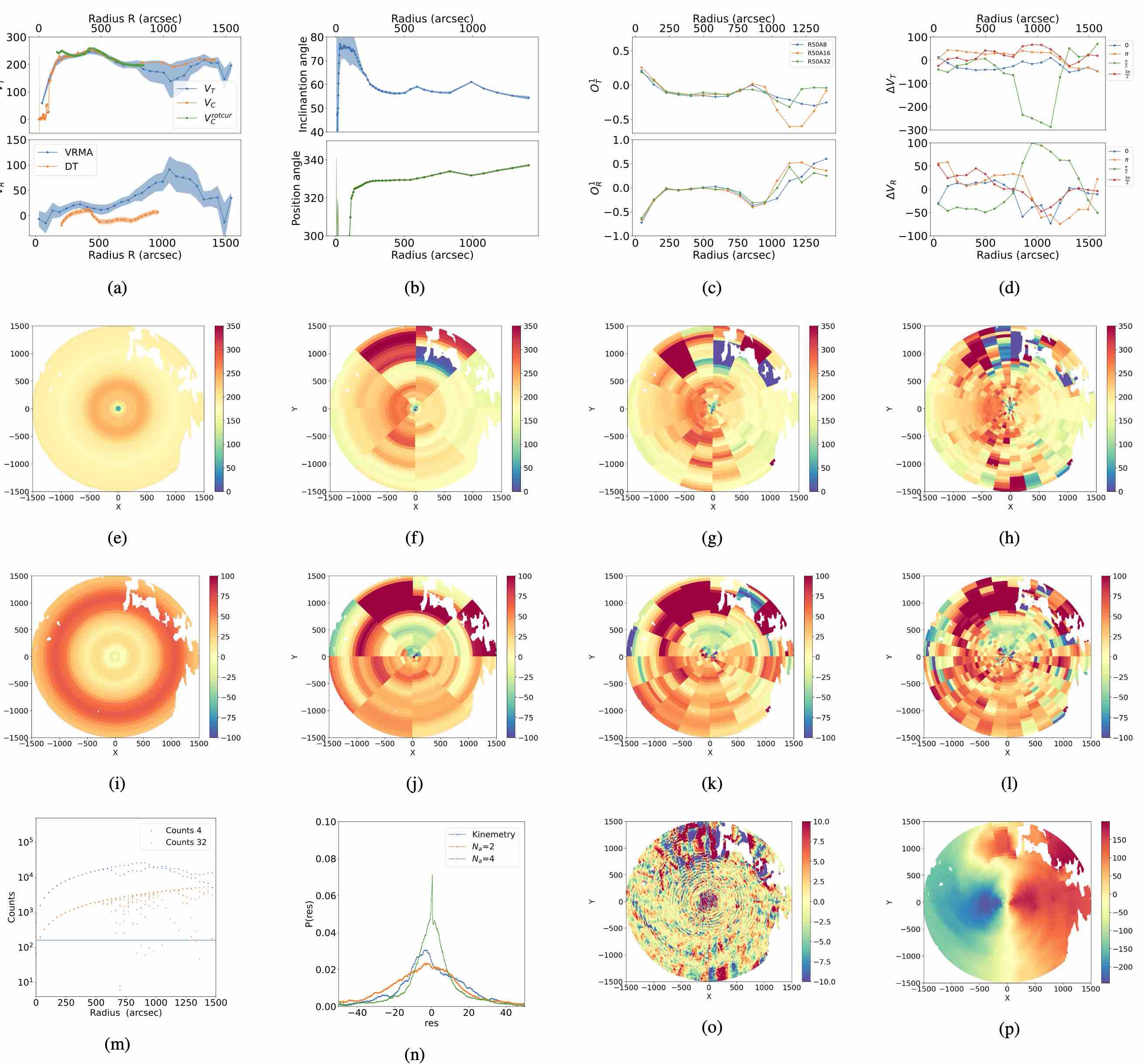}
\caption{Analysis of NGC 3031: symbols and lines are as  Fig.\ref{NGC628}. }
\label{NGC3031} 
\end{figure*}


\subsubsection*{NGC 3184}

{  NGC 3184 is a spiral galaxy with a global inclination angle of $i=29^\circ$ \citep{deBlok_etal_2008}: this is the value measured by the TRM at large enough radii, whereas in the very inner disk the inclination angle results to be highly fluctuating  (see Fig.\ref{NGC3184}).} The P.A. determined by the TRM is constant for $R>150''$,    while it shows some large fluctuations at smaller radii.   Both of these fluctuating behaviors are due to the TRM's inability to treat galaxies with small inclination angles as they occur inside the optical  radius $R_{25}\approx 220''$.  Note that \cite{DiTeodoro+Peek_2021} found instead that both the inclination angle and P.A.  display small variations in function of radius and they are not affected by large fluctuations: these difference are may be due to the different weighting schemes used by the different algorithms. Thus in this case the additional control given by the TRM is less reliable and the co-planarity of disk must be assumed. 
{  We find that $v_t(R)$ and $v_c(R)$ ({  measured with {\tt Barolo} by  \cite{DiTeodoro+Peek_2021}})  have approximately the same dependence on radius but they show a different amplitude that reflects a different value of the global inclination angle adopted in the VRMA analysis: the determination of $v_c(R)$  is less noisy and agrees with ours but for a small offset due to the fact that the  inclination angle was, different from that we used and estimated to be $i=25^\circ$. Instead, the determination with {\tt kinemesty} does not converge well and it is highly fluctuating.} The radial velocity is small in the inner disk, $v_r<10$ km/s, and it shows different sign but similar amplitude with the measurement by \cite{DiTeodoro+Peek_2021}. 
{  As mentioned above, these differences are due to the different values of the P.A. and inclination angles adopted by the TRM and VRMA.
In the periphery of the galaxy, the radial velocity measured by the VRM presents a larger (in absolute value) amplitude. For $R>300''$, there are large fluctuations in the behaviors of $O^i_T(R)$ and $O^i_R(R)$. They converge only for $R< 300''$ where fluctuations are $<20\%$.  We note that the velocity field is characterized by small-amplitude anisotropy patterns. Only in the outermost regions are there some significant fluctuations that can be an artifact of the VRMA method as shown by the behaviors of the octopole moments discussed above.}

\begin{figure*}
\includegraphics[width=\linewidth,angle=0]{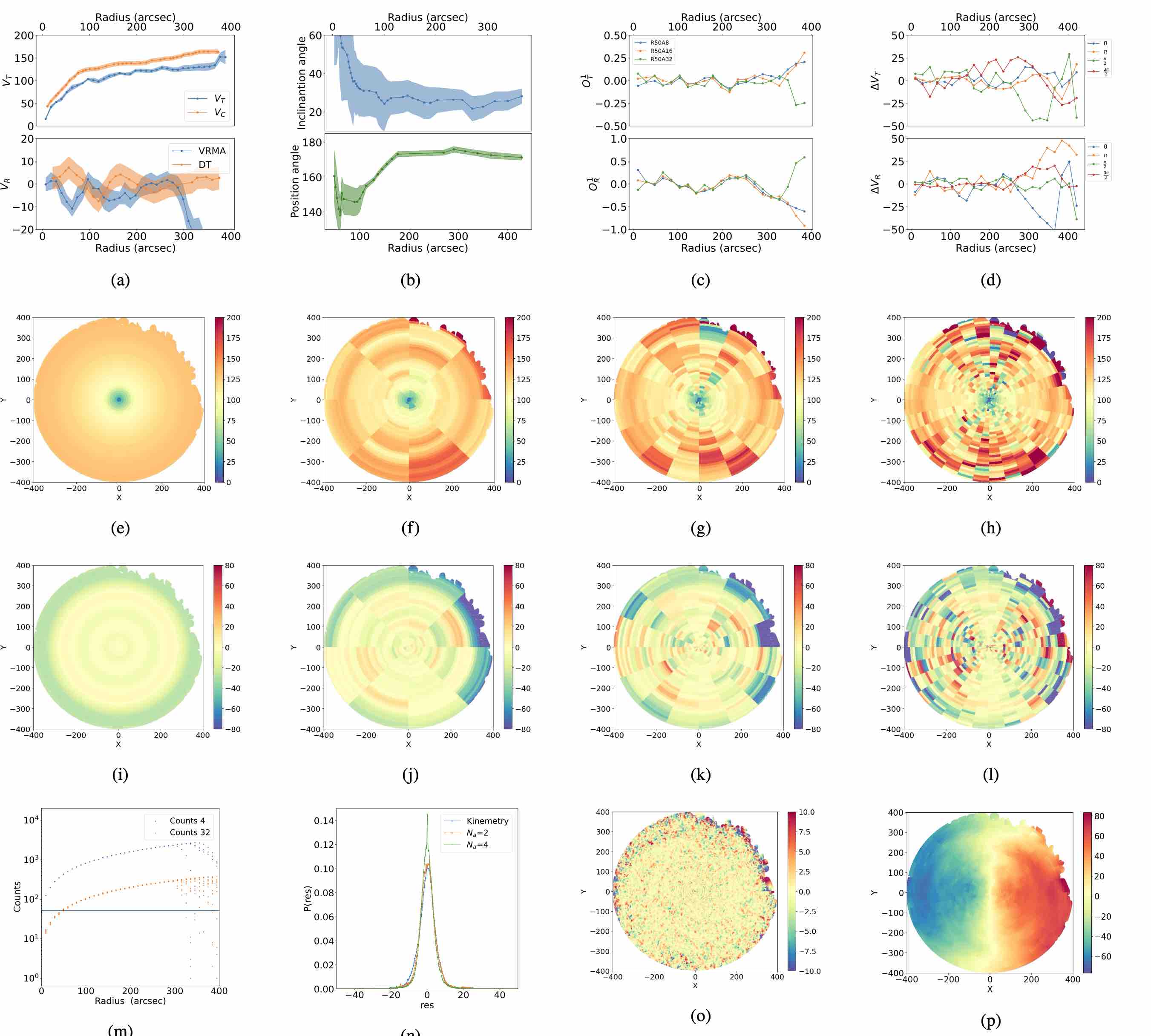}
\caption{Analysis of NGC 3184: symbols and lines are as  Fig.\ref{NGC628}.}
\label{NGC3184} 
\end{figure*}


\subsubsection*{NGC 3198}

{   NGC 3198 is a barred spiral galaxy. According to the TRM, the inclination angle is close to constant at all radii and the P.A. also shows very little variation across the galactic disk, both inside and outside the optical radius is $R_{25}\approx 220''$  (see Fig.\ref{NGC3198}). } Because of these small variations, we find that $v_t(R) \approx v_c(R)$ at all radii. Additionally, the determination of the radial velocity in the inner disk by means of the VRM is very similar to that obtained by \cite{DiTeodoro+Peek_2021}. The behaviors of $O^i_T(R)$ and $O^i_R(R)$ converge at all radii, showing moderate fluctuations of less than $20\%$. Correspondingly, the 2D maps of $v_t(R)$ and $v_r(R)$ reveal a very quiet velocity field. The maps also show coherent positive fluctuations in the direction perpendicular to the kinematic axis, which is approximately along the bar \citep{Daigle_etal_2006}.

\begin{figure*}
\includegraphics[width=\linewidth,angle=0]{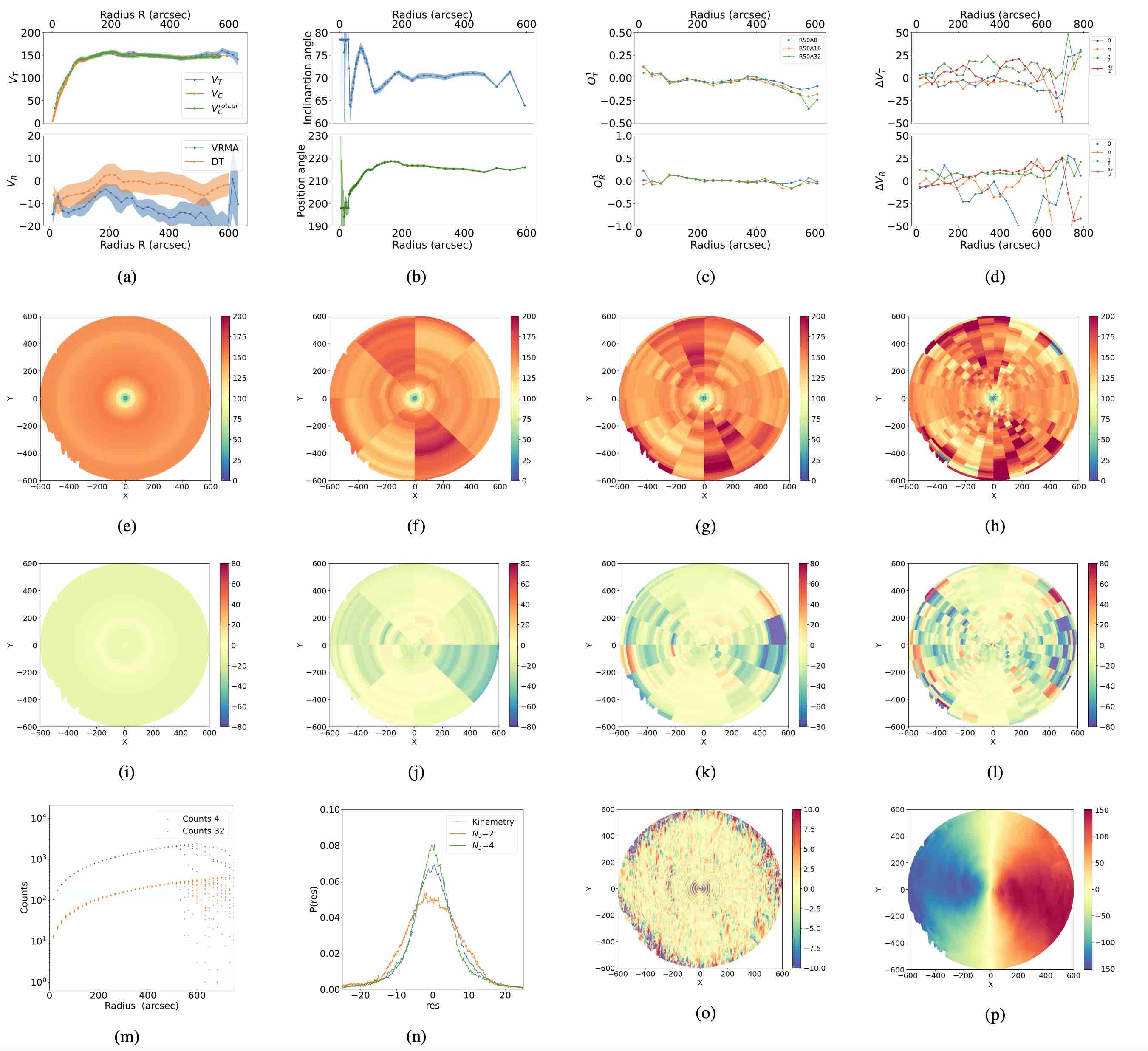}
\caption{Analysis of NGC 3198: symbols and lines are as  Fig.\ref{NGC628}.}
\label{NGC3198} 
\end{figure*}


\subsubsection*{NGC 3351}

{  NGC 3351 is a barred spiral galaxy.} The measurement of $i(R)$ through the TRM shows fluctuations in the very inner disk.   The same occurs for the P.A., {  which is then close to constant for $R>50''$ (see Fig.\ref{NGC3351}). } These fluctuations are an artifact of the TRM due to the low global inclination angle of this galaxy. Indeed, \cite{DiTeodoro+Peek_2021} finds a less fluctuating behavior for both angles.  Due to these moderate variations both inside and outside the optical radius $R_{25}\approx 220''$, we find $v_t(R) \approx v_c(R)$. The radial velocity in the inner disk determined by the VRM is very similar to that obtained by \cite{DiTeodoro+Peek_2021}. In the outermost regions of the galaxy, our determination of $v_r$ becomes more negative. Due to the variation of the P.A. in this same region, it is inconclusive whether larger radial velocities are present in the peripheries of the galaxy or whether a small warp of $< 10^\circ$ is present. The behaviors of $O^i_T(R)$ and $O^i_R(R)$ converge, and the 2D maps of $v_t(R)$ and $v_r(R)$ show a quiet velocity field. The 2D maps of $v_t(R)$ and $v_r(R)$ display an isotropic fluctuation field, with only small anisotropic fluctuations at small radii, which may correspond to the presence of the bar.

\begin{figure*}
 \includegraphics[width=\linewidth,angle=0]{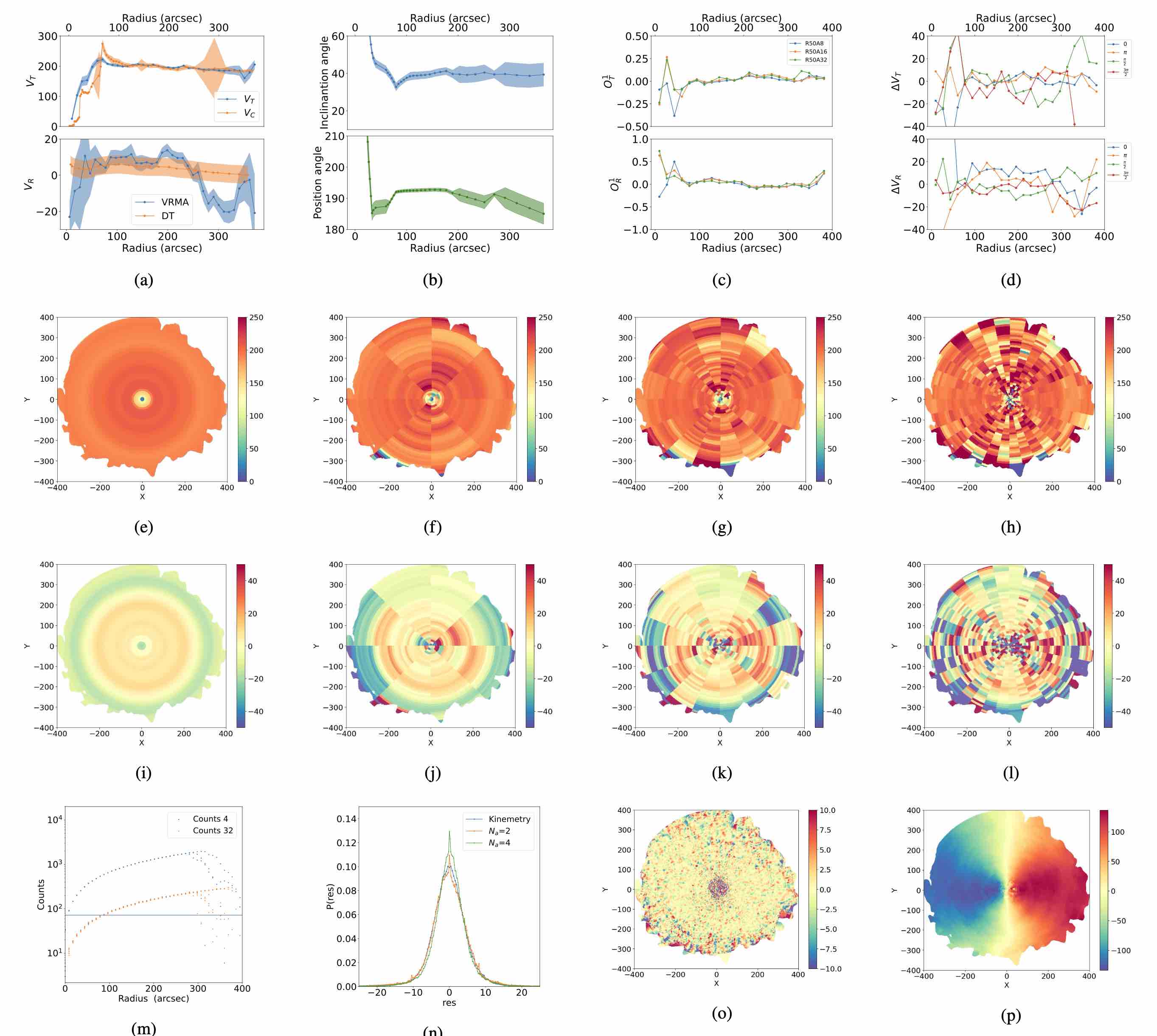}
\caption{Analysis of NGC 3351: symbols and lines are as  Fig.\ref{NGC628}.}
\label{NGC3351} 
\end{figure*}


\subsubsection*{NGC 3521}

{  NGC 3521 is a flocculent spiral galaxy}. The TRM measurements of the inclination angle and P.A. show small fluctuations, $<10^\circ$, across the disk (see Fig.\ref{NGC3521}). Due to these moderate variations, we find that $v_t(R) \approx v_c(R)$. The radial velocity remains small, $v_r<10$ km/s, for $R<300''$, and the determination with the VRM is similar to that by \cite{DiTeodoro+Peek_2021}. However, in the outer disk,
i.e. for $R>R_{25}\approx 250''$,  these two determinations show a different behavior that is induced by the small variations in both the P.A. and the inclination angle. The monotonic decrease of the radial velocity seems to be a genuine feature as the P.A. does not show a large enough smooth change in the external regions. Indeed, only in that case there can be a degeneracy between the effect of a large enough radial velocity and the presence of a warp. The octopole moments $O^i_T(R)$ and $O^i_R(R)$ converge well for $R<600''$. The 2D maps of $v_t(R)$ and $v_r(R)$ reveal a rough velocity field with significant anisotropies both in the inner and outer disk, in particular, the outermost regions are characterized by large scales streaming motions of significant amplitude that break circular symmetry. 

\begin{figure*}
 \includegraphics[width=\linewidth,angle=0]{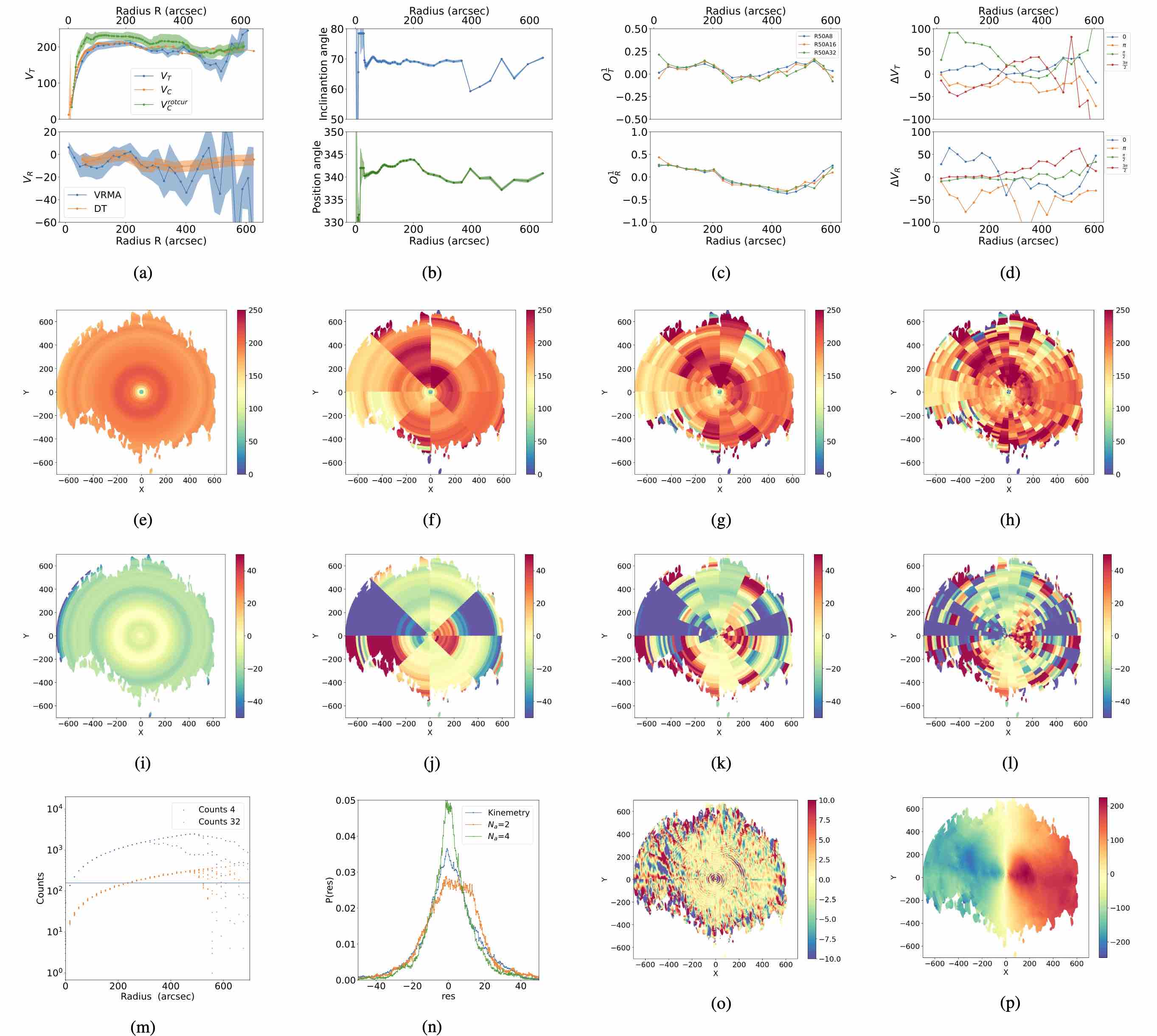}
\caption{Analysis of NGC 3521: symbols and lines are as  Fig.\ref{NGC628}.}
\label{NGC3521} 
\end{figure*}


\subsubsection*{NGC 3621}
{   NGC 3621 is a late-type spiral galaxy}. The TRM detects that the inclination angle is almost constant but with some small fluctuations; the P.A. increases by $10^\circ$ going from the inner disk to the outermost regions of the galaxy, i.e., for $R>500''$ (see Fig.\ref{NGC3621}). As a result, for $R<500''$ we find $v_t(R) \approx v_c(R)$ and the radial velocity, with an amplitude of $v_r<10$ km s$^{-1}$, agrees well with the measurements by \cite{DiTeodoro+Peek_2021}. However, for $R>R_{25}\approx 300''$ , the VRM may be unable to correctly measure the velocity field due to the presence of a warp or an increase in radial velocity. This degeneracy cannot be resolved without further analysis. The moments $O^i_T(R)$ and $O^i_R(R)$ converge well everywhere except in the outermost region, $R>500''$. The 2D maps of $v_t(R)$ and $v_r(R)$ reveal a quiet velocity field in the inner disk, but there are large asymmetric structures in the galaxy's peripheries velocity fields which cannot be easily associated with spatial structures.

\begin{figure*}
 \includegraphics[width=\linewidth,angle=0]{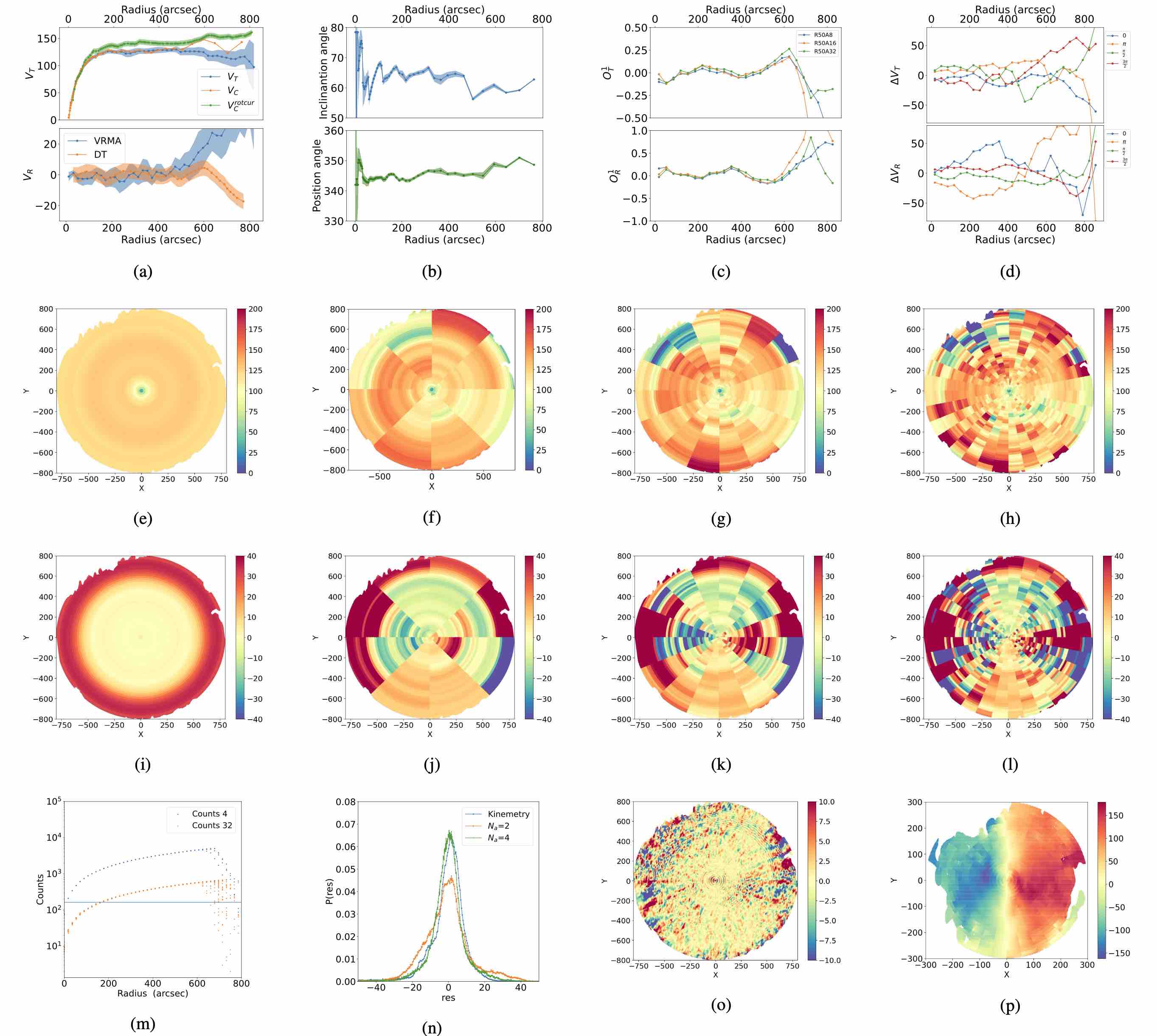}
\caption{Analysis of NGC 3621: symbols and lines are as  Fig.\ref{NGC628}.}
\label{NGC3621} 
\end{figure*}


\subsubsection*{NGC 3627}

NGC 3627 is a late-type spiral galaxy with weak bar features and loosely wound arms forming an asymmetric spiral structure. It is also interacting with other galaxies in the Leo Triplet. {  The TRM detects that the inclination angle shows significant changes with radius, indicating that the galaxy maybe  affected by a significant warp:  however in that case the warp would be present inside the optical radius $R_{25}\approx 310''$. } The P.A. shows  a radial oscillation of moderate amplitude. We find that $v_t(R) \approx v_c(R)$ for $R \le 150''$ while at large radii the circular velocity shows a significant fluctuation. We note that  both \cite{deBlok_etal_2008} and  \cite{DiTeodoro+Peek_2021} find a smoother behavior for the angles and $v_c(R)$. These differences probably arise because the galaxy  has very elongated structures such as the bar and the very prominent spiral arms: these might significantly affect the fitting algorithm depending on how it rejects or uses outliers.   The radial velocity is small and the VRM determination is similar to the measurement by \cite{DiTeodoro+Peek_2021}. The octopole moments $O^i_T(R)$ and $O^i_R(R)$ converge well, but for the outermost regions of the galaxy where considerable fluctuations are present. The 2D maps of $v_t(R)$ and $v_r(R)$ reveal a rough velocity field. The transversal velocity presents positive symmetric fluctuation patterns in the direction orthogonal to the kinematic axis (and thus approximately to the bar), while the radial velocity field remains approximately isotropic but with some significant fluctuations.
\begin{figure*}
\includegraphics[width=\linewidth,angle=0]{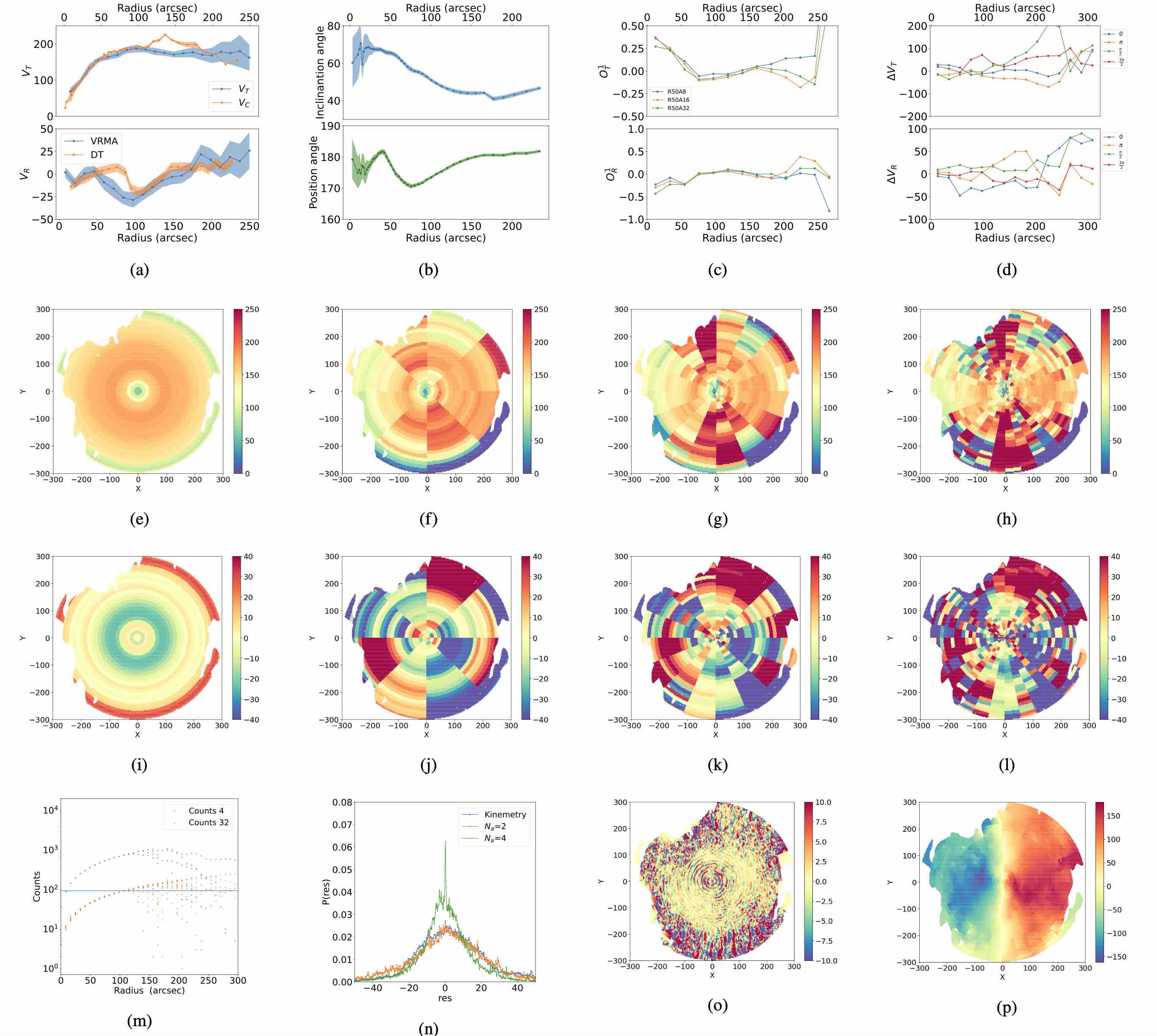}
\caption{Analysis of NGC 3627: symbols and lines are as  Fig.\ref{NGC628}.}
\label{NGC3627} 
\end{figure*}


\subsubsection*{NGC 4214}

{  NGC 4214 is a dwarf barred irregular galaxy}. The TRM measurement of the inclination angle is highly fluctuating at small radii and becomes uncertain at large radii, resulting in large error bars (see Fig.\ref{NGC4214}).  {  This uncertainty also highly affects the {\tt kinemetry} measurement of $v_c(R)$: the errors are too large to be meaningful. } Even the P.A.  shows an increasing trend with radius: however, even in this case the low inclination may affect the measurement. Fluctuations and indetermination in the measurements of the inclination angle cause the differences between $v_t(R) $ and $ v_c(R)$. In this situation, as for other galaxies with small inclination angle, results from the TRM are unreliable and thus cannot be used to constrain  the co-planarity of the disk, which thus must be assumed. However, the optical radius is $R_{25}\approx 210''$ and the variation of the P.A is $<10^circ$ for $R>R_{25}$. The radial velocity is small in the inner disk and increases from -10 km s$^{-1}$ to 20 km s$^{-1}$ in the outer disk. The octopole moments $O^i_T(R)$ and $O^i_R(R)$ agree within large fluctuations, and the 2D maps of $v_t(R)$ and $v_r(R)$ display a rough velocity field with coherent patterns of anisotropies, particularly along the direction of the galaxy's bar. In that particular direction, the radial velocity shows small fluctuations compared to adjacent regions.
\begin{figure*}
 \includegraphics[width=\linewidth,angle=0]{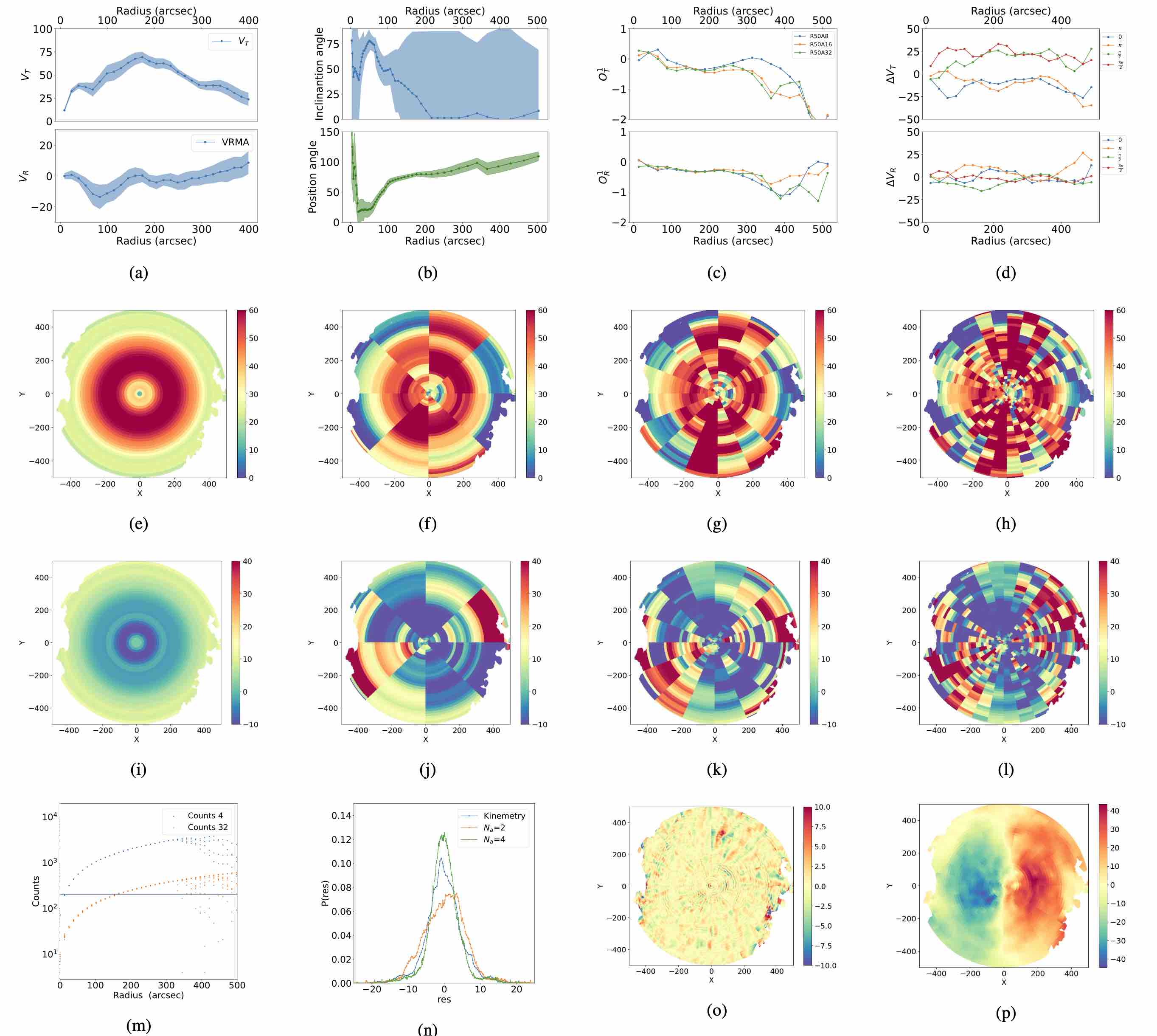}
\caption{Analysis of NGC 4214: symbols and lines are as  Fig.\ref{NGC628}.}
\label{NGC4214} 
\end{figure*}


\subsubsection*{NGC 4736}
NGC 4736 is a spiral galaxy with a bar within the inner disk. 
{  The inclination angle measured by the TRM is $i=44^\circ$ for  $R>50''$; at smaller radii, both the inclination angle and the P.A. show large variations, likely due to the TRM's difficulty in handling galaxies with small inclination angles  (see Fig.\ref{NGC4736}).  } These fluctuations are well inside the optical radius $R_{25}\approx 230''$, cause the transversal velocity from the VRM to differ from the circular velocity from the TRM in the inner disk, while they are closer at larger radii. The radial velocity is small, with an amplitude similar to the measurement by \cite{DiTeodoro+Peek_2021}, but the behaviors do not agree due to different assumptions on the angles used by the two methods. The octopole moments display small fluctuations and converge well. The 2D maps show coherent anisotropies of considerable amplitude, but with a spatial distribution that does not break circular symmetry.

\begin{figure*}
 \includegraphics[width=\linewidth,angle=0]{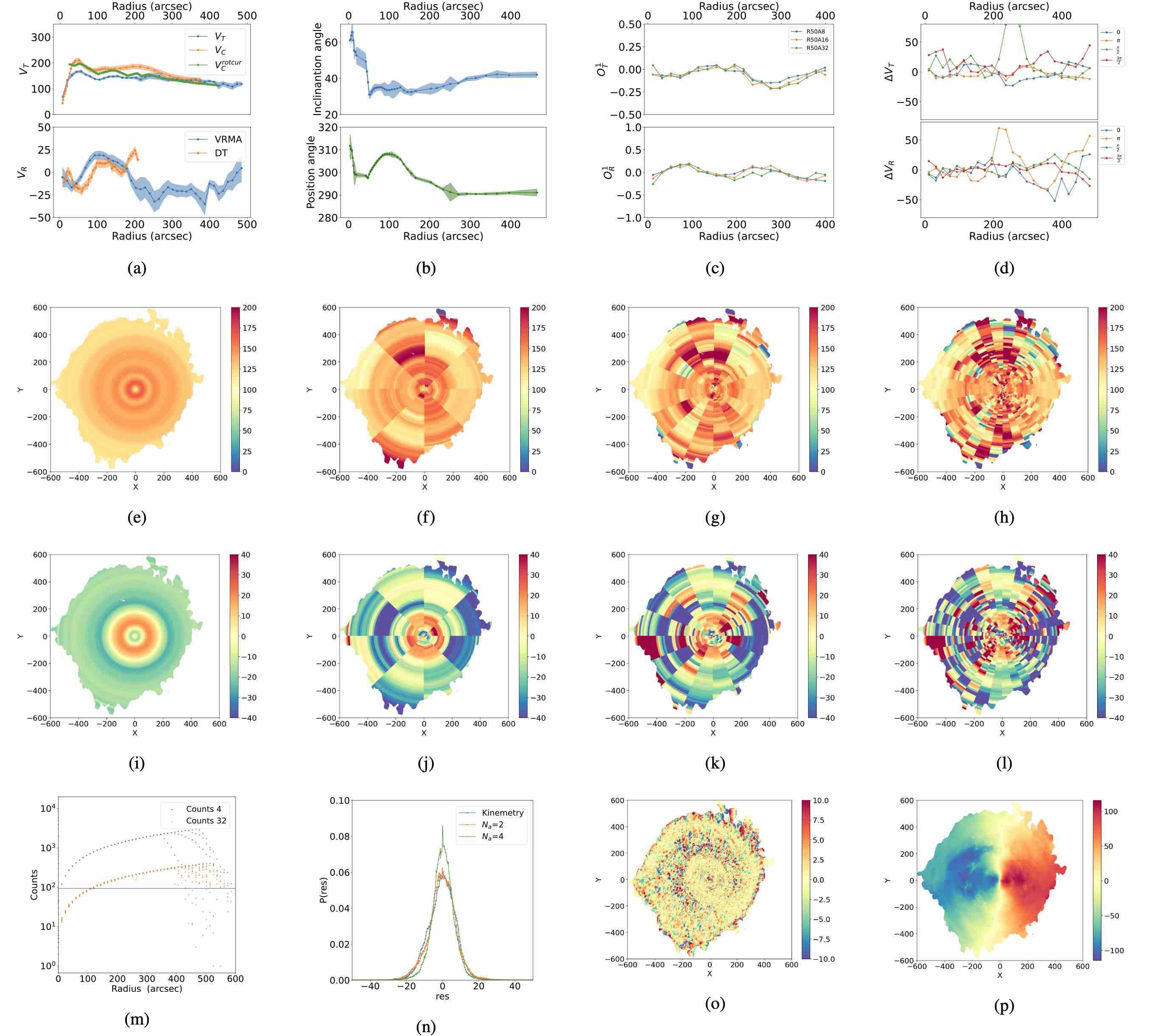}
\caption{Analysis of NGC 4736: symbols and lines are as  Fig.\ref{NGC628}.}
\label{NGC4736} 
\end{figure*}

%

\subsubsection*{NGC 4826}

NGC 4826 is an early-type spiral galaxy known for its counter-rotating gas disks, with a high-column density inner \HI{}    disk rotating in the same direction as the stars, and a low-column density outer \HI{}    disk rotating in the opposite direction \citep{Braun_etal_1994}. The inner \HI{}    disk has an high-column density \HI{}    disk and it is associated with the bright stellar disk and rotates in the same direction. {  The outer, much lower column-density \HI{}  disk, rotates in the direction opposite to that of the stars.  The transition between the two disks occurs around $\approx 100''$ radius that is about 1/3 of the optical radius $R_{25}\approx 310''$. }The TRM determines a behavior of the inclination angle such that there is a large bump of about $40^\circ$, followed by a smoother and monotonic decay of $\approx 30^\circ$. Even the P.A. displays a large change at about $R \approx 100''$ and then it flattens to a plateau (see Fig.\ref{NGC4826}).  The determination of both angles for $R<200''$ is affected by large fluctuations and for this is reason this region is avoided by \cite{deBlok_etal_2008}. Thus even in this case the co-planarity of the disk must be assumed for the interpretation of the VRM results. The behaviors of $v_t(R)$ and $v_c(R)$ are very similar, although the latter velocity is more affected by fluctuations. Note that there is a change of sign at $R\approx 100''$, corresponding to the fact that the inner disk is counter-rotating with respect to the external one. At the same scales where the change of rotation occurs, the radial velocity shows a negative peak of about $v_r\approx 50$ km s$^{-1}$, corresponding to the fact that a very peculiar kinematic structure is present at those scales. At larger radii, the radial velocity has small amplitude, i.e. $v_r \le 20$ km s$^{-1}$. The octopole moments $O^i_T(R)$ and $O^i_R(R)$ converge and, beyond the large fluctuations for $R \approx 100''$ associated with the transition from the inner to the outer disk, present small amplitude fluctuations, i.e. $< 20\%$. The 2D map of $v_t(R)$ shows relatively small amplitude anisotropies, while the $v_r(R)$ map displays a more complex anisotropy structure: along the axis orthogonal to the kinematic one, anisotropies display the smallest amplitude.

\begin{figure*}
 \includegraphics[width=\linewidth,angle=0]{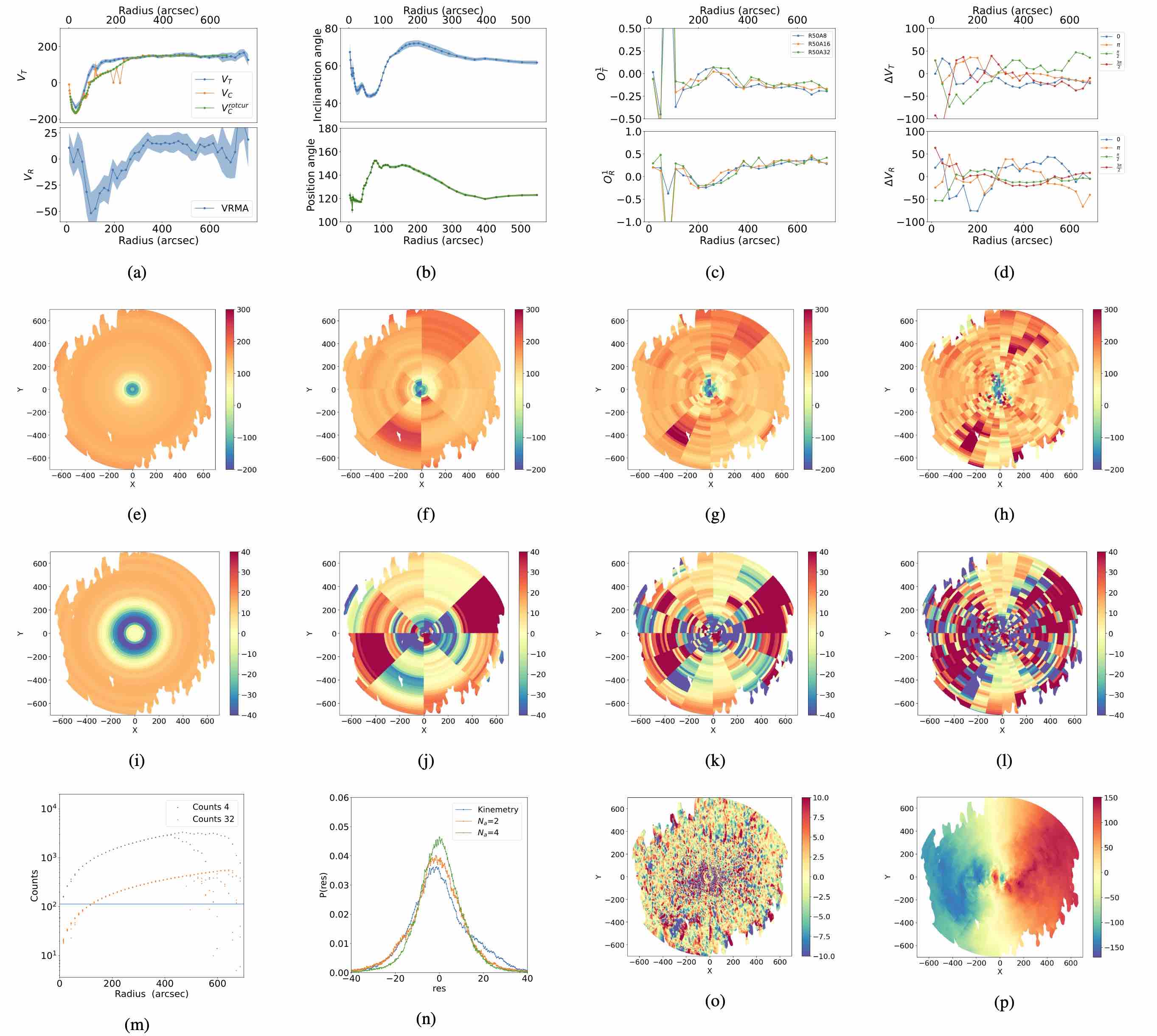}
\caption{Analysis of NGC 4826: symbols and lines are as  Fig.\ref{NGC628}.}
\label{NGC4826} 
\end{figure*}


\subsubsection*{NGC 5055}

{  NGC 5055 is a spiral galaxy}. The TRM measurements of the inclination angle and P.A.  show moderate variations across the disk, with fluctuations of less than $15^\circ$ (see Fig.\ref{NGC5055}). In the inner disk, for $R< R_{25}\approx 350''$,  the transversal velocity determined by the VRM is similar to the circular velocity determined by the TRM. The radial velocity determined by the VRM is also similar to the measurements by \cite{DiTeodoro+Peek_2021} at small radii. However, in the outer disk for $R> R_{25}$, there are differences in the sign and amplitude of the radial velocity due to changes in the inclination angle and the P.A.. The octopole moments $O^i_T(R)$ and $O^i_R(R)$ converge well and have small fluctuations. The 2D maps of $v_t(R)$ and $v_r(R)$ show coherent anisotropies, with large amplitude anisotropies along the kinematic axis extending over the entire disk.

\begin{figure*}
 \includegraphics[width=\linewidth,angle=0]{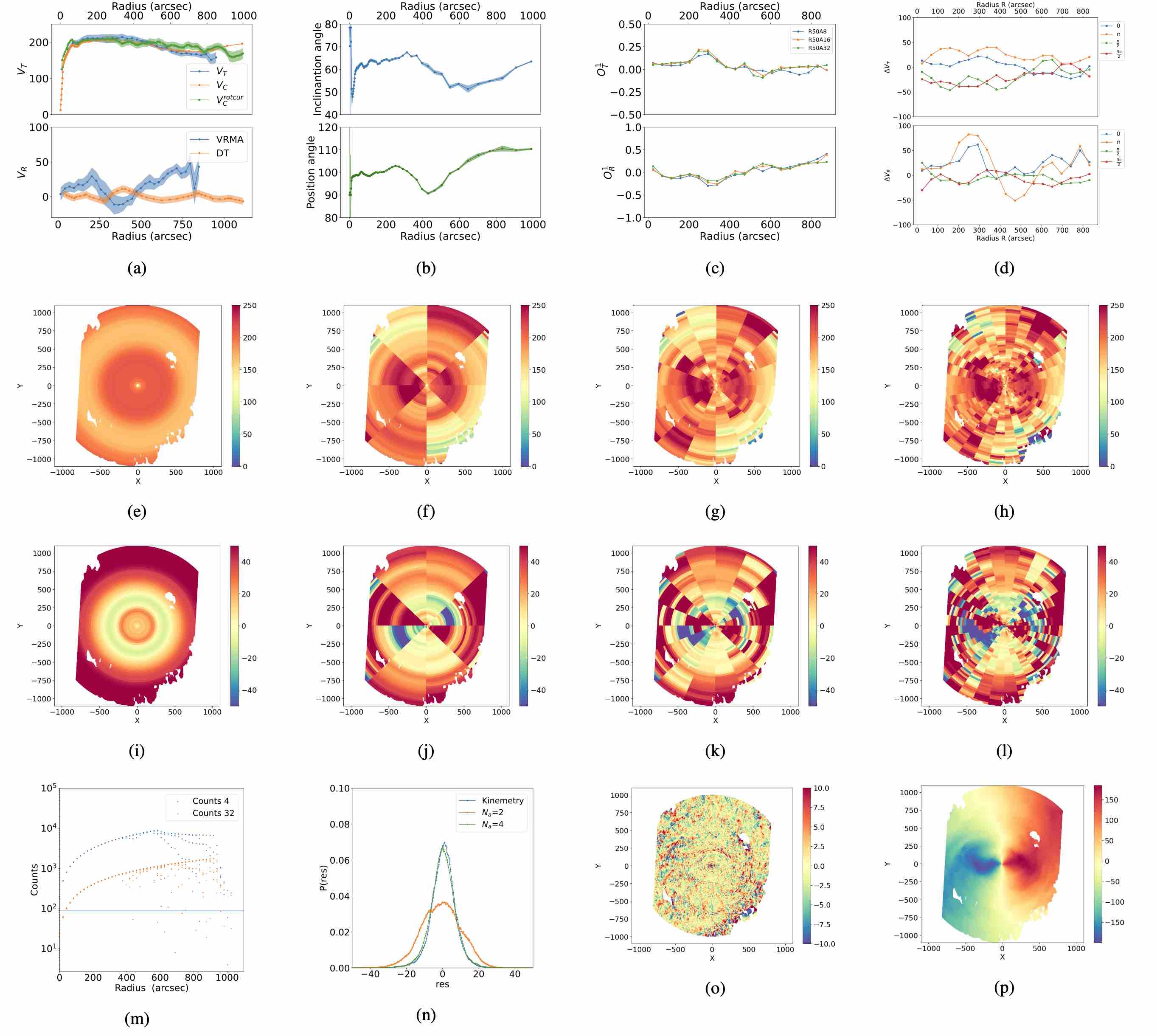}
\caption{Analysis of NGC 5055: symbols and lines are as  Fig.\ref{NGC628}.}
\label{NGC5055} 
\end{figure*}


\subsubsection*{NGC 5194}

This is a grand design spiral galaxy with a close satellite with which it interacts. The VRM analysis is limited to the galactic disk, $R<300"$, a region that extends beyond the optical radius $R_{25} \approx 230''$. The galaxy at larger radii is characterized by the presence of the satellite and a long tail of HI, making the velocity field complex and dominated by large anisotropic fluctuations due to strong tidal effects. The TRM finds that the inclination angle is in the range of $20^\circ < i <50^\circ$, with a peculiar "bump" at small radii (see Fig.\ref{NGC5194}). We take $i=30^\circ$ as the value of the global inclination angle, in agreement with \cite{deBlok_etal_2008}. The P.A. is mostly constant, but shows a fluctuation in correspondence to a local bump of $i(R)$, which is likely an artifact of the TRM due to the small inclination of the galaxy.    Given the small value of the inclination angle, the measuring of the orientation angles through the TRM gives large uncertainties. For this reason in this case we assume that the inner disk $R<R_{25}$ is flat. However, given the grand design spiral symmetry of the inner disk we think this is a reasonable assumption for $R<300''$. The circular velocity, as measured by TRM, has large error bars due to the small inclination angle, but its radial dependence is similar to the transversal velocity measured by the VRM in the inner disk. The rotation curve measured by the TRM, agrees with the one in \cite{Oikawa+Sofue_2014} obtained from a compilation of observations and is nearly flat in the inner disk at $R<140"$ and then bends suddenly at $R<150"$, beyond which the velocity decreases faster than the Keplerian law. This peculiar behavior is interpreted as being due to warping of the galactic disk well inside the optical radius \citep{Oikawa+Sofue_2014,Colombo_etal_2014}. According to the VRM, the warp is small enough that its effect on the rotation curve is limited. However, the radial velocity is small across the disk but raises in its peripheries, which may be an artifact due to the rise of the P.A. In this case, we would not expect an anisotropies in the velocity field, which instead we observe, as we discuss in what follows. The behaviors of the octopole moments $O^i_T(R)$ and $O^i_R(R)$ converge well but show some fluctuations, particularly in the disk's peripheries. These fluctuations have a straightforward interpretation in terms of coherent anisotropies patterns. In particular, 2D maps of $v_t(R)$ and $v_r(R)$ show an increase in the transversal and radial velocities in the direction of the satellite. This is clearly shown by the two-dimensional map of $v_r(R)$, which displays a coherent and large amplitude pattern of positive anisotropies in the direction $\approx \theta \approx 5/4 \pi$, where the satellite is located. Even the transversal velocity displays similar anisotropies in the same direction. The method used by \cite{Shetty_etal_2007} to fit the two-dimensional CO and \HI{}     velocity fields assumes that radial variations of azimuthally averaged quantities are negligible, and that $v_r$ and $v_t$ vary primarily with spiral-arm phase. However, the VRMA, which avoids making any hypotheses on the radial dependence of azimuthally averaged quantities, shows that variations in the azimuthal variation of both velocity components, of the order of 40 km s$^{-1}$, are similar to what is observed using this method.

\begin{figure*}
 \includegraphics[width=\linewidth,angle=0]{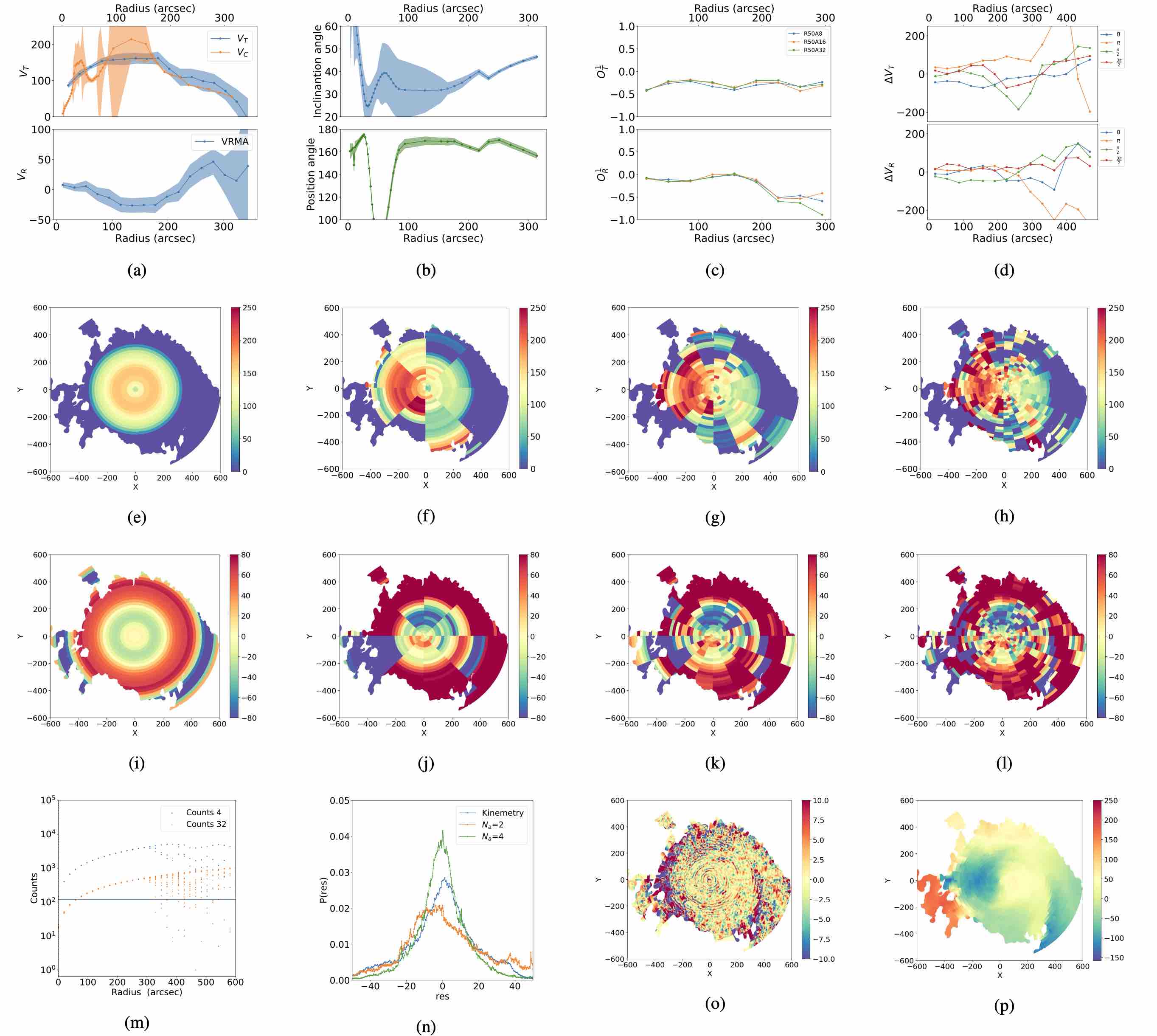}
\caption{Analysis of NGC 5194: symbols and lines are as  Fig.\ref{NGC628}.}
\label{NGC5194} 
\end{figure*}
%


\subsubsection*{NGC 5236}

{  NGC 5236 is a barred spiral galaxy with a close satellite (NGC 5253) that likely interacts with it. The kinematic structure of this galaxy is complex, so we limit our analysis to radii of 500'' as on larger radii the velocity field becomes too fluctuating. As $R_{25} \approx 380''$, thus the analysis stops beyond the optical disc.  } The TRM detects an inclination angle $i(R)$ with large variations, while the P.A. has a large fluctuation only at small radii and then flattens to a constant value (see Fig.\ref{NGC5236}). Because of the small inclination angle, we expect the TRM to perform poorly, resulting in large errors on the circular velocity and large fluctuations in the behavior of $v_c(R)$. In agreement with \cite{Lundgren_etal_2004}, we find that the circular velocity rises linearly in the central regions and reach a plateau at larger radii. The behavior of $v_t(R)$ is smoother and characterized by smaller errors. The radial velocity has a low amplitude at small radii and agrees approximately with the measurement by \cite{DiTeodoro+Peek_2021}, but then grows in the peripheries of the system. This growth may be interpreted as an artifact due to a warp, but there are also coherent patterns of anisotropies that cannot be easily explained in the TRM picture unless the warp breaks axi-symmetry. The octopole moments show convergent behaviors: while the transversal component has limited fluctuations, i.e. $< 20\%$, the radial component reaches $50\%$ variations in the peripheries of the disk. The 2D maps of $v_t(R)$ and $v_r(R)$ show that the inner disk is quiet, while in the region for $200''<R<500''$ there are large amplitude anisotropies that form coherent patterns not clearly correlated with galaxy's structures. The rise of $v_t(R)$ in the direction $\theta \approx 7/4 \pi$ may correspond to the tidal effect of the companion NGC 5253.
\begin{figure*}
 \includegraphics[width=\linewidth,angle=0]{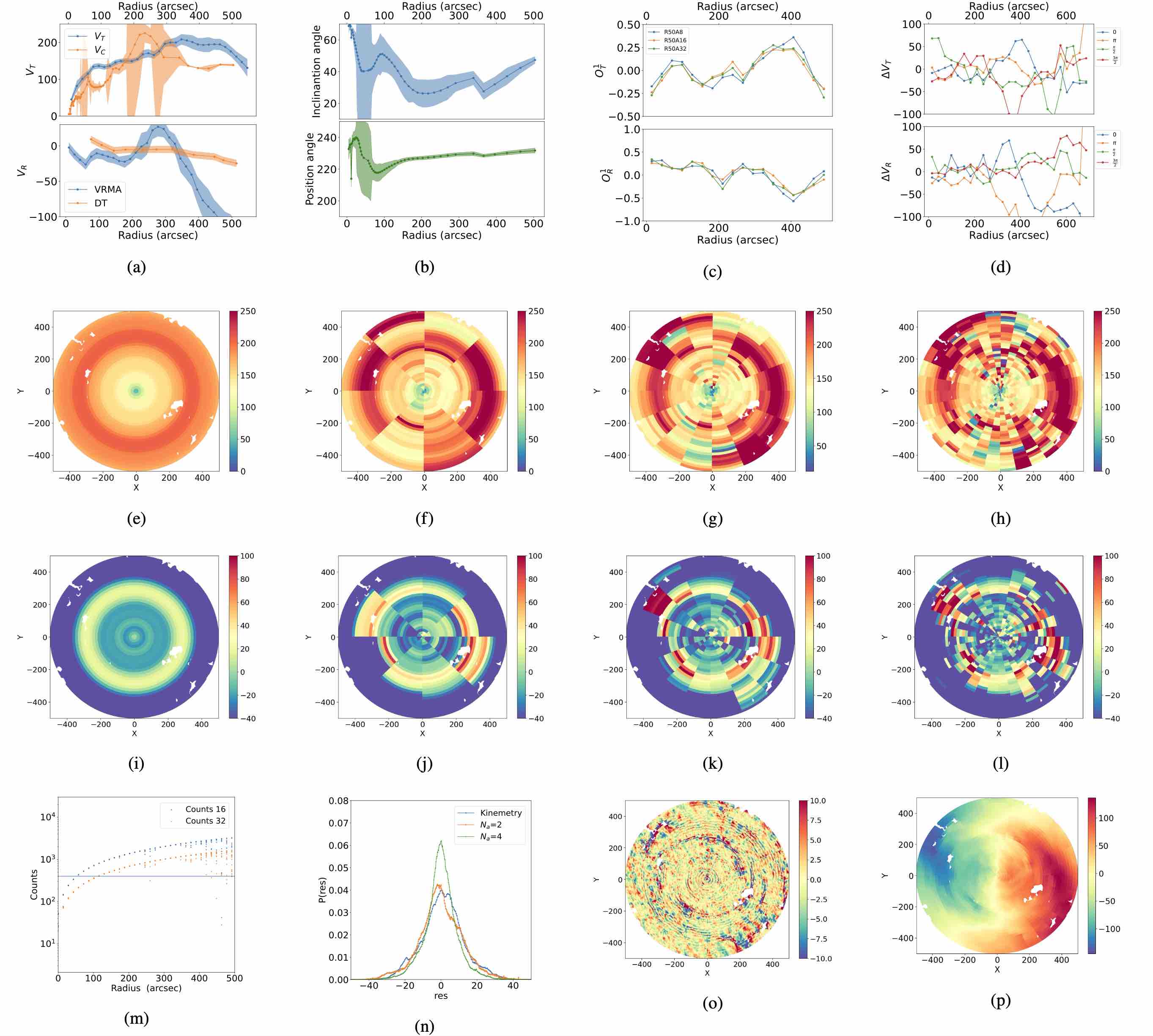}
\caption{Analysis of NGC 5236: symbols and lines are as  Fig.\ref{NGC628}.}
\label{NGC5236} 
\end{figure*}


\subsubsection*{NGC 5457}
{  NGC 5457 is a spiral galaxy}. The small inclination angle can cause large errors in the TRM analysis, as seen in the variations of the inclination angle and P.A.  with $10^\circ$ and $20^\circ$, respectively (as shown in Fig.\ref{NGC5457}).   As for the case of NGC 5194 we assume that the inner disk is co-planar: this hypothesis is consistent with the symmetric spiral structure observed and with the fact that $R_{25} \approx 720''$ whereas the analysis is extended to $\approx 1000''$.   The circular velocity is similar to the transversal velocity only at small radii, but there are significant differences in the outer disk where the inclination angle changes. The radial velocity remains small in amplitude, with values less than 20 km/s, across the entire disk. The octopole moments only approximately converge, displaying significant fluctuations. The 2D maps of $v_t(R)$ and $v_r(R)$ show large amplitude and coherent anisotropy patterns, particularly a large amplitude anisotropy pattern (with different signs) in the direction of $\theta \approx \pi/4$.
\begin{figure*}
 \includegraphics[width=\linewidth,angle=0]{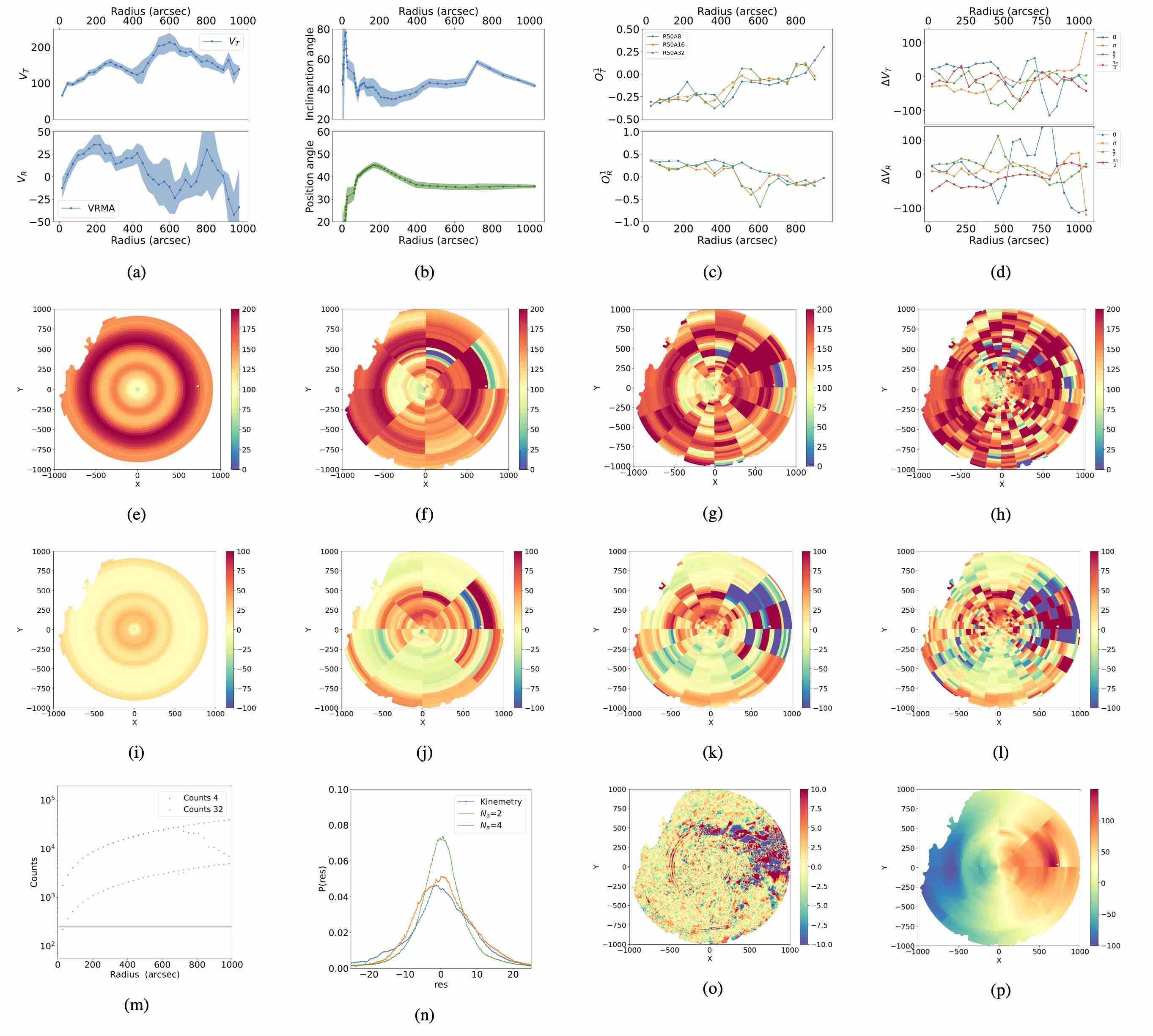}
\caption{Analysis of NGC 5457: symbols and lines are as  Fig.\ref{NGC628}.}
\label{NGC5457} 
\end{figure*}


\subsubsection*{NGC 6946}

{  NGC 6946 is an intermediate spiral galaxy}.  The inclination angle and the P.A. display moderate variations that are compatible with the flat disk hypothesis even beyond the optical radius $R_{25} \approx 340''$  (see Fig.\ref{NGC6946}).  We find that $v_t(R)$ has a behavior that is  similar to that of $v_c(R)$. The radial velocity has a similar behavior and amplitude to that reported by \cite{DiTeodoro+Peek_2021}. The octopole moments have converging behaviors, but the radial velocity shows significant fluctuations. The 2D transversal velocity map is rather quiet, while the radial one has coherent patterns of anisotropies of considerable amplitude in the galaxy's peripheries
which may be related to the spiral arms. 

\begin{figure*}
\includegraphics[width=\linewidth,angle=0]{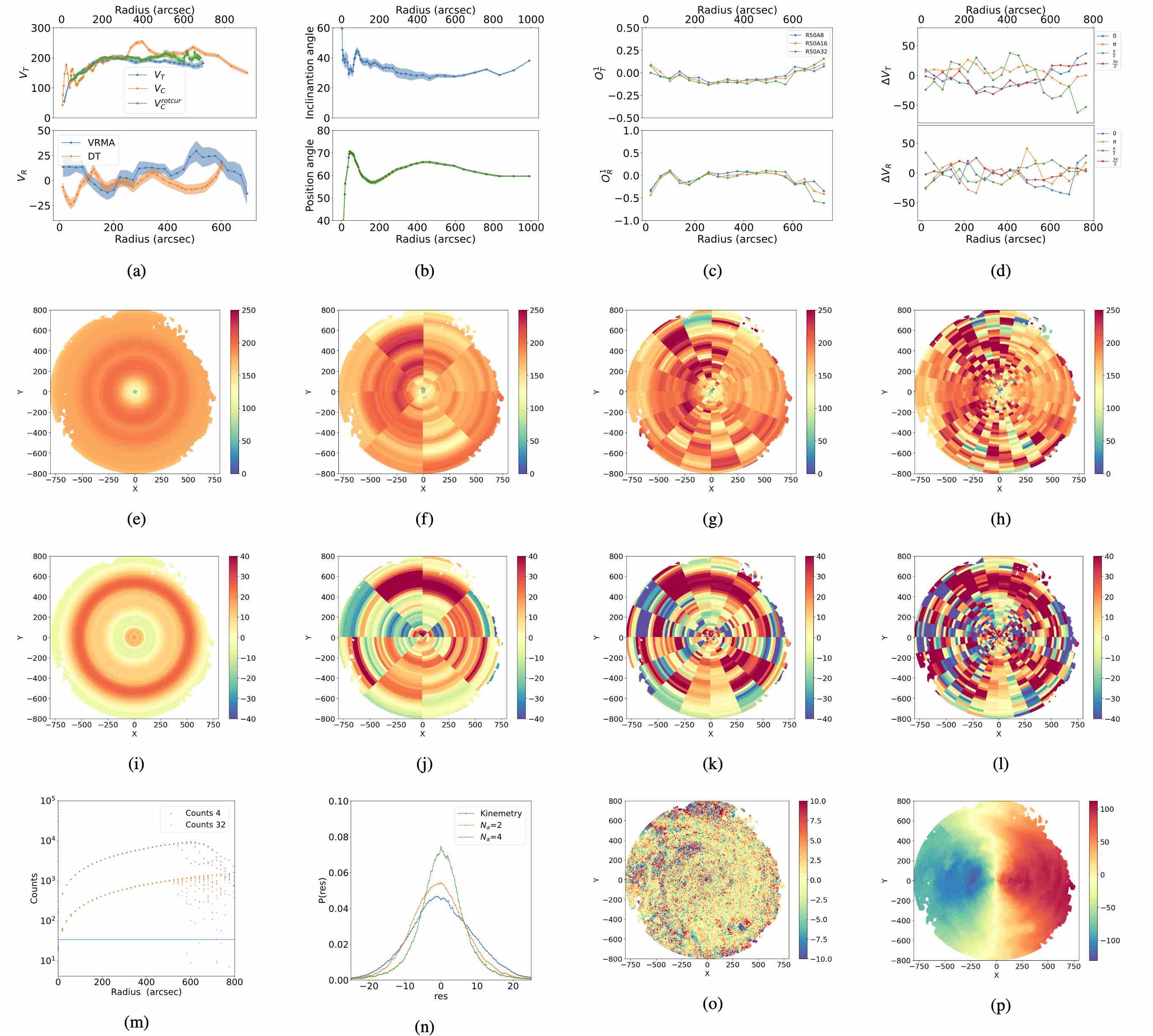}
\caption{Analysis of NGC 6946: symbols and lines are as  Fig.\ref{NGC628}.}
\label{NGC6946} 
\end{figure*}


\subsubsection*{NGC 7331}

{  NGC 7331 is an unbarred spiral galaxy.} The TRM detects an inclination angle $i(R)$ with a large variation of about $20^\circ$ in the middle of the disk. In correspondence to this, the P.A also displays a local change but of smaller amplitude, i.e. $10^\circ$ (see Fig.\ref{NGC7331}). Given these moderate variations, both inside and outside the optical radius $R_{25} \approx 270''$, we have that $v_t(R) \approx v_c(R)$. The radial velocity has an amplitude of $<20$ km s$^{-1}$ and agrees with the determination of \cite{DiTeodoro+Peek_2021} across the whole disk except for very small radii. The octopole moments converge well and show moderate variations across the disk, except for the outermost regions. Correspondingly, the 2D maps of the transversal and radial components show coherent anisotropy patterns which do not seem to break circular symmetry.
\begin{figure*}
\includegraphics[width=\linewidth,angle=0]{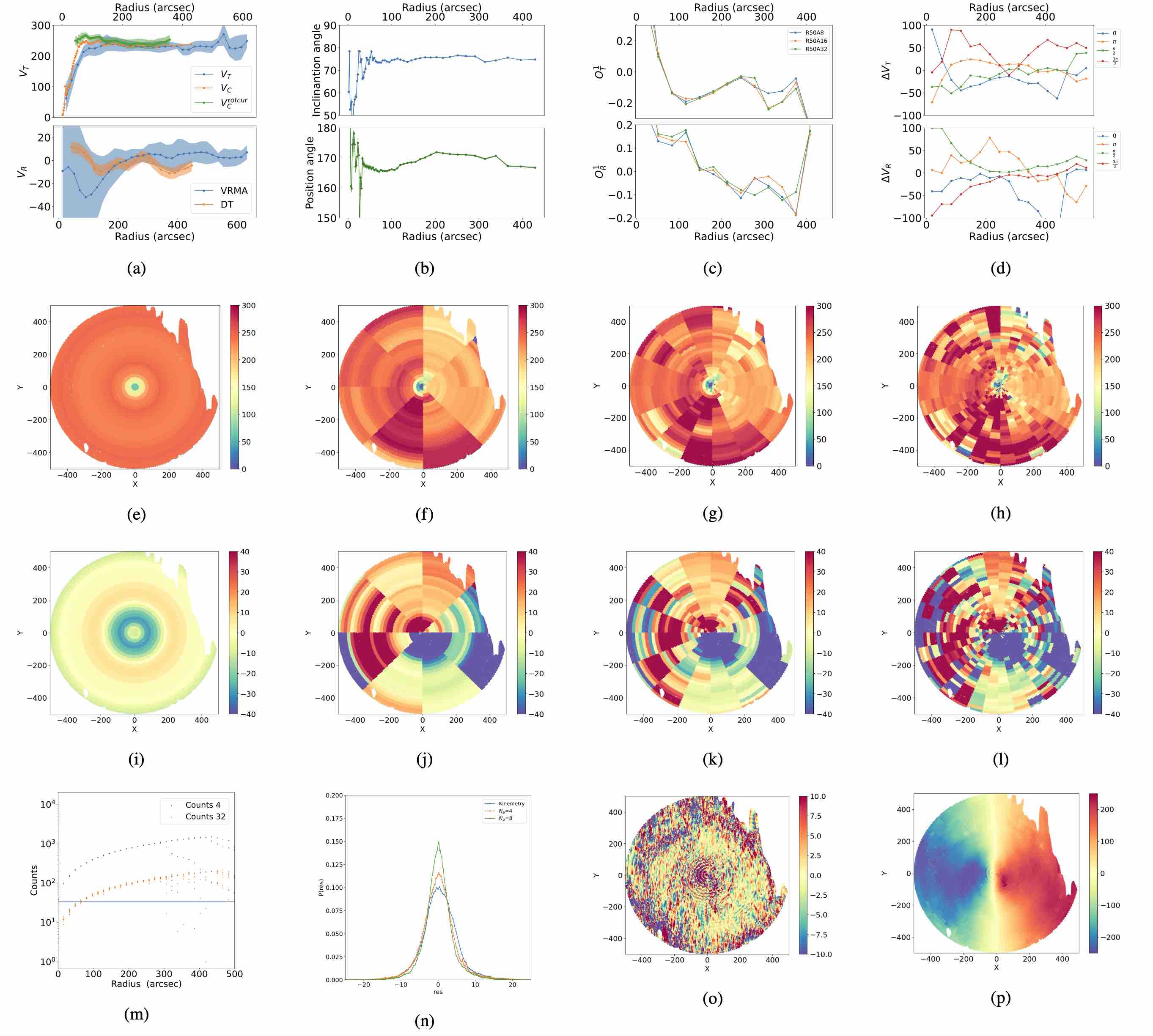}
\caption{Analysis of NGC 7331: symbols and lines are as  Fig.\ref{NGC628}.}
\label{NGC7331} 
\end{figure*}


\subsubsection*{NGC 7793}
{  NGC 7793 is a flocculent spiral galaxy}. The TRM analysis shows a monotonic decrease of the inclination angle by approximately $20^\circ$ from the inner to the outer disk (where $R_{25} \approx 315''$) and a corresponding monotonic increase in the P.A. by the same amount (as seen in Fig.\ref{NGC7793}) that may correspond to a warped geometry. As a result of these moderate changes, the transversal and circular velocities are similar, but only in the outermost region. The radial velocity is small in the inner region, with $v_r<10$ km s$^{-1}$, and is in agreement with the measurements of \cite{DiTeodoro+Peek_2021}. However, in the outermost regions, the radial velocity increases, which may be due to either a large radial velocity or a warp. The octopole moments converge well and have small fluctuations. This is reflected in the 2D maps of $v_t(R)$ and $v_r(R)$, which show coherent anisotropy patterns, particularly in the radial velocity.

\begin{figure*}
 \includegraphics[width=\linewidth,angle=0]{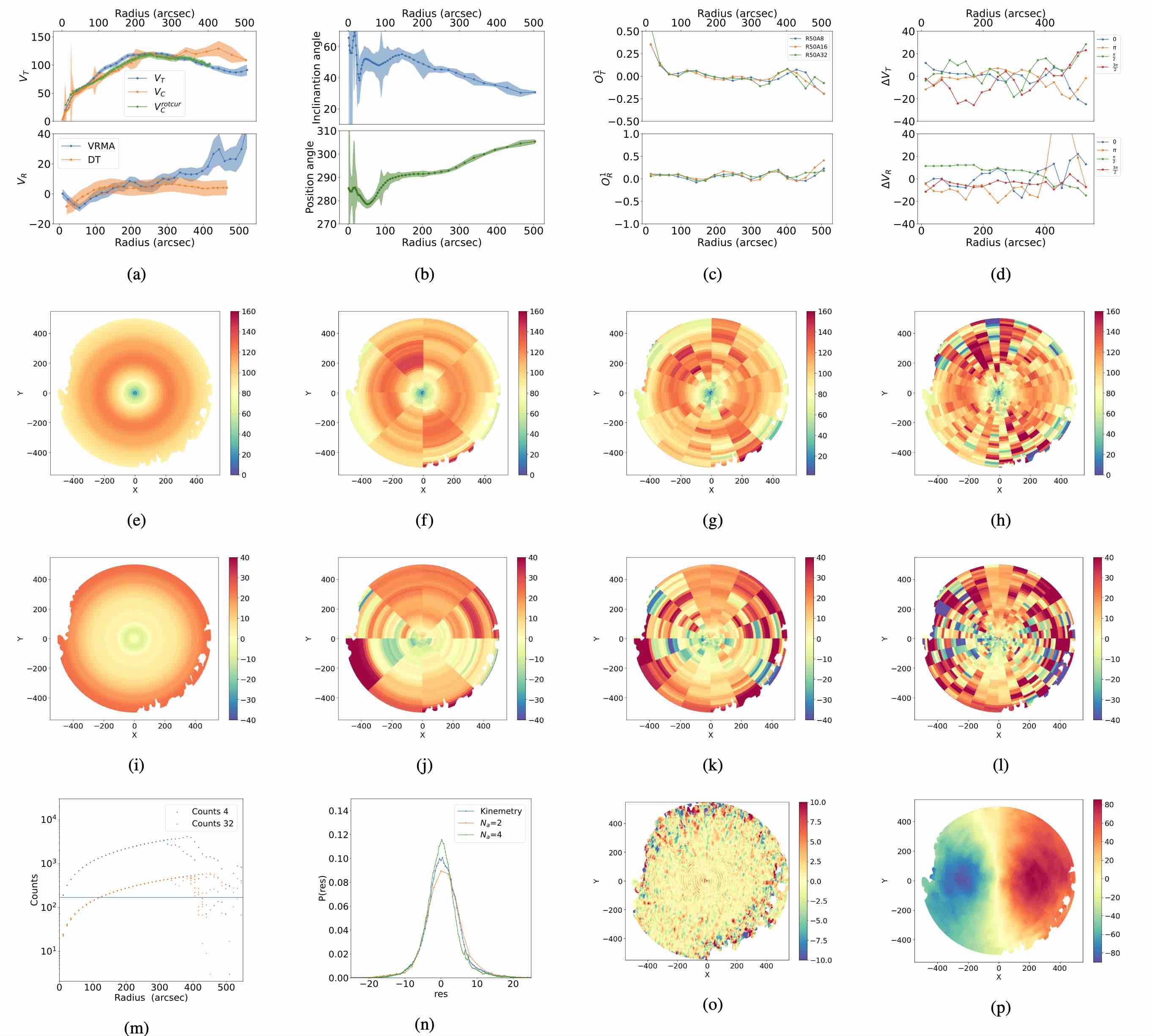}
\caption{Analysis of NGC 7793: symbols and lines are as  Fig.\ref{NGC628}.}
\label{NGC7793} 
\end{figure*}


\subsubsection*{DDO 154}
 
{  The TRM analysis shows a clear discontinuity in the outer disk at $R \approx 2 R_{25} \approx 200''$, with a large variation of $\approx 30^\circ$ in the inclination angle.} The P.A. is relatively constant, but with some fluctuations in the very inner regions (as seen in Fig.\ref{DDO154}). As a result of this, the transversal velocity ($v_t$) is similar to the circular velocity ($v_c$) with a slight difference in amplitude due to the varying inclination angle in the outer disk. The radial velocity is of the order of a few km s$^{-1}$:   this is consistent with a nearly constant P.A. . The octopole moments converge and present some significant fluctuations in the outermost regions. The 2D maps of $v_t(R)$ and $v_r(R)$ reveal a quiet velocity field with some coherent patterns.

\begin{figure*}
 \includegraphics[width=\linewidth,angle=0]{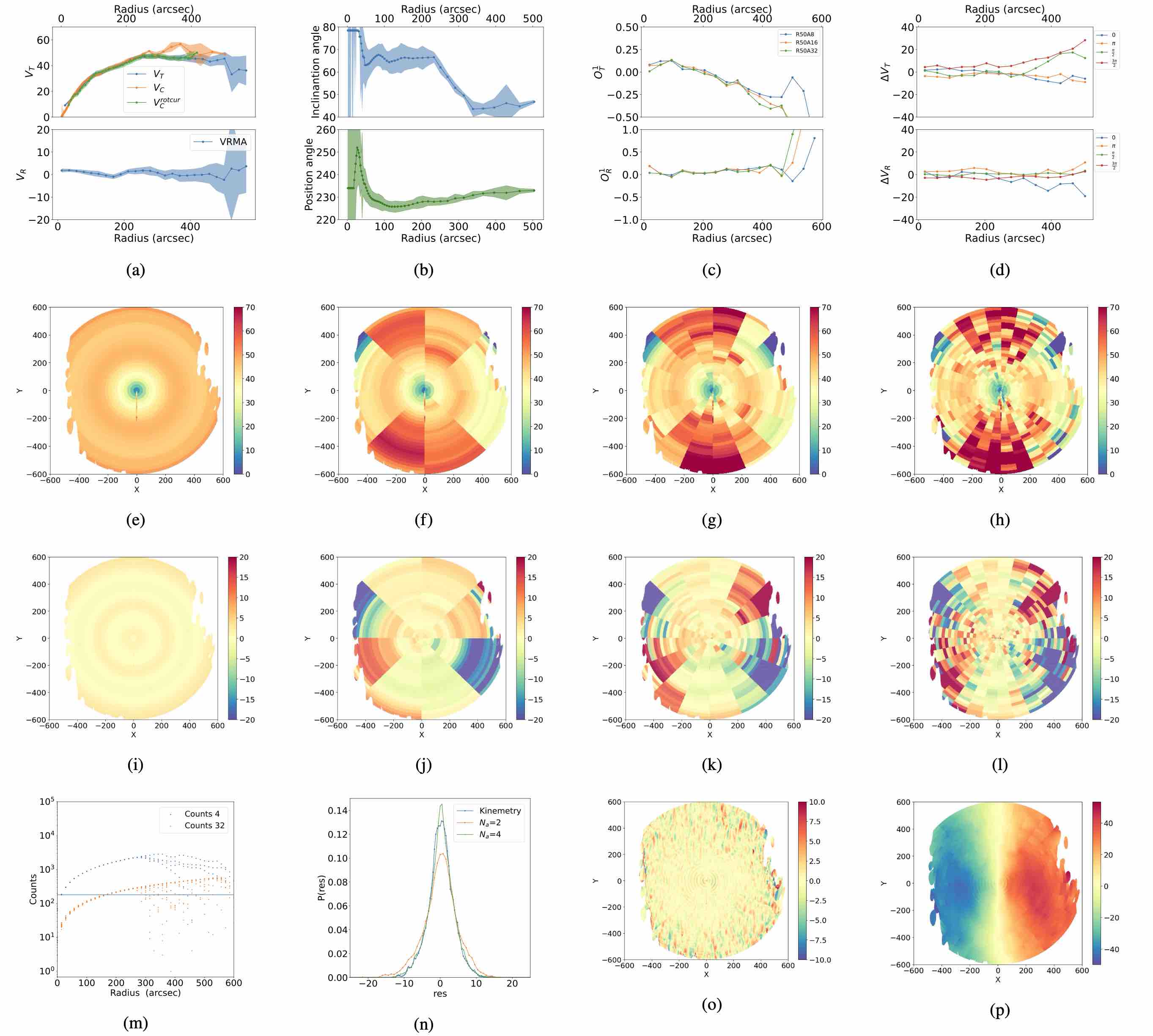}
\caption{Analysis of DDO 154: symbols and lines are as  Fig.\ref{NGC628}.}
\label{DDO154} 
\end{figure*}

\end{document}